\documentclass[fleqn,usenatbib]{mnras}

\usepackage{newtxtext,newtxmath}
\usepackage[T1]{fontenc}

\DeclareRobustCommand{\VAN}[3]{#2}
\let\VANthebibliography\thebibliography
\def\thebibliography{\DeclareRobustCommand{\VAN}[3]{##3}\VANthebibliography}

\usepackage{graphicx}	% Including figure files
\usepackage{amsmath}	% Advanced maths commands

\usepackage{threeparttable}
\usepackage{pdflscape}
\usepackage{xcolor}

\newcommand{\RNum}[1]{\uppercase\expandafter{\romannumeral #1\relax}}

\title[Supernova Host Galaxies]{Linking Transients to their Host Galaxies: 
II. A Comparison of Host Galaxy Properties and Rate Dependencies across Supernova Types}
\author[Y.-J. Qin et al.]{
Yu-Jing Qin,$^{1}$\thanks{E-mail: yujingq@caltech.edu}
and Ann Zabludoff$^{2}$
\\
$^{1}$California Institute of Technology, 1200 E California Blvd, Caltech MC 249-17, Pasadena, CA 91125\\
$^{2}$Department of Astronomy and Steward Observatory, University of Arizona, 933 N Cherry Ave, Tucson, AZ 85721
}

\date{Accepted XXX. Received YYY; in original form ZZZ}

\pubyear{2024}

\begin{document}
\label{firstpage}
\pagerange{\pageref{firstpage}--\pageref{lastpage}}
\maketitle

% Abstract of the paper
\begin{abstract}
We use the latest dataset of supernova (SN) host galaxies to investigate how the host properties -- stellar mass, star formation rate, metallicity, absolute magnitude, and colour -- differ across SN types, with redshift-driven selection effects controlled.
SN Ib and Ic host galaxies, on average, are more massive, metal-rich, and redder than SN II hosts.
For subtypes, SN Ibn and Ic-BL have bluer hosts than their normal SN Ib and Ic siblings;
SN IIb has consistent host properties with SN Ib, while hosts of SN IIn are more metal-rich than those of SN II.
Hydrogen-deficient superluminous supernovae feature bluer and lower luminosity hosts than most subtypes of core-collapse supernova (CC SN).
Assuming simple proportionality of CC SN rates and host star formation rates (SFRs) does not recover the observed mean host properties;
either a population of long-lived progenitors or a metallicity-dependent SN production efficiency better reproduces the observed host properties.
Assuming the latter case, the rates of SN II are insensitive to host metallicity, but the rates of SN Ib and Ic are substantially enhanced in metal-rich hosts by a factor of $\sim10$ per dex increase in metallicity.
Hosts of SN Ia are diverse in their observed properties;
subtypes including SN Ia-91T, Ia-02cx, and Ia-CSM prefer star-forming hosts, while subtypes like SN Ia-91bg and Ca-rich prefer quiescent hosts.
The rates of SN Ia-91T, Ia-02cx, and Ia-CSM are closely dependent on, or even proportional to, their host SFRs, indicating relatively short-lived progenitors.
Conversely, the rates of SN Ia-91bg and Ca-rich transients are proportional to the total stellar mass, favoring long-lived progenitors.
\end{abstract}

% Select between one and six entries from the list of approved keywords.
% Don't make up new ones.

%%%%%%%%%%%%%%%%%%%%%%%%%%%%%%%%%%%%%%%%%%%%%%%%%%

%%%%%%%%%%%%%%%%% BODY OF PAPER %%%%%%%%%%%%%%%%%%

\section{Introduction}

The cataclysmic deaths of stars are observed as supernovae (SNe).
The diverse spectroscopic and photometric signatures of SNe indicate a wide variety of potential progenitor stars.
Connecting the observed SN subtypes to their progenitors is a pivotal task in the study of these stellar explosions.

Constraints on or direct detections of progenitors in archival, pre-explosion images provide the most compelling evidence of their properties.
Core-collapse supernovae (CC SNe) are supposedly the explosions of stars with zero-age main sequence (ZAMS) mass above $\simeq8\,\mathrm{M}_{\odot}$.
Despite the similar physical mechanism, due to the different structures and evolution pathways of their progenitors, CC SNe exhibit heterogeneous photometric and spectroscopic signatures, by which they are further classified into subtypes.
Only a few dozens of nearby CC SNe have possible progenitors identified or constrained using pre-explosion \textit{HST} and occasionally ground-based images so far \citep{Smartt09ARAA, Smartt15, VanDyk17}.
Except for the indicative progenitor detections of SN Ib \citep{Cao13bvn, Kilpatrick21yvr}, IIb \citep{Crockett08, Maund11, VanDyk11, VanDyk14}, and IIn \citep{GalYam07_05gl, Smith11_10jl}, the only CC SN subtype with definitive progenitor association is SN II-P with red supergiants \citep{Smartt09IIP}.

Type Ia supernovae (SNe Ia), on the contrary, are presumably the thermonuclear explosions of white dwarfs (WDs) near the Chandrasekhar mass limit ($M_{\text{Ch}}$) and are thus homogeneous in the observed properties. 
Their photometric and spectroscopic signatures, albeit similar, outline several subgroups featuring extreme behaviors or properties that clearly distinguish themselves from the majority of normal SN Ia \citep{Taubenberger17}.
To reach $M_{\text{Ch}}$, the WD progenitor needs to accrete mass from a donor star -- either a non-degenerate companion or another WD \citep{Maoz14ARAA}.
To date, there is no direct progenitor detection for normal SN Ia; the most rigorous upper limit of progenitor luminosity comes from SN 2011fe \citep{Li11_11fe}, which excludes red giant as the donor.
Among SN Ia subtypes, only SN Ia-02cx\footnote{We use abbreviated names (e.g., Ia-91T, Ia-91b, and Ia-02cx) for subtypes named after prototype events through this paper.} (or Iax) has one progenitor detection \citep[SN 2012Z;][]{McCully14}, suggesting a helium star donor.

Most SNe occur at cosmological distances where individual progenitors are beyond the resolution or sensitivity limits of space telescopes.
Besides direct photometric and spectroscopic observations of SNe themselves, evidence of SN progenitors is only available indirectly from host galaxy properties, representing the stellar populations from which SNe arise.

For example, hosts of stripped-envelope supernovae (SE SNe) are, on average, more massive, luminous, and metal-rich than SN II hosts \citep{Arcavi10, Li11_frac, Graur17_relrates, Schulze21}; the rate ratio of SE SN to SN II also increases with host stellar mass and metallicity \citep{Li11_rates, Graur17_rates}.
SN Ic-BL hosts feature lower stellar mass and metallicity than SN Ic hosts \citep{Arcavi10, Sanders12, Modjaz20, Schulze21}, indicating different progenitors of these two sibling subtypes.
Hydrogen-deficient superluminous supernovae (SLSN-I) almost exclusively occur in star-forming, low-metallicity dwarf galaxies \citep{Neill11, Leloudas15, Perley16, Schulze18}, hinting at unique or extreme progenitors among CC SN subtypes.

SNe Ia have a broad distribution of delay times (i.e., the time from progenitor star formation to SN explosion) because their WD progenitors are the end products of low-mass star evolution, which need further time to accrete mass before reaching the condition of a thermonuclear explosion.
The wide range of delay times ($50$ Myr to several Gyrs; \citealt{Maoz12Review, Maoz14ARAA}), coupled with the evolution of host galaxies themselves, lead to more heterogeneous host galaxy properties than CC SNe.
SNe Ia-91T are predominantly hosted by star-forming galaxies \citep{Howell01, Li11_frac}, so are those under-luminous and likely pure-deflagration SNe Ia-02cx \citep{Jha17Iax, McCully14};
conversely, SNe Ia-91bg are preferably observed in massive and quiescent galaxies \citep{Neill09, Howell01, Li11_frac}.
Besides those subtypes, the light curve parameters of normal SNe Ia also correlate with host properties; brighter and slowly-fading ones occur more commonly in late-type hosts \citep[e.g.,][]{Lampeitl10, Sullivan10, Childress13_host}.

The examples above are certainly not a comprehensive summary of systematic differences in host properties across SN subtypes; other subtle differences in host properties can be revealed with larger datasets using advanced techniques \citep[e.g.,][]{Kisley22}.

Limited by the sample sizes and selection effects of SNe and galaxies, characterizing and further interpreting the connection of SN to host galaxy properties could be challenging.
In this paper, we use our database of transient host galaxies \citep{Qin22} and new statistical techniques to explore the systematic differences of host properties across SN subtypes.
In Section \ref{sec:Data}, we present our SN host galaxy sample and the derivation of host photometric properties.
In Section \ref{sec:StatMethod}, we present our technique to control for selection biases and characterize the differences of host properties across SN types;
In Section \ref{sec:ResultsAndDiscussion}, we present our results and discuss the physical implications.
Finally, we summarize in Section \ref{sec:Conclusions}.
Details of feature derivation and statistical methods are summarized in the Appendices.
Throughout the paper, unless noted otherwise, we use a flat $\Lambda$CDM cosmology with $\Omega_\text{M}=0.3$ and $H_0=100h$ $\text{km}\,\text{s}^{-1}\,\text{Mpc}^{-1}$.
Magnitudes are given in the AB Magnitude System.
Sky coordinates are in the J2000 equinox.

\section{Supernova and Host Data} \label{sec:Data}

The sample of SN host galaxies we analyze here is based on the value-added database in \citet{Qin22} (hereafter `database paper'). Here, we briefly outline the construction of the dataset and then describe the sample selection criteria.
For the analysis here, we further derive some ancillary host properties which are not part of the original data release in our database paper. These derived data are provided as online electronic tables in this paper.

\begin{figure*}
\includegraphics[width=\linewidth]{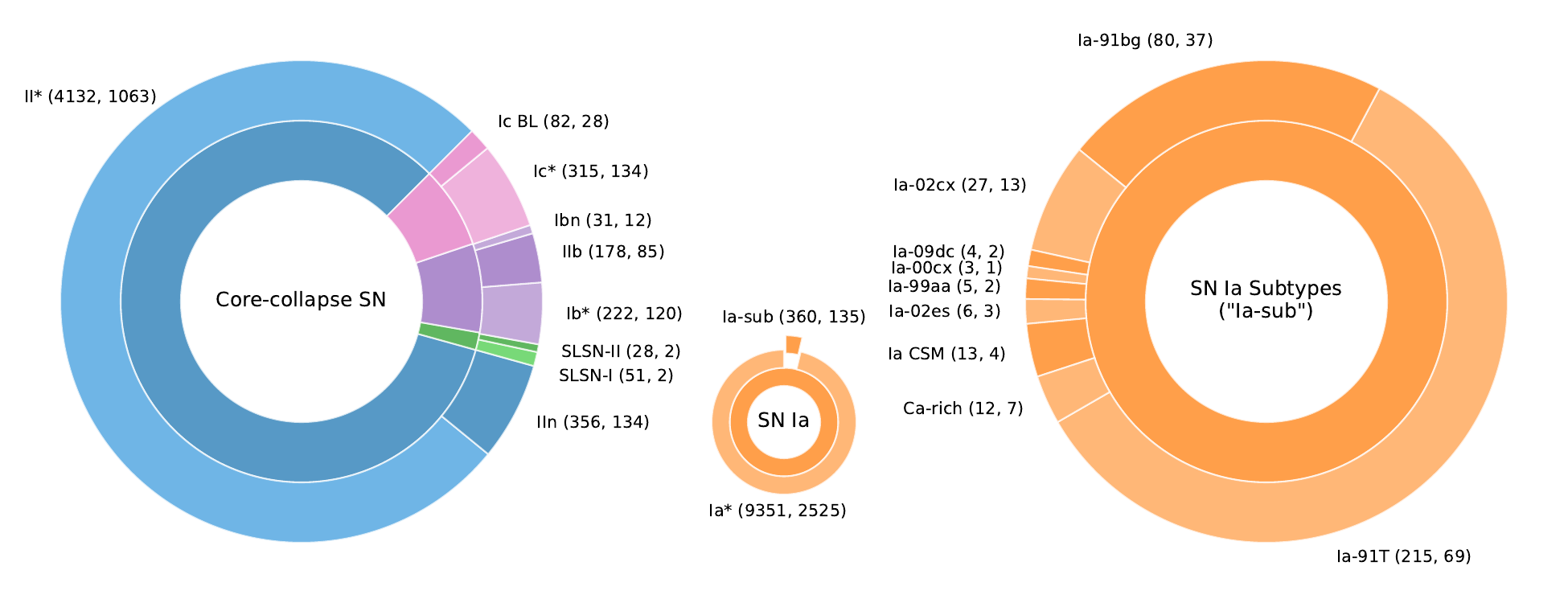}
\caption{\label{fig:TypePieChart}
Breakdown of SN subtypes in our sample. The numbers of SNe in the larger photometric and smaller spectroscopic samples are indicated in parentheses.
The left and right panels show CC SN subtypes and subtypes of SN Ia (`Ia-sub'), respectively. The smaller middle panel compares the sample sizes of `normal' SN Ia and subtypes of SN Ia.
See Sections \ref{sec:Subtypes} and \ref{sec:SubtypeNotes} for a discussion of these subtypes.
}
\end{figure*}

\subsection{Sample selection}

We constructed a database of extragalactic transients previously detected in the literature from various surveys (until June 1, 2021) and compiled their host galaxy properties across survey catalogues and value-added databases.
We use the galaxy names or coordinates reported in upstream data sources to match known hosts. For transients without host galaxies reported in those data sources, we use our machine learning-based host candidate ranking algorithm, trained with known transient-host pairs, to identify potential hosts in survey catalogues and databases.
We further visually inspected our host galaxies to assess the performance of our methods and to correct occasionally mismatched or misidentified hosts.
Our database contains $36333$ unique transients, where $18233$ have known host galaxies in upstream data sources, and $18100$ have new host galaxies identified by us.
For the analysis of this work, we select a subset of SNe based on their subtypes, robustness of host association, and availability of host properties.
As a starting point, we select transients that have been classified into any subtype of SNe based on their spectra (Table 2 in the database paper; see also Section \ref{sec:Subtypes}). SNe with only photometric, light curve-based classification are excluded by their reference sources (Table 7 in the database paper).
This work does not analyze the host galaxy properties of other non-SN transients, such as tidal disruption events and fast radio bursts.

We visually inspected the host galaxy association in our database and re-assigned a minor fraction of obviously mismatched or misidentified hosts (Appendix J in the database paper).
For this work, we selected a subset of records based on the robustness of host galaxy association.
We only choose (1) hosts that are cross-matched using known galaxy names or coordinates, that either passed our inspection (Case `A1'), or have been manually re-assigned due to obvious mismatches (Case `B1'),
and (2) hosts that are identified using our machine learning-based algorithm, that either passed our inspection (Case `F1') or have been manually re-assigned if obviously misidentified (Case `G1').
Records of other cases, such as potentially hostless transients and ambiguous hosts, are not included.
The sample analyzed in this work combines SNe detected by multiple surveys, with an overall selection function that cannot be easily characterized. Assuming that redshift is the primary biasing factor to be controlled for in this work (Section \ref{sec:StatMethod}), we further exclude records without any redshift values (either from SN or host) in the database.
We focus on host galaxy properties based on Sloan Digital Sky Survey (SDSS) photometric and spectroscopic catalogues \citep{York00, Strauss02}. Hosts without SDSS photometric or spectroscopic targets cross-matched are excluded.

The selection criteria here resulted in $12943$ records. We summarize the selected SN-host pairs in Table \ref{tab:sourcetable}.
% 

% raw MNRAS table

\begin{table*}
\caption{List of supernovae and host galaxies.}
\begin{threeparttable}
\label{tab:sourcetable}
\begin{tabular}{llllcllr}
\hline
Name & R.A. & Decl. & $z$ & Type & Host R.A. & Host Decl. & $E(B-V)$ \\
\hline
SN 2021jzw  & 13:15:24.85 & +01:07:43.2 & 0.047 & Ia*    & 13:15:24.84 & +01:07:43.1 & $0.035$ \\
SN 2012by   & 13:15:28.90 & +62:07:47.8 & 0.031 & II*    & 13:15:30.59 & +62:07:45.3 & $0.022$ \\
SN 2020gzy  & 13:15:31.40 & +62:07:44.8 & 0.031 & II*    & 13:15:30.59 & +62:07:45.3 & $0.022$ \\
% ASASSN-15lg & 13:15:32.88 & +03:30:02.3 & 0.020 & Ia*    & 13:15:32.70 & +03:30:02.8 & $0.031$ \\
SN 2018cvu  & 13:15:34.32 & +28:20:34.5 & 0.102 & Ia*    & 13:15:34.06 & +28:20:29.4 & $0.011$ \\
SN 2019ecw  & 13:15:46.78 & +33:31:16.3 & 0.071 & Ia-91T & 13:15:46.86 & +33:31:16.5 & $0.012$ \\
SN 2021apn  & 13:15:52.00 & +15:51:14.8 & 0.084 & Ia*    & 13:15:51.95 & +15:51:16.2 & $0.026$ \\
% SN 2020abcp & 13:15:54.86 & +08:01:39.4 & 0.025 & Ia*    & 13:15:54.78 & +08:01:37.7 & $0.028$ \\
% SN 2020gam  & 13:16:03.55 & +32:15:14.8 & 0.066 & II*    & 13:16:03.49 & +32:15:19.9 & $0.009$ \\
SN 2019boh  & 13:16:06.14 & +26:40:09.5 & 0.034 & II*    & 13:16:06.17 & +26:40:03.4 & $0.012$ \\
% SN 2019avx  & 13:16:16.24 & +62:32:51.5 & 0.041 & Ia*    & 13:16:15.43 & +62:32:53.3 & $0.019$ \\
SN 2016jdw  & 13:16:19.62 & +30:40:32.7 & 0.019 & Ib*    & 13:16:20.54 & +30:40:42.0 & $0.012$ \\
% SN 2019uwd  & 13:16:20.74 & +30:40:48.8 & 0.019 & II*    & 13:16:20.54 & +30:40:42.0 & $0.012$ \\
% SN 2017deu  & 13:16:21.59 & +27:41:35.3 & 0.050 & Ia*    & 13:16:21.64 & +27:41:37.2 & $0.011$ \\
SN 2021bgd  & 13:16:33.52 & +31:32:51.6 & 0.025 & II*    & 13:16:33.36 & +31:32:47.1 & $0.011$ \\
% SN 2019cem  & 13:16:36.30 & +35:30:57.8 & 0.045 & II*    & 13:16:36.39 & +35:30:55.9 & $0.010$ \\
% PTF11bas    & 13:16:47.78 & +43:31:13.4 & 0.086 & Ia*    & 13:16:47.79 & +43:31:13.3 & $0.018$ \\
SN 2021jmg  & 13:16:51.93 & +14:24:54.2 & 0.059 & Ic*    & 13:16:51.74 & +14:24:55.3 & $0.027$ \\
% SN 2018baf  & 13:17:01.58 & -02:16:36.8 & 0.120 & Ia*    & 13:17:01.56 & -02:16:36.8 & $0.030$ \\
% SN 2019ewu  & 13:17:12.26 & +53:54:57.4 & 0.028 & II*    & 22:06:33.66 & -08:02:06.7 & $0.035$ \\
% SN 2018fhd  & 13:17:26.43 & +44:56:50.5 & 0.091 & II*    & 13:17:26.42 & +44:56:50.7 & $0.023$ \\
% SN 2017dia  & 13:17:28.85 & +03:02:54.2 & 0.111 & Ia*    & 13:17:28.72 & +03:02:53.0 & $0.033$ \\
% SN 2013bo   & 13:17:29.19 & +42:44:29.6 & 0.036 & Ia*    & 13:17:29.23 & +42:44:29.7 & $0.015$ \\
% PTF11dzm    & 13:18:03.69 & +42:10:34.1 & 0.041 & Ia*    & 13:18:03.93 & +42:10:34.0 & $0.012$ \\
% SN 2018bfx  & 13:18:06.63 & +27:52:24.2 & 0.075 & Ia*    & 13:18:06.63 & +27:52:23.4 & $0.015$ \\
% PTF12dwm    & 13:18:17.76 & +35:42:25.7 & 0.053 & Ia*    & 13:18:17.76 & +35:42:25.8 & $0.011$ \\
% SN 2020drt  & 13:18:21.65 & +55:50:48.3 & 0.080 & Ia*    & 13:18:21.87 & +55:50:46.5 & $0.019$ \\
% SN 2016cld  & 13:18:22.88 & +31:35:16.6 & 0.211 & Ia*    & 13:18:22.81 & +31:35:16.8 & $0.011$ \\
% SN 2019cyz  & 13:18:40.11 & +01:54:51.0 & 0.051 & Ia*    & 13:18:40.06 & +01:54:50.9 & $0.028$ \\
% SN 2018adf  & 13:18:42.77 & -00:45:33.0 & 0.113 & Ia*    & 13:18:42.78 & -00:45:33.9 & $0.029$ \\
% SN 2020exr  & 13:18:45.03 & +50:37:26.8 & 0.090 & Ia*    & 13:18:45.02 & +50:37:26.9 & $0.009$ \\
% SN 2020flg  & 13:18:56.18 & +01:13:38.6 & 0.119 & Ia*    & 13:18:56.19 & +01:13:38.6 & $0.028$ \\
% iPTF15aps   & 13:18:56.24 & +48:25:56.1 & 0.119 & Ia*    & 13:18:56.51 & +48:25:56.8 & $0.010$ \\
% SN 2019abk  & 13:18:59.68 & +32:58:28.7 & 0.038 & Ia*    & 13:18:59.51 & +32:58:26.3 & $0.015$ \\
\hline
\end{tabular}
\begin{tablenotes}
% \small 
\item Columns: \textit{Name}: SN name; \textit{R.A.} and \textit{Decl.}: SN coordinate; $z$: SN redshift as reported in our database paper;
\textit{Type}: SN type, where type with an asterisk indicates that the event is not further classified as a subtype of the indicated type;
\textit{Host R.A.} and \textit{Host Decl.}: best host galaxy coordinate in our database paper;
$E(B-V)$: Galactic foreground reddening at the host position, based on \citet{Schlegel98} and \citet{Schlafly11}. \\
This table is available in its entirety in machine-readable form.
\end{tablenotes}
\end{threeparttable}
\end{table*}

% SN2005au & 13:16:12.42 & +30:56:40.5 & 0.019 & II* & 13:16:12.32 & +30:57:01.2 & $0.012$ \\
% SN2004bj & 13:16:18.03 & +07:02:52.7 & 0.050 & Ia* & 13:16:17.02 & +07:02:46.6 & $0.031$ \\
% SN2004E & 13:16:51.40 & +31:35:11.8 & 0.030 & Ia* & 13:16:51.16 & +31:34:52.4 & $0.015$ \\
% SN2001F & 13:17:20.05 & +20:38:09.8 & 0.023 & Ia* & 13:17:19.79 & +20:38:17.1 & $0.028$ \\
% AT2013ka & 13:18:17.71 & -02:51:43.9 & 0.085 & II* & 13:18:18.08 & -02:51:29.6 & $0.029$ \\
% SN2019eoe & 13:19:02.08 & +45:01:34.9 & 0.060 & II* & 13:19:02.10 & +45:01:35.1 & $0.025$ \\
% SN2017faa & 13:19:03.90 & -02:30:45.8 & 0.018 & II* & 13:19:05.04 & -02:30:55.1 & $0.033$ \\
% SN2007bi & 13:19:20.19 & +08:55:44.3 & 0.128 & Ic*, SLSN & 13:19:20.19 & +08:55:44.3 & $0.027$ \\
% SN2019dfv & 13:19:30.34 & -00:59:32.1 & 0.090 & Ia* & 13:19:30.32 & -00:59:31.7 & $0.025$ \\
% PS1-14gj & 13:19:32.99 & +16:38:40.8 & 0.023 & Ia* & 13:19:32.78 & +16:38:40.3 & $0.021$ \\
% SN2019wzd & 13:19:39.23 & +05:06:30.7 & 0.074 & Ia* & 13:19:39.17 & +05:06:31.6 & $0.034$ \\
% PTF10hml & 13:19:49.70 & +41:59:01.6 & 0.053 & Ia* & 13:19:50.65 & +41:58:57.2 & $0.014$ \\
% SN2021kjh & 13:19:52.01 & +07:45:38.1 & 0.049 & Ia* & 13:19:51.87 & +07:45:37.0 & $0.027$ \\
% SN2021coe & 13:19:54.19 & +00:30:37.8 & 0.054 & Ia* & 13:19:54.16 & +00:30:43.0 & $0.033$ \\
% \tablecomments{}

%
\subsection{Supernova subtypes} \label{sec:Subtypes}

SNe are classified into subtypes\footnote{Major SN types and SN subgroups are collectively referred to as `subtypes' in this work. See Table 2 in the database paper for the hierarchy of SN subtypes.} based on their photometric or spectroscopic signatures. We focus on familiar and well-defined subtypes in the literature, as described below. SNe beyond those subtypes are excluded from our analysis.

We include CC SN and SN Ia at the top level. CC SN includes hydrogen-deficient SN Ib and Ic, which are collectively referred to as SE SN, as well as hydrogen-rich SN II. (See \citealt{Filippenko97} for a review of these SN types.)
SLSN, including hydrogen-deficient (SLSN-I) and hydrogen-rich (SLSN-II) subtypes \citep{GalYam12SLSNReview, GalYam19ARAA}, is also considered a CC SN subtype.
Note that SLSN-I and SLSN-II are defined by peak luminosity \textit{and} spectroscopic signature and may thus overlap with other CC SN subtypes by definition. To compare the host properties of SLSNe and non-SLSN, `normal' CC SNe, in this work, we do not consider SLSN-I as a subtype of SE SN or SN Ic, nor do we list SLSN-II as a subtype of SN II or IIn.

We also include a variety of SN subgroups, namely subsets of events under a major SN type with specific signatures or extreme properties.
CC SN with interacting signatures (see \citealt{Smith17review} for a review), mainly SN IIn \citep[e.g.,][]{Schlegel90, GalYam07_05gl, Smith07_06gy} and Ibn \citep[e.g.,][]{Foley07, Pastorello07}, are included as subtypes of SN II and Ib, respectively.
Shock-excited narrow emission lines in their circumstellar material (CSM) make these subtypes distinct from their parent types.\footnote{
Recently, SN Ic with CSM interaction signatures was discovered and designated as Type Icn \citep{GalYam22Icn, Perley22}.
We exclude SN Icn due to the limited sample size and the fact that the database we use was constructed before the discovery.
Notably, there is evidence for a non-homogeneous host galaxy population \citep{Pellegrino22}.}
We also include SN Ia with CSM-interacting signatures \citep[SN Ia-CSM;][]{Chugai04_csm, Wang04_02ic, Ofek07_06gy, Aldering06_05gj, Silverman13}.
However, there might be confusion in those interacting subtypes due to the challenges in distinguishing the true SN type veiled by the shock-excited emission (e.g., \citealt{Smith17review}) and contaminated by host galaxy emission lines \citep{Ransome21}.
Besides interacting subtypes, we also include SN Ic-BL, i.e., SN Ic with broad spectral features related to high-velocity outflows -- a subtype commonly associated with gamma-ray bursts and X-ray flashes \citep{Galama98_98bw, Kulkarni98, Hjorth03, Stanek03, Woosley06ARAA}.
Furthermore, we included SN IIb, an intermediate subtype whose spectral evolution resembles a transition from SN II to SN Ib \citep{Filippenko94}. 
Because progenitors have lost part of their hydrogen envelopes, we consider SN IIb as a SE SN subtype under SN Ib rather than SN II \citep{Shivvers17}

Based on the post-maximum light curve shape, SN II is further classified into two subtypes, namely II-P (`plateau') and II-L (`linear'). SN II-P is more common than II-L by frequency \citep[e.g.,][]{Li11_frac, Shivvers17}.
However, recent observations and theoretical models hint at a continuum, rather than a dichotomy, of light curve properties and progenitor models \citep[e.g.,][]{Anderson14, Valenti16, Morozova17}. Meanwhile, many SNe in transient circulars are classified into Type II-P or II-L by matching spectral templates  \citep[e.g.,][]{Howell05Superfit, Blondin07, Harutyunyan08} rather than using post-maximum light curves as the two subtypes are defined.
Therefore, we consider SN II-P and II-L collectively as SN II without further distinguishing these two subtypes.

Thermonuclear SNe are homogeneous in the observed properties, yet, their photometric and spectroscopic signatures outline several subtypes.
We discuss the host galaxy properties for the subtypes summarized in the review of \citet{Taubenberger17}.
We include SN Ia-91T, an overluminous subtype featuring prominent Fe \RNum{3} lines in the pre-maximum spectrum \citep{Filippenko92, Phillips92}, as well as SN Ia-91bg, a subluminous subtype with low ionization temperatures \citep{Leibundgut93, Turatto96}.
Besides these two extreme subtypes, we also include subluminous and slowly-fading SN Ia-02cx \citep[or Iax;][]{Li03, Jha06, Foley13Iax, McCully14}, in which a deflagration does not disintegrate the entire WD \citep{Jha17Iax}.
SN Ia-91T, Ia-91bg, and Ia-02cx are included in most spectroscopic classification tools \citep[e.g.,][]{Howell05Superfit, Blondin07, Harutyunyan08}.
Other SN Ia subtypes are either rare in nature or require complete photometric data, multi-epoch spectroscopy, or human expertise to identify; they are thus less commonly seen in transient circulars or in the literature.
Nonetheless, we include SN Ia-99aa \citep{Li01Pec, Strolger02, Garavini04, Matheson08} and Ia-00cx \citep{Li0100cx, Candia03} in our sample. SN Ia-99aa represents an intermediate subtype of SN Ia and Ia-91T, while the spectroscopic properties of SN Ia-00cx may represent extreme cases of SN Ia-91T.
We also include SN Ia-09dc (or Ia-06gz, Ia-03fg; \citealt{Howell06, Hicken07, Yamanaka09}), sometimes referred to as `super-Chandrasekhar' SN Ia, in our analysis. The high peak luminosity and slow light curve evolution of SN Ia-09dc clearly stand out among other SN Ia subtypes.
At the fainter side of SN Ia luminosity, we include SN Ia-02es \citealt[e.g.,][]{Maguire11, White15}, a group of subluminous but slowly-fading outliers in the SN Ia parameter space \citep{Phillips93}.
Furthermore, subluminous Ca-rich gap transients \cite[e.g.,][]{Perets10, Kasliwal12}, which feature strong Ca\RNum{2} emission in their nebular-phase spectra, are also included in our sample due to the proposed thermonuclear origin \citep[e.g.,][]{Waldman11, De20}.

Together, all these subtypes only represent $\lesssim 4$ per cent SNe Ia in our sample; most SNe Ia are not classified into any detailed subtype.
Luminous SN Ia subtypes are rare; only $2$ to $6$ per cent SNe Ia are Ia-91T/99aa in a cosmic volume \citep{Blondin12, Silverman12}.
Fainter subtypes are common in a cosmic volume; about $6$ to $11$ per cent SNe Ia are Ia-91bg \citep{Ganeshalingam10, Silverman12}, and nearly one-third SNe Ia are Ia-02cx \citep{Jha17Iax}.
Despite the higher volumetric fractions, the frequencies of these fainter subtypes are suppressed by the Malmquist bias in magnitude-limited SN samples, which are closer to most surveys.
Therefore, we consider those unspecified SNe Ia as `normal' ones and compare other subtypes with them.

\subsection{Notes on supernova classification} \label{sec:SubtypeNotes}

The taxonomy of transients is constantly updated as we learn about their demographics. New notations and subtypes are continuously being introduced as subgroups with coherent observational signatures are identified by modern time-domain surveys.
We inherit the transient type hierarchy and the existing subtype classification of individual transients in our database.
These subtypes are compiled across a wide variety of reference sources, mainly published catalogues and reports in transient circulars, as listed in Table 7 in the database paper.
As aforementioned, we only use classification based on spectroscopic data for its better robustness over photometric, light curve-based SN classification.
Using spectroscopic classification does not guarantee the consistency of classification across reference sources due to the possibly different classification techniques and spectral templates (or training datasets) used.
A thorough reclassification of these transients would resolve the consistency issue, but this requires original spectra that are sometimes inaccessible.
Therefore, we adhere to the existing classification in our database. 
As a result, the purity of subtypes could be affected by the possible confusion of spectroscopically similar subtypes (e.g., SN Ib and Ic).
Also, certain detailed subtypes could be incomplete as some events under this subtype are classified into the parent type. 
A transient in our database may have several inconsistent subtypes assigned by different reference sources. We do not choose one of them based on any criterion. Instead, we include the event in every subtype it was classified into.
As a result, there could be overlaps of records across subtypes. We characterize the overlap in Appendix \ref{appendix:ConfusionMatrix}.
For SN types with refined (or lower-level) subtypes in our hierarchy, we also create a `unspecified' subset (noted with an asterisk) for events that are \textit{not} further classified into any subtype under this type by other reference sources.
Meanwhile, SNe classified into any subtype will also be included in the corresponding parent type but not the `unspecified' subset of the parent type.

\subsection{Host redshifts and distances} \label{sec:GalRedshift}

Finding the best-available redshift is vital for our analysis. Sometimes, multiple redshift values are available for the same SN-host pair.
For example, the SN data source may report a redshift value (either from SN or host), and the cross-matched galaxy catalogues or databases may also have redshift values available.
We use galaxy redshifts reported in NASA/IPAC Extragalactic Database (NED), SIMBAD Astronomical Database, HyperLEDA, and SDSS spectroscopic catalogue, in the preferred order here.
Note that some galaxy redshifts in NED or SIMBAD are actually measured using SN spectra. We exclude such redshift values whenever indicated.
Finally, when there is no redshift in those catalogues or databases, we use the reported redshift from our SN data source, regardless of its source (SN or host).

We further derive host luminosity distances assuming that the redshift values are heliocentric. The assumption is valid for galaxy redshifts but may not hold for SN redshifts.
We correct the observed galaxy recession velocities to the CMB frame using the peculiar motion measured in \citet{Fixsen96}.
For galaxies in the local universe (within $200$ $\text{Mpc}\,\text{h}^{-1}$), we further correct for their peculiar motions based on the velocity field of \citet{Carrick15}.
Using the CMB dipole and peculiar motion-corrected recession velocities, we calculate the `true redshift' and distance modulus of our host galaxies, as listed in Table \ref{tab:sourcetable}.

\subsection{Host spectroscopic properties} \label{sec:SpecProperties}

There are $4128$ SNe with host spectra cross-matched in the Main Galaxy Sample \citep[MGS;][]{Strauss02} of the SDSS spectroscopic catalogue.
These hosts may have fibre aperture-corrected stellar mass \citep{Kauffmann03} and star formation rate \cite[SFR;][]{Brinchmann04}\footnote{Both stellar mass and SFR are measured assuming a flat $\Lambda$CDM cosmology with $\Omega_\text{M}=0.3$ and $H_0=70\,\mathrm{km\,s^{-1}\,Mpc^{-1}}$.}, as well as gas-phase metallicity within the $3$ arcsec-diameter fibre aperture \citep{Tremonti04}, derived from spectral indices and line measurements.

The SDSS MGS targets are selected primarily based on an extinction-corrected limiting Petrosian magnitude of $r\sim17.8$ \citep{Strauss02}.
Due to fibre collision, incompleteness of target coverage may occur in galaxy pairs or regions with higher area densities of galaxies (e.g., galaxy clusters). 
Meanwhile, not every MGS target has the complete set of physical properties listed above, as deriving these properties requires the coverage of certain rest-frame spectral regions and measurable spectral features. The incompleteness of measured properties may thus depend on target redshift, spectral data quality, and galaxy type.

The derived physical properties we analyze here are based on the legacy SDSS-I/II surveys, not including the later, cosmology-focused Baryon Oscillation Spectroscopic Survey and its extended survey (BOSS and eBOSS; \citealt{Dawson13, Abolfathi18}). There are also catalogues of galaxy properties based on later data releases, including both legacy and BOSS/eBOSS targets.
However, we found a limited overlap of BOSS/eBOSS targets with our host galaxies ($352$ cross-matched), and BOSS/eBOSS have more complicated and biased target selection functions that may challenge our later analysis. Therefore, we adhere to the legacy SDSS catalogues above.

Host physical properties from the best-matching spectroscopic targets are listed in Table \ref{tab:sourcetablespec}.
\renewcommand{\arraystretch}{1.25}
\begin{table*}
\caption{Host physical properties derived from SDSS spectra}
\begin{threeparttable}
\label{tab:sourcetablespec}
\begin{tabular}{lcccrrrr}
\hline
Name & Plate & MJD & FiberID & $z$ & $\log M_{*}$ & $\log$ SFR & 12+[O/H] \\
\hline
%SN 2017glk & 461 & 51910 & 270 & 0.071 & $9.66_{-0.06}^{+0.08}$ & $0.10_{-0.26}^{+0.26}$ & $0.10_{-0.26}^{+0.26}$ \\
SN 2019vin & 461 & 51910 & 171 & 0.039 & $10.99_{-0.09}^{+0.09}$ & $-0.03_{-0.98}^{+0.45}$ & $-0.03_{-0.98}^{+0.45}$ \\
%SN 2018jma & 462 & 51909 & 553 & 0.116 & $11.35_{-0.10}^{+0.11}$ & $-0.71_{-1.31}^{+0.87}$ & $-0.71_{-1.31}^{+0.87}$ \\
%SN 2018kpo & 462 & 51909 & 493 & 0.017 & $10.29_{-0.10}^{+0.11}$ & $-0.92_{-1.38}^{+0.61}$ & $-0.92_{-1.38}^{+0.61}$ \\
%SN 2005gm & 462 & 51909 & 517 & 0.021 & $10.66_{-0.10}^{+0.09}$ & $-1.15_{-1.37}^{+0.77}$ & $-1.15_{-1.37}^{+0.77}$ \\
%LSQ12fmd & 462 & 51909 & 204 & 0.119 & $10.52_{-0.09}^{+0.10}$ & $0.55_{-0.31}^{+0.33}$ & $0.55_{-0.31}^{+0.33}$ \\
%LSQ13cve & 464 & 51908 & 484 & 0.065 & $11.12_{-0.09}^{+0.09}$ & $-0.88_{-1.31}^{+0.85}$ & $-0.88_{-1.31}^{+0.85}$ \\
ASASSN-15pn & 465 & 51910 & 328 & 0.038 & $9.32_{-0.07}^{+0.09}$ & $-0.18_{-0.09}^{+0.13}$ & $-0.18_{-0.09}^{+0.13}$ \\
%LSQ12hxg & 465 & 51910 & 59 & 0.039 & $9.39_{-0.07}^{+0.09}$ & $-0.36_{-0.25}^{+0.28}$ & $-0.36_{-0.25}^{+0.28}$ \\
SN 2021dov & 468 & 51912 & 14 & 0.013 & $9.68_{-0.08}^{+0.10}$ & $-0.41_{-0.29}^{+0.33}$ & $-0.41_{-0.29}^{+0.33}$ \\
%SN 2019tyg & 469 & 51913 & 230 & 0.041 & $10.96_{-0.09}^{+0.10}$ & $0.63_{-0.31}^{+0.38}$ & $0.63_{-0.31}^{+0.38}$ \\
%SN 2019tve & 471 & 51924 & 153 & 0.102 & $10.50_{-0.09}^{+0.09}$ & $-1.55_{-1.29}^{+0.82}$ & $-1.55_{-1.29}^{+0.82}$ \\
SN 2020dcv & 472 & 51955 & 85 & 0.054 & $10.90_{-0.11}^{+0.13}$ & $0.89_{-0.79}^{+1.02}$ & $0.89_{-0.79}^{+1.02}$ \\
%SN 2019tpp & 473 & 51929 & 621 & 0.070 & $10.46_{-0.10}^{+0.10}$ & $0.24_{-0.46}^{+0.43}$ & $0.24_{-0.46}^{+0.43}$ \\
%SN 2019wyw & 473 & 51929 & 610 & 0.089 & $10.72_{-0.09}^{+0.09}$ & $-1.18_{-1.32}^{+0.81}$ & $-1.18_{-1.32}^{+0.81}$ \\
%SN 2013gf & 483 & 51902 & 107 & 0.085 & $10.66_{-0.09}^{+0.08}$ & $-1.58_{-1.24}^{+0.81}$ & $-1.58_{-1.24}^{+0.81}$ \\
SN 2021cjc & 487 & 51943 & 161 & 0.075 & $10.86_{-0.10}^{+0.10}$ & $-0.48_{-1.36}^{+0.67}$ & $-0.48_{-1.36}^{+0.67}$ \\
%SN 2017hmi & 488 & 51914 & 182 & 0.040 & $9.64_{-0.10}^{+0.09}$ & $-1.19_{-0.86}^{+0.44}$ & $-1.19_{-0.86}^{+0.44}$ \\
%SN 2018jfn & 489 & 51930 & 272 & 0.037 & $9.22_{-0.05}^{+0.08}$ & $-0.14_{-0.15}^{+0.22}$ & $-0.14_{-0.15}^{+0.22}$ \\
SN 2019nkf & 489 & 51930 & 554 & 0.011 & $7.86_{-0.06}^{+0.09}$ & $-1.91_{-0.26}^{+0.27}$ & $-1.91_{-0.26}^{+0.27}$ \\
%SN 2010Y & 490 & 51929 & 273 & 0.011 & $10.22_{-0.09}^{+0.09}$ & $-1.90_{-1.26}^{+0.77}$ & $-1.90_{-1.26}^{+0.77}$ \\
SN 2020nu & 490 & 51929 & 268 & 0.043 & $10.14_{-0.12}^{+0.13}$ & $-1.07_{-1.50}^{+0.87}$ & $-1.07_{-1.50}^{+0.87}$ \\
%SN 2016xb & 491 & 51942 & 380 & 0.021 & $9.03_{-0.06}^{+0.08}$ & $-0.67_{-0.36}^{+0.30}$ & $-0.67_{-0.36}^{+0.30}$ \\
%SN 2020abqx & 492 & 51955 & 589 & 0.063 & $9.76_{-0.07}^{+0.09}$ & $0.24_{-0.27}^{+0.32}$ & $0.24_{-0.27}^{+0.32}$ \\
%SN 2020mwl & 494 & 51915 & 104 & 0.061 & $10.17_{-0.09}^{+0.09}$ & $0.02_{-0.32}^{+0.33}$ & $0.02_{-0.32}^{+0.33}$ \\
%SN 2020exs & 496 & 51988 & 633 & 0.067 & $9.48_{-0.06}^{+0.08}$ & $-0.03_{-0.35}^{+0.32}$ & $-0.03_{-0.35}^{+0.32}$ \\
%SN 2018fru & 497 & 51989 & 383 & 0.010 & $9.08_{-0.08}^{+0.09}$ & $-1.68_{-0.66}^{+0.52}$ & $-1.68_{-0.66}^{+0.52}$ \\
PTF10fxp & 497 & 51989 & 130 & 0.104 & $11.31_{-0.09}^{+0.10}$ & $-1.33_{-1.16}^{+0.97}$ & $-1.33_{-1.16}^{+0.97}$ \\
%SN 2019gaj & 498 & 51984 & 60 & 0.036 & $10.63_{-0.10}^{+0.11}$ & $-1.16_{-1.46}^{+0.87}$ & $-1.16_{-1.46}^{+0.87}$ \\
%SN 2020hvy & 498 & 51984 & 533 & 0.068 & $10.59_{-0.09}^{+0.09}$ & $0.50_{-0.18}^{+0.22}$ & $0.50_{-0.18}^{+0.22}$ \\
PS1-11wj & 500 & 51994 & 606 & 0.077 & $9.94_{-0.08}^{+0.09}$ & $-0.02_{-0.28}^{+0.30}$ & $-0.02_{-0.28}^{+0.30}$ \\
iPTF13acq & 500 & 51994 & 547 & 0.107 & $10.11_{-0.09}^{+0.10}$ & $-0.11_{-0.26}^{+0.25}$ & $-0.11_{-0.26}^{+0.25}$ \\
%PS1-10t & 500 & 51994 & 225 & 0.093 & $10.05_{-0.08}^{+0.09}$ & $0.31_{-0.23}^{+0.30}$ & $0.31_{-0.23}^{+0.30}$ \\
%PTF12awi & 501 & 52235 & 198 & 0.046 & $10.08_{-0.08}^{+0.09}$ & $0.22_{-0.22}^{+0.25}$ & $0.22_{-0.22}^{+0.25}$ \\
\hline
\end{tabular}
\begin{tablenotes}
% \small 
\item Columns: \textit{Name}: SN name; \textit{Plate}, \textit{MJD}, \textit{FiberID}: Best-matching host spectrum in SDSS spectroscopic catalogue, in `plate-MJD-fiberID' format;
$\log M^*$: Host stellar mass (in $\mathrm{M}_{\odot}$) from \citet{Kauffmann03};
$\log$ SFR: Host star formation rate (in $\mathrm{M}_{\odot}\,yr^{-1}$) from \citet{Brinchmann04};
12+[O/H]: Host gas-phase metallicity (in $dex$) from \citet{Tremonti04}.
\\
This table is available in its entirety in machine-readable form.
\end{tablenotes}
\end{threeparttable}
\end{table*}

\subsection{Host photometric properties} \label{sec:PhotoProperties}

Most host galaxies in our database do not have physical properties based on SDSS spectra. We use their derived photometric properties (absolute magnitudes and rest-frame colours) as a proxy of their physical properties.

The SDSS photometric survey has a limiting magnitude of about $r\sim21.9$ for extended sources (based on the $50$ per cent completeness magnitude estimated in \citealt{Rykoff15}), considerably deeper than the spectroscopic survey (extinction-corrected $r\sim17.8$). As a result, there are more host galaxies with photometric data than with spectroscopic data.
Although absolute magnitudes and rest-frame colours do not monotonically map to stellar mass and SFR (or specific SFR) due to factors like dust, mass-to-light ratio, and the diverse formation histories of galaxies, they still serve as meaningful indicators for their physical properties.

We use the measurements in the SDSS DR16 photometric object catalogue \citep{Ahumada20}.
We choose Petrosian magnitudes instead of profile-fitting magnitudes as our hosts are often well-resolved low-redshift galaxies (median $R_{90}\sim5.2$ arcsec; $R_{90}$ is the radius enclosing $90$ per cent of the Petrosian flux), where aperture-based photometry better represents their fluxes and colours. 
For a fair comparison of photometric properties over the redshift range of the hosts, we transform the measured apparent magnitudes to their rest-frame absolute magnitudes using a standard $k$-correction \citep{Blanton03kcorr, Blanton07} so that the photometric properties are intrinsic and redshift-independent.
We first calculate the Galactic extinction-corrected magnitudes in SDSS filters using the extinction map and the coefficients in \citet{Schlegel98} and \citet{Schlafly11}.
We then calculate the rest-frame colours and absolute magnitudes using the \textsc{kcorrect} package (v4.2, \citealt{Blanton07}) and the luminosity distances we derived earlier for each host (Section \ref{sec:GalRedshift}).

The measured and derived photometric properties of host galaxies are summarized in Table \ref{tab:sourcetablephoto}.

\subsection{Sample summary} \label{sec:SampleSummary}

We defined two samples in our analysis, a spectroscopic sample with any physical parameter (stellar mass, SFR, and gas-phase metallicity) available based on SDSS spectra and a photometric sample with $k$-corrected rest-frame colours and absolute magnitudes based on SDSS photometry.
Given the target selection criteria of the SDSS spectroscopic survey, the spectroscopic sample is essentially a subset of the larger photometric sample. 
There are $4128$ hosts in the spectroscopic sample and $12943$ hosts in the photometric sample. Figure \ref{fig:TypePieChart} summarizes the type distribution of SNe in our sample.

\section{Statistical Methods} \label{sec:StatMethod}

The primary goal of this work is to characterize the systematic differences of host properties across SN subtypes, which indicate the delay times and chemical abundances of their progenitors and provide valuable insights into their nature.
Fundamentally, the systematic differences of host properties reflect the different relationships of the observed SN rates and galaxy properties across subtypes.
SN rate dependence relationships connect progenitor models to the stellar populations and past formation histories of galaxies and are thus more revealing.
We also test which model of rate dependence relationship best reproduces the observed host properties.
Addressing either question could be a non-trivial task given the intricate and coupled selection effects in \textit{both} the SN sample and galaxy catalogues.
The SN sample comprises detections of several surveys, each with distinct sensitivity limits, cadences, and target selection strategies, as well as serendipitous discoveries outside these surveys. This mix of data sources leads to a complex selection function that cannot be easily characterized. Therefore, we assume that the selection effect is predominantly driven by distance or redshift, similar to the classical Malmquist bias, i.e., more luminous objects can be detected out to a greater distance.
Factors such as host galaxy extinction and the timescales of transients are considered secondary. For galaxies, we assume that the selection effect is Malmquist-like; biases such as fiber collisions in the SDSS spectroscopic sample are considered negligible.
However, even with this assumption, two significant biasing factors persist: one is driven by the different redshift distributions of SN subtypes, and the other is driven by the intrinsically higher SN rates in massive, high-SFR, or luminous galaxies.
Therefore, establishing robust baselines is critical to reveal the systematic differences in host properties and make meaningful comparisons and discussions.
% 

% raw MNRAS table

\begin{landscape}
\begin{table}
\caption{Host absolute magnitudes and colors derived from SDSS photometry.}
\begin{threeparttable}
\label{tab:sourcetablephoto}
\begin{tabular}{@{\extracolsep{4pt}}lcccccccccc@{}}
\hline
Name & \multicolumn{5}{c}{Petrosian Magnitude} & \multicolumn{5}{c}{Absolute Magnitude} \\
\cline{2-6} \cline{7-11}
     & $u$ & $g$ & $r$ & $i$ & $z$             & $M_u$ & $M_g$ & $M_r$ & $M_i$ & $M_z$  \\
%SN 2000ft & $13.55 \pm 0.01$ & $12.78 \pm 0.02$ & $12.26 \pm 0.02$ & $12.19 \pm 0.02$ & $11.85 \pm 0.02$ & $-19.91$ & $-20.65$ & $-21.09$ & $-21.24$ & $-21.52$ \\
%SN 2009md & $13.58 \pm 0.01$ & $12.63 \pm 0.01$ & $12.52 \pm 0.02$ & $12.30 \pm 0.02$ & $12.26 \pm 0.02$ & $-17.82$ & $-18.63$ & $-18.90$ & $-18.99$ & $-19.04$ \\
SN 2017jmk & $13.61 \pm 0.01$ & $12.20 \pm 0.01$ & $11.51 \pm 0.06$ & $11.16 \pm 0.01$ & $10.91 \pm 0.02$ & $-18.42$ & $-19.80$ & $-20.48$ & $-20.81$ & $-21.07$ \\
%SN 1998dh & $13.61 \pm 0.01$ & $12.20 \pm 0.01$ & $11.51 \pm 0.06$ & $11.16 \pm 0.01$ & $10.91 \pm 0.02$ & $-18.42$ & $-19.79$ & $-20.48$ & $-20.81$ & $-21.07$ \\
%SN 2009at & $13.62 \pm 0.04$ & $12.42 \pm < 0.01$ & $11.76 \pm 0.01$ & $11.38 \pm 0.01$ & $11.34 \pm 0.02$ & $-17.67$ & $-19.04$ & $-19.70$ & $-20.03$ & $-20.28$ \\
%SN 2011eh & $13.64 \pm < 0.01$ & $11.64 \pm 0.02$ & $10.85 \pm 0.02$ & $10.48 \pm 0.02$ & $10.23 \pm 0.02$ & $-18.45$ & $-20.22$ & $-21.08$ & $-21.51$ & $-21.86$ \\
SN 2019iex & $13.64 \pm 0.01$ & $12.38 \pm < 0.01$ & $11.78 \pm < 0.01$ & $11.46 \pm < 0.01$ & $11.37 \pm < 0.01$ & $-19.38$ & $-20.59$ & $-21.15$ & $-21.41$ & $-21.62$ \\
SN 2012Z & $13.64 \pm 0.01$ & $12.15 \pm < 0.01$ & $11.56 \pm < 0.01$ & $11.29 \pm < 0.01$ & $11.13 \pm < 0.01$ & $-18.24$ & $-19.59$ & $-20.19$ & $-20.42$ & $-20.64$ \\
%SN 2002fk & $13.64 \pm 0.01$ & $12.15 \pm < 0.01$ & $11.56 \pm < 0.01$ & $11.29 \pm < 0.01$ & $11.13 \pm < 0.01$ & $-18.24$ & $-19.59$ & $-20.19$ & $-20.42$ & $-20.64$ \\
SN 1999ev & $13.66 \pm 0.01$ & $11.33 \pm < 0.01$ & $10.44 \pm < 0.01$ & $10.12 \pm < 0.01$ & $9.70 \pm < 0.01$ & $-16.05$ & $-17.79$ & $-18.62$ & $-19.03$ & $-19.37$ \\
SN 2004fc & $13.67 \pm 0.01$ & $12.44 \pm 0.01$ & $11.84 \pm 0.01$ & $11.56 \pm 0.01$ & $11.40 \pm 0.01$ & $-17.68$ & $-18.88$ & $-19.44$ & $-19.70$ & $-19.90$ \\
%SN 2014G & $13.68 \pm 0.02$ & $12.71 \pm 0.01$ & $12.34 \pm 0.05$ & $12.09 \pm 0.01$ & $12.14 \pm 0.04$ & $-17.71$ & $-18.62$ & $-19.02$ & $-19.21$ & $-19.34$ \\
%SN 2009dd & $13.69 \pm 0.02$ & $12.19 \pm 0.01$ & $11.39 \pm 0.01$ & $10.95 \pm < 0.01$ & $10.65 \pm < 0.01$ & $-15.92$ & $-17.45$ & $-18.24$ & $-18.63$ & $-18.96$ \\
%SN 2012cc & $13.70 \pm 0.01$ & $11.74 \pm 0.01$ & $10.85 \pm 0.01$ & $10.41 \pm 0.01$ & $10.20 \pm 0.01$ & $-10.45$ & $-12.21$ & $-13.07$ & $-13.49$ & $-13.84$ \\
SN 2019ein & $13.70 \pm 0.03$ & $11.73 \pm 0.02$ & $10.90 \pm 0.02$ & $10.42 \pm 0.02$ & $10.20 \pm 0.01$ & $-18.64$ & $-20.39$ & $-21.23$ & $-21.65$ & $-21.99$ \\
%SN 2007le & $13.72 \pm 0.02$ & $12.18 \pm 0.01$ & $11.64 \pm < 0.01$ & $11.27 \pm < 0.01$ & $11.24 \pm 0.01$ & $-17.84$ & $-19.19$ & $-19.83$ & $-20.14$ & $-20.37$ \\
SN 2017jfs & $13.73 \pm 0.01$ & $12.76 \pm < 0.01$ & $12.33 \pm < 0.01$ & $12.06 \pm < 0.01$ & $11.99 \pm < 0.01$ & $-18.46$ & $-19.39$ & $-19.82$ & $-20.04$ & $-20.18$ \\
%SN 2003iq & $13.73 \pm 0.01$ & $11.92 \pm < 0.01$ & $11.10 \pm < 0.01$ & $10.68 \pm < 0.01$ & $10.40 \pm < 0.01$ & $-18.09$ & $-19.80$ & $-20.61$ & $-20.99$ & $-21.31$ \\
%SN 2003hl & $13.73 \pm 0.01$ & $11.92 \pm < 0.01$ & $11.10 \pm < 0.01$ & $10.68 \pm < 0.01$ & $10.40 \pm < 0.01$ & $-18.09$ & $-19.80$ & $-20.61$ & $-20.99$ & $-21.31$ \\
%SN 2009jf & $13.74 \pm 0.03$ & $12.04 \pm 0.04$ & $11.20 \pm 0.01$ & $11.07 \pm 0.06$ & $10.83 \pm 0.04$ & $-18.13$ & $-19.76$ & $-20.49$ & $-20.81$ & $-21.07$ \\
SN 2020scc & $13.76 \pm < 0.01$ & $12.74 \pm < 0.01$ & $12.10 \pm < 0.01$ & $11.83 \pm < 0.01$ & $11.57 \pm < 0.01$ & $-18.66$ & $-19.68$ & $-20.29$ & $-20.56$ & $-20.82$ \\
SN 2013dy & $13.76 \pm 0.01$ & $13.00 \pm 0.01$ & $12.82 \pm 0.01$ & $13.02 \pm 0.01$ & $12.95 \pm 0.02$ & $-16.99$ & $-17.58$ & $-17.82$ & $-17.70$ & $-17.79$ \\
SN 2019upv & $13.77 \pm 0.01$ & $13.20 \pm 0.02$ & $12.51 \pm 0.02$ & $12.26 \pm 0.02$ & $11.82 \pm 0.01$ & $-18.62$ & $-19.19$ & $-19.78$ & $-20.13$ & $-20.52$ \\
\hline
\end{tabular}
\begin{tablenotes}
% \small 
\item Columns: \textit{Name}: SN name; \textit{R.A.} and \textit{Decl.}: SN coordinate; $z$: SN redshift as reported in our database paper;
\textit{Type}: SN type, where type with an asterisk indicates that the event is not further classified as a subtype of the indicated type;
\textit{Host R.A.} and \textit{Host Decl.}: best host galaxy coordinate in our database paper;
$E(B-V)$: Galactic foreground reddening at the host position, based on \citet{Schlegel98} and \citet{Schlafly11}. \\
This table is available in its entirety in machine-readable form.
\end{tablenotes}
\end{threeparttable}
\end{table}
\end{landscape}

\subsection{Matched resampling} \label{sec:zMatchedResampling}

The primary biasing factor is that the different redshift distributions of SN subtypes, coupled with the selection effects in the parent galaxy catalogue, could lead to inaccurate estimates of mean host properties.
Assuming a specific SN subtype is detectable to a higher redshift limit than other subtypes. In that case, the classical Malmquist bias (i.e., luminous objects are detectable to a greater distance) in the parent galaxy catalogue could lead to higher mean host luminosity or mass than other subtypes, as those lower luminosity or mass hosts, if they exist, are beyond the galaxy survey detection limit.
The bias is not a primary concern if \textit{all} hosts are detected in the galaxy survey (i.e., $100$ per cent host recovery ratio) and host properties are always complete.
However, our spectroscopic sample has a relatively bright host magnitude limit ($r\simeq17.8$ after extinction correction; \citealt{Strauss02}); only $32$ per cent hosts in the photometric sample have matched SDSS spectra (Section \ref{sec:SampleSummary}).
This redshift-driven bias, if not properly controlled for, can lead to inaccurate estimates of mean host properties.
We correct or compensate for this bias by calculating the offsets of mean host properties with respect to the \textit{reference sample} of each subtype.
The reference sample here is drawn from the same parent galaxy catalogue and has the same redshift distribution as the observed hosts, which supposedly inherit those redshift-related selection effects but do not carry any information about SN-host connections.
As a baseline or `zero point' of comparison, any offsets of mean host properties from the reference sample indicate systematic differences driven by SN-host connections rather than selection effects.
Setting separate baselines for each subtype also allows us to compare the offsets of host properties across subtypes in a less biased way. 
To generate a reference sample that follows the same redshift distribution as the observed hosts, we calculate the resampling weights of individual galaxies in the parent galaxy catalogue.
The weights are proportional to the probabilities of galaxies to be resampled, with which we randomly draw galaxies from the parent catalogue with replacement and assemble the reference sample.

First, we construct distance modulus distributions for the entire parent galaxy catalogue, and the observed SNe under the subtype of interest with hosts matched in this specific catalogue, approximated using Gaussian Kernel Density Estimation (KDE).
We choose distance modulus instead of redshift because galaxy properties of concern, such as absolute magnitude, stellar mass, and SFR, are represented in logarithmic scale like distance modulus.
Then, we derive the probability density ratio of these two distributions (observed hosts to the entire parent catalogue) as a function of redshift and calculate the ratio for the redshift of each galaxy in the parent catalogue.
These redshift-dependent ratios, normalized over the entire parent catalogue, serve as the resampling weights of individual galaxies.
For each subtype, we randomly draw $16384$ galaxies with replacement by the derived weights. The resampled galaxies follow the same redshift distribution as the observed SNe, \textit{without} any physical information about SN-host connections.
For the spectroscopic sample, the parent galaxy catalogue (SDSS MGS; \citealt{Strauss02}) is already a complete sample of galaxies that satisfies the target selection criteria and bears the selection effects.
We directly resample this parent catalogue so that the reference samples match the redshift distributions of SN subtypes.
As we noted earlier (Section \ref{sec:SpecProperties}), not every host has a SDSS spectrum matched; even for hosts with SDSS spectra, not every spectrum has the complete set of physical properties (mass, SFR, and metallicity).
Therefore, to compare hosts in the mass-SFR plane, we draw galaxies in the subset of the parent catalogue with mass and SFR measurements to match the redshifts of the observed hosts with mass and SFR measurements.
Similarly, to compare hosts in the mass-metallicity plane, we construct reference samples using galaxies in the parent catalogue with mass and metallicity measurements to match the observed hosts with mass and metallicity measurements.
The selection effects related to the completeness of galaxy properties can also be compensated or controlled for in this approach. 
There is a clear advantage of our redshift-matched resampling technique for the spectroscopic sample.
SDSS spectra measure galaxy properties within their $3$ arcsec-diameter fibre aperture, an aperture size that only covers the central regions of low-redshift galaxies.
Aperture corrections are required to estimate the integrated mass and SFR over the entire galaxy using fibre spectra (\citealt{Kauffmann03, Brinchmann04}), but metallicity cannot be corrected and is only measured in the central regions covered by the fibre aperture, where gas tends to be more metal-rich than in the outskirts. (e.g., \citealt{Sanchez14, Ho15}).
When coupled with the metallicity gradients of galaxies, the redshift-dependent fibre aperture bias could overestimate the mean metallicity of low-redshift or large angular-size galaxies.
Matching the redshift distributions can partly compensate for this bias and lead to more accurate comparisons, as the baseline is also affected this bias.
The photometric sample does not have a parent galaxy catalogue like the spectroscopic sample.
Indeed, our host galaxies represent a tiny subset of the SDSS photometric catalogue, but the entire catalogue cannot serve as the parent catalogue for the purpose here.
Most galaxies in the SDSS photometric catalogue do not have spectroscopic redshifts available anywhere.
Photometric redshifts, although estimated for a substantial fraction of observed galaxies (e.g., \citealt{Beck16}), suffer from greater errors and possible biases, hindering the calculation of absolute magnitudes and rest-frame colours, and leading to inaccurate matching of redshifts.
Therefore, we construct a `parent catalogue' for our photometric sample in an alternative approach resembling the construction of galaxy luminosity functions using redshift surveys (e.g., \citealt{Blanton03lumfunc}).
We first construct a `fair population' of galaxies in a comoving cosmic volume using the photometric magnitudes and spectroscopic redshifts of galaxies in the SDSS MGS.
Assuming that the population evolution of galaxies is negligible within the redshift interval of interest (which is reasonable given the redshift range of MGS targets), we then resample this fair population while imposing a redshift- and colour-dependent absolute magnitude cutoff.
The detailed procedure is discussed in Appendix \ref{appendix:Mockcatalogue}.

\subsection{Weighted reference samples} \label{sec:WeightedResampling}

We correct and compensate for the redshift-driven biases of host properties by setting reference samples as fair baselines of comparison.
Even if the reference samples reveal how the observed hosts differ from other galaxies with similar selection effects in the same redshift interval, the comparison here remains intrinsically biased.

Stellar transients usually prefer luminous or massive galaxies, as there tend to (but not always) be more progenitors, boosting the observed rates. However, the number density of galaxies in a cosmic volume is always dominated by low-luminosity or low-mass ones.
Host galaxies of stellar transients, as a result, are often (but not always) at the massive or luminous end of the galaxy distribution, but redshift-matched reference samples are heavily biased toward the low-luminosity or low-mass end.
Further complicating matters, the mean galaxy properties in the reference sample are also sensitive to the cutoff luminosity or mass of the galaxy survey -- a function of redshift which depends on the usually indistinct fainter edge of the selection function.
As a result, the mean galaxy properties of the reference samples can shift substantially with redshift, making them unstable baselines of comparison.
To establish robust yet physically meaningful baselines of comparison, we calculate the mean galaxy properties of redshift-matched and \textit{weighted} reference samples.
These reference samples are also drawn to match the redshift distributions of the observed hosts; meanwhile, at any redshift, the resampling probabilities of galaxies in the parent catalogue are also proportional to (or weighted by) their mass, SFR, or luminosity.
Choosing the mean properties of weighted reference samples as the baselines carries two clear benefits. 
First, these weighted baselines directly compare the host properties of SNe (i.e., the stellar populations from which they arise) to the `host properties' of average stars or average young, massive stars in the parent catalogue, within the same redshift interval.
Second, such baselines allow us to test the simple hypothesis that SN rates scale with the stellar mass, SFR, or luminosity of the host.
Assuming that the mean mass of individual stars is similar across galaxies, the total stellar mass-weighted mean galaxy properties eventually reflect the average `host properties' of a typical star in the parent catalogue, within the redshift interval.
Similarly, assuming similar initial mass function and binary star fraction across galaxies, the SFR-weighted mean galaxy properties reflect the average `host properties' of a newly formed massive star in the parent catalogue, within the redshift interval.
Choosing the mean properties of the weighted reference samples, in this perspective, sets the `host properties' (or integrated stellar population properties) of a typical star or a typical newly formed massive star as the baseline of comparison. 
From another point of view, either present-day stellar mass or current SFR could serve as an approximate proxy for the number of progenitors with imminent explosions inside a galaxy; which host property better traces the number of such progenitors depends on the typical delay time of the SN subtype.
SN subtypes that arise from newly born massive stars (e.g., most CC SN subtypes) have short delay times (e.g., $\lesssim 50$ Myr for single stars, up to $200$ Myr for binaries; \citealt{Zapartas17}), and their rates may thus follow the current SFR.
In contrast, SN subtypes with long-lived progenitors (e.g., WDs) may have delay times long enough on par with the Hubble timescale or the timescale of galaxy evolution (i.e., the timescale on which galaxies change their properties substantially); the present-day stellar mass may better reflect the number of progenitors that are about to explode, and hence the observed rates.
We choose the expected host properties of a hypothetical SN subtype, whose observed rates are proportional to either stellar mass or SFR, as the reference standard of comparison.
If the observed mean host properties deviate from the baseline of mass-weighted or SFR-weighted reference samples, the SN rates are not naively proportional to the present-day stellar mass or current SFR.
Even for the photometric sample, we may include the rest-frame luminosity in a specific band in the resampling weights.
The rest-frame luminosity of a galaxy depends not only on its total stellar mass but also on its star formation history and dust content.
Despite the complex dependency here, the mean photometric properties of a luminosity-weighted reference sample could still reveal the different host properties of SN progenitors compared to the `host properties' of `average stellar light' and serve as a baseline for the comparison across SN subtypes.
To generate a redshift-matched, weighted reference sample in which the resampling weights at a fixed redshift also scale with mass, SFR, or luminosity, we extended our existing matched resampling technique.
We first construct the distance modulus distribution of the observed hosts under a specific subtype using equal weights for the Gaussian KDE.
We also construct the distance modulus distribution of the parent galaxy catalogue using mass, SFR, or luminosity as the weights of galaxies.
We then divide the equally weighted distance modulus distribution of the observed hosts by the weighted distance modulus distribution of the parent galaxy catalogue to derive a redshift-dependent weighting factor.
We further normalize the product of this weighting factor with mass, SFR, or luminosity over the entire parent galaxy catalogue to find the proper resampling weights.
Randomly drawing galaxies from the parent catalogue with replacement based on these resampling weights yields a reference sample that follows the redshift distribution of the observed hosts; meanwhile, at a given redshift, the probability of a galaxy to be drawn is also proportional to its mass, SFR, or luminosity.
Host properties in the spectroscopic sample can be incomplete, as discussed earlier (Section \ref{sec:SpecProperties}).
To compare a given set of host properties, we construct redshift distribution for the observed hosts with these properties measured; we also resample the subset of SDSS galaxies with these properties available to match the redshift distribution.
Host properties are complete for the photometric sample; therefore, we always resample the entire mock parent catalogue (Appendix \ref{appendix:Mockcatalogue}) with luminosity weights applied. 

\subsection{General rate dependence models} \label{sec:RateScaling}

The preference for certain host galaxies essentially reflects the dependence of observed SN rates (i.e., number of detections per unit time, per galaxy) on galaxy properties, namely the rate dependence relationship.
Assuming environment-independent evolution of progenitors (i.e., only isolated single star evolution or close binary interaction) and constant detection efficiency, SN rates scale directly with the number of progenitors close to the condition or stage of explosion.
How the observed SN rates (hence numbers of pre-explosion progenitors) scale with galaxy properties is a more fundamental question that connects host properties to progenitor models.

Characterizing the dependence of SN rates on galaxy properties requires time-controlled surveys and careful modeling of the selection function (e.g., \citealt{Leaman11, Li11_rates, Graur17_rates}), which are unfeasible for the literature-compiled sample of this work.
Nonetheless, we could still test or fit a hypothetical rate dependence model by comparing the observed mean host properties with those of a reference sample weighted by the model-predicted rates.
With weighted reference samples (Section \ref{sec:WeightedResampling}), we test the simple hypothesis that the observed SN rates are proportional to the SFR or stellar mass of the host.
These represent the two extreme cases in which SN progenitors are either exclusively short-lived massive stars or long-lived stars whose lifetimes are comparable to the Hubble timescale or the timescale of galaxy evolution.
Certain SN subtypes, particularly SN Ia \citep{Maoz11, Maoz12SDSS, Maoz12DTD}, have a wide dynamical range of delay times; their rates may follow neither the stellar mass nor SFR of the host.
Still, there are empirical models to connect SN Ia rates to galaxy properties, such as the classical `A+B' model in which the rate is a linear combination of mass ($M^*$) and SFR with $A$ and $B$ coefficients, representing the `delayed' and `prompt' components of SN Ia rates \citep{Mannucci05, Scannapieco05, Sullivan06}.
As a further generalization of this `A+B' model, the observed rate of any SN subtype $c$ in a galaxy ($R_{c}$, in $\text{yr}^{-1}$) could be written as
\begin{equation} \label{eq:kmodel}
R_{c}=\mathcal{R}_{c,0} \times [k\, M_{*} + (1-k) w \, \text{SFR}]
\end{equation}
where $\mathcal{R}_{c,0}$ is a specific SN rate in per unit stellar mass and per unit time; $k$ is a dimensionless free parameter within $0$ and $1$; $w$ is a time-like parameter to offset the order-of-magnitude difference between typical stellar mass and SFR of galaxies.\footnote{We choose $w=10^{10}\,\text{yr}$ because the difference of median $\log M_{*}$ and median $\log\text{SFR}$ in the SDSS MGS is close to $10$. Also, SFR is measured in $\mathrm{M}_{\odot}\text{yr}^{-1}$, and $w=10^{10}\,\text{yr}$ is close to the age of the universe.}
Here, a lower value of $k$ represents stronger dependence of SN rates on SFR than stellar mass and implies short-lived progenitors; conversely, a higher value of $k$ indicates closer connections of SN rates to stellar mass than SFR and favors long-lived progenitors.
The $A$ and $B$ coefficients in \citet{Sullivan06} imply $k=0.576$ and $\mathcal{R}_0=9.20\times 10^{-14}\text{yr}^{-1}\mathrm{M}_{\odot}^{-1}$ or $0.092$ SNuM, following the definition of \citet{Mannucci05}.
Minimizing the differences of mean galaxy properties between the observed hosts and rate-weighted reference samples gives a best-fitting $k$ parameter that reveals how SN rates, and hence the number of pre-explosion progenitors, depend on host properties. 
% k=0.576 for w=10^10; k=0.931 for w=10^11
%

%
There is a similar generalization for the photometric sample. We may assume that SN rates are proportional to the luminosity in a specific rest-frame band.
Luminosity in bluer rest-frame bands is sensitive to the amount of young stars and hence the recent SFR; in contrast, luminosity in redder rest-frame bands is sensitive to both young and old stars and may better trace the total stellar mass, despite the different mass-to-light ratios of young and old stellar populations (see the review of \citealt{Conroy13}).
Assuming progenitors are short-lived, SN rates should follow closer with the luminosity in bluer bands; conversely, SN rates may scale better with the luminosity in redder bands if the progenitors are long-lived.
The rest-frame band that best minimizes the difference in mean galaxy properties between the observed hosts and a luminosity-weighed reference sample could qualitatively reveal the delay time of progenitors. 
Furthermore, for CC SN subtypes, a critical question is how their relative frequencies depend on the host metallicity (e.g., \citealt{Anderson10, Modjaz11, Kelly12, Sanders12, Galbany16}).
We assume that at a constant metallicity, CC SN rates are proportional to their host SFR; meanwhile, the ratio of SN rate to SFR is metallicity-dependent.
Specifically, the rate of a CC SN subtype $c$ in a galaxy ($R_{c}$, in $\text{yr}^{-1}$) could be written as
\begin{equation} \label{eq:tmodel}
R_{c} = \mathcal{Y}_{c,0} \times \text{SFR} \times 10^{\,t \Delta \log Z }
\end{equation} 
in which $\Delta \log Z$ is the relative abundance of the host\footnote{We define $\Delta \log Z = (12+[\text{O/H}]) - 8.7$ here, where the constant $8.7$ is the solar abundance \citep{Asplund09}.
The normalized weights are independent of this constant.}, $t$ (`metallicity tilt') is a dimensionless free parameter, and $\mathcal{Y}_{c,0}$ is the SN production efficiency (i.e., number of SN explosions per unit stellar mass formed) at $\Delta \log Z=0$, in units of $\mathrm{M}_{\odot}^{-1}$.
This $t$ parameter is the increase in SN production efficiency or SN rate-to-SFR ratio (assuming negligible delay times), in logarithmic scale, per unit increase in host metallicity.
A positive $t$ boosts the weights of metal-rich galaxies and suppresses the weights of metal-deficient galaxies, while a negative $t$ boosts the weights of metal-rich galaxies and suppresses the weights of metal-deficient galaxies.
Similarly, we may minimize the difference in mean galaxy properties between the observed hosts and a rate-weighted reference sample to find a best-fitting $t$ that describes the metallicity dependence of CC SN rates.

\subsection{Statistical tests}

We compare the mean host properties of the observed hosts to the weighted reference samples and the offsets of mean host properties across subtypes under the baseline of weighted reference samples.
Here, we choose a two-step approach to assess the statistical significance of the differences and offsets.
First, we check whether the differences with the weighed reference samples or the offsets across subtypes are significant (confidence level reaches $2\sigma$) using Hotelling's $t^2$-test \citep{Hotelling92}, a multivariate generalization of the Student's $t$-test.
If the offsets across subtypes are significant, we then examine which host property shows a significant difference ($2\sigma$ level) under the classical Student's $t$-test.

Notably, the $t^2$-test is more conservative (or less sensitive) than the $t$-test in detecting the difference; there are cases in which we detect significantly different offsets in one variable in a $t$-test while the $t^2$ statistic indicates an insignificant difference in the multivariate mean. We note such cases in our discussion.
Finally, if we see tentative evidence for a difference with low statistical significance, we refer to it as a marginal or possible difference.

\subsection{Limitations of the technique} \label{sec:limitations}

We describe two techniques that are motivated by different science questions but shared a common basis.
For both techniques, we assume that the selection effects of SNe are presumably redshift-driven (i.e., a classical Malmquist bias) and are independent of other factors.
One consideration is the timescale, as slowly-evolving transients are more easily identified in a photometric survey under a specific cadence and also have a broader time window for spectroscopic classification. However, since we do not discuss the relative frequencies of SN subtypes, this does not affect our analysis.
The Galactic foreground extinction might also be a concern, though we assume the systematic differences across SN subtypes are minor.
Finally, the joint SN-host selection effect could be an issue. Assuming a simple Malmquist bias, the hosts should not affect the detection limit or sample completeness of SNe from a results-oriented perspective.
This assumption, however, might overlook host-related secondary selection effects throughout the entire workflow of SN candidate detection, follow-up observation, and classification.
For example, SNe close to galaxy nuclei tend to be undermined by the strong background light (`Shaw effect,' \citealt{Shaw79}).
Even in modern transient surveys, successful detection and classification could still suffer from the reduced signal-to-noise ratio under strong host light, confusion with AGN variability for pipelines or human observers, and failed classification with host-contaminated SN spectra.
Host galaxy dust extinction could also reduce the observed brightness of SNe and hence their detectability. Such secondary selection effects may lead to a biased comparison of host properties if SN detectability is host-dependent.
We leave the assessment of such effects in our future work. 

The SN sample we analyze here is a collection of several surveys with vastly different facilities, detection and classification strategy, and, more importantly --- scientific motivation.
Most time-domain surveys identify transients over a large sky area, independent of host properties.
Although the strategy still suffers from detection biases as discussed above, such ``untargeted" surveys are expected to produce less biased samples in host properties.
In contrast, galaxy-targeted surveys monitor selected galaxies for new transients; the selection of these galaxies, therefore, could impose biases in the comparison of host properties.
For example, low-mass galaxies are underrepresented in the galaxy sample monitored by the Lick Observatory Supernova Search (LOSS, \citealt{Li11_rates}); the survey will thus miss SNe in those low-mass galaxies, and the resulting sample is likely not representative in host properties.
Although the impact could be limited, as only $\sim15\%$ of CC SNe and $\sim10\%$ of SNe Ia are missed due to the incompleteness of low-mass host galaxies in the LOSS sample \citep{Leaman11}, combining these two types of surveys may still introduce inherent biases when comparing host properties across SN subtypes.
Therefore, characterizing the true `selection function' of the entire sample is likely not practical. Simple, redshift-driven selection bias is only an approximation.

\section{Results and Discussion} \label{sec:ResultsAndDiscussion}

In this Section, we analyze and discuss the host properties of CC SN and SN Ia subtypes.
We first focus on the host stellar mass and SFR in the smaller spectroscopic sample; then, we discuss their absolute magnitude and rest-frame colours in the larger photometric sample.
For CC SN hosts, we further discuss their gas-phase metallicity.
We present results for CC SN and SN Ia separately, while the discussion is also made across these major types.
Basic statistics of host properties for each subtype are summarized in Tables \ref{tab:summarytablespec} and \ref{tab:summarytablephoto}.
The differences of mean host properties across subtypes and their statistical significance are summarized in Tables \ref{tab:typewisetablespec} and \ref{tab:typewisetablephot}.
Besides a comprehensive discussion on host properties, we also investigate how SN rates scale with galaxy properties.

We assess how the selection of SN surveys affects the results for some of our key conclusions by reproducing the analysis using only SNe detected by several wide-field surveys --- the Panoramic Survey Telescope and Rapid Response System (Pan-STARRS; \citealt{Chambers16}), All Sky Automated Survey for SuperNovae (ASAS-SN), Asteroid Terrestrial-impact Last Alert System (ATLAS; \citealt{Smith20}), Palomar Transient Factory (PTF; \citealt{Law09}), and Zwicky Transient Facility (ZTF; \citealt{Bellm19}) --- an incomplete ``wide-field" sample which represent $43\%$ of the original dataset used here and are relatively unbiased with respect to host stellar mass.
We also limited the redshift of the entire sample to $z<0.05$ (the ``low-redshift" sample), a range where most SN subtypes and our galaxy catalogs are relatively more complete.
This allows us to test whether our key conclusions are sensitive to the redshift range of our sample and whether redshift-dependent biases are properly controlled.

\subsection{Stellar mass and SFR of CC SN hosts} \label{sec:CCMassSFR}

\begin{figure*}
\centering
\includegraphics[width=\linewidth]{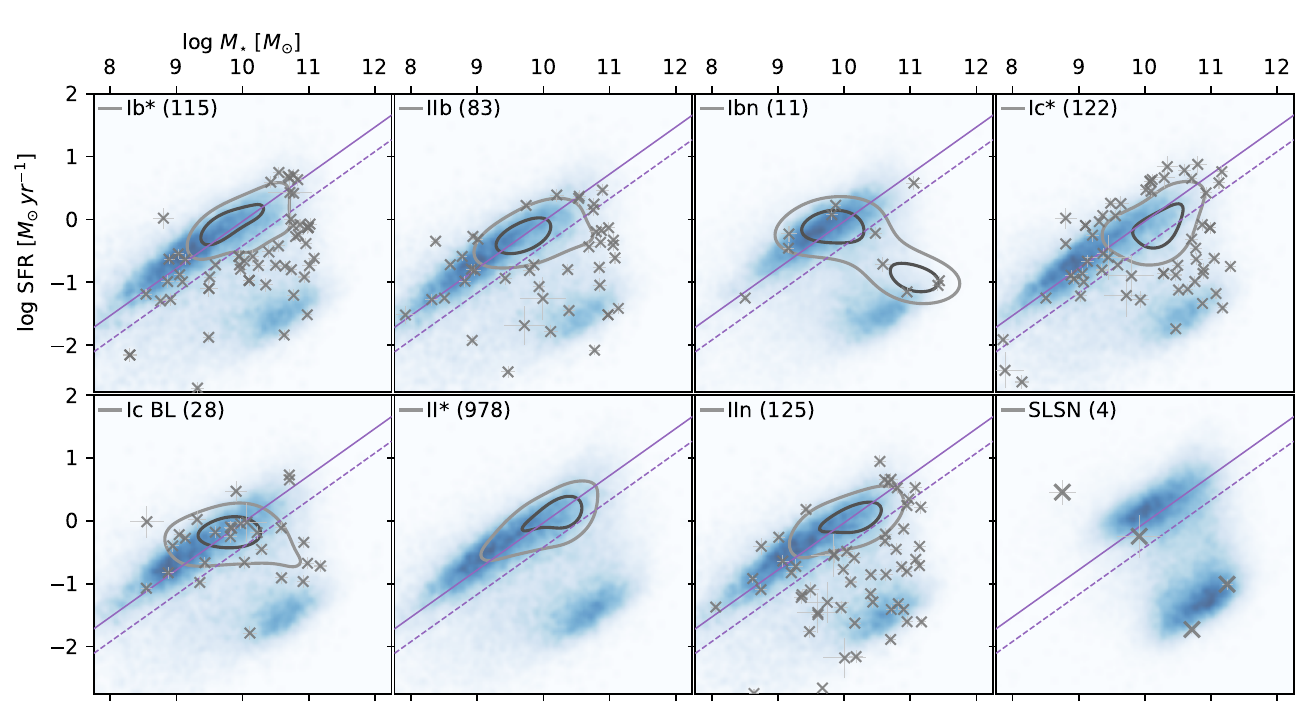}
\caption{\label{fig:CCMassSFR}
Host galaxy stellar mass and SFR of CC SN subtypes in the spectroscopic sample. 
Contours enclose probability density corresponding to 1$\sigma$ (inner) and 2$\sigma$ (outer) levels of a bivariate normal distribution.
We plot outliers (beyond 2$\sigma$ contour) for panels with less than $200$ data points and all data points for panels less than $30$ data points.
Magenta lines indicate the SFMS (solid) and its $1\sigma$ lower limit (dashed) in \citet{Chang15}, while background density maps show the redshift-matched reference samples drawn from the parent SDSS spectroscopic catalogue.
CC SN hosts prefer star-forming galaxies in general, and host properties differ only subtly across subtypes.
See Section \ref{sec:CCMassSFR} for discussion.  
The direct comparison of hosts across CC SN subtypes, and between CC SN hosts and unweighted, redshift-matched samples here may suffer from selection biases, as discussed earlier in Section \ref{sec:limitations}. We use redshift-matched, SFR- or mass-weighted reference samples for comparison in Figure \ref{fig:CCMassSFROffset}.
}
\end{figure*}

To begin with, we directly compare the host properties across CC SN subtypes, and between CC SN hosts and SDSS galaxies within the same redshift interval.
Figure \ref{fig:CCMassSFR} shows the stellar mass and SFR of CC SN hosts, grouped by subtypes. For reference, we also show the redshift-matched reference sample of each subtype and the SFMS fit in \citet{Chang15} (Equation 5), based on the same SDSS spectroscopic sample.
Generally, CC SN hosts populate the region of star-forming galaxies in the mass-SFR plane and follow the SFMS,
consistent with the expectation that CC SN progenitors are generally massive stars contributed by recent star formation activities (e.g., \citealt{Smartt09ARAA, Smartt15, VanDyk17}).
Massive star-forming galaxies may have higher CC SN rates due to the greater number of massive star progenitors therein, but most CC SNe occurred in intermediate-mass galaxies around $\log M^* \sim 10$, which are more abundant than massive ones in the universe.
Despite favouring star-forming galaxies, CC SNe do not exclusively occur in star-forming galaxies.
Some CC SN hosts are inside the `green valley' (i.e., the low-density region between star-forming and quiescent galaxies), and even quiescent galaxies can host CC SNe, albeit rare (e.g., \citealt{Irani22}).
Quiescent galaxies have low-level star formation activities, and the trace amount of massive stars may also give rise to CC SN explosions; even the rates may reflect the SFR in such environments \citep{Sedgwick21}. 

\begin{figure*}
\centering
\includegraphics[width=0.75\linewidth]{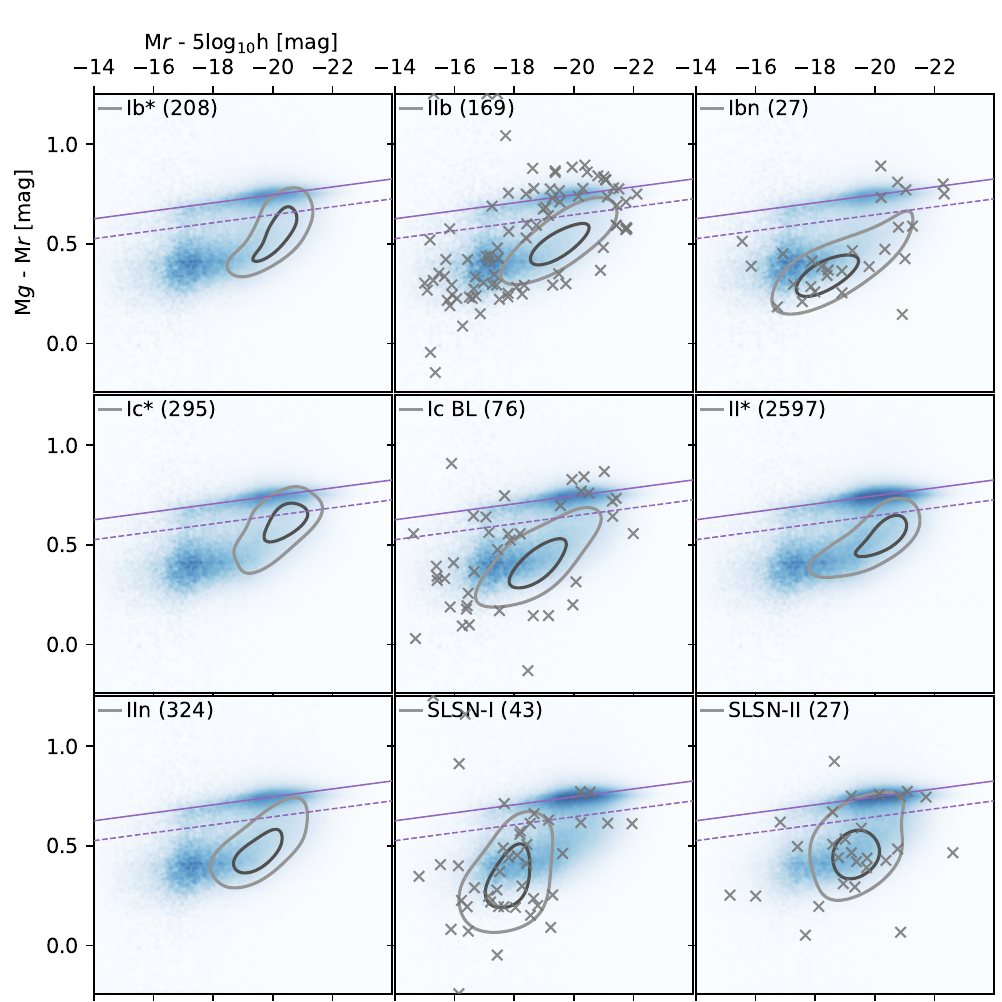}
\caption{ \label{fig:CCcolourMag}
Host galaxy rest-frame $r$-band absolute magnitude and $g$-$r$ colour of CC SNe in the photometric sample, grouped by the subtypes.
Contours and scatter points follow the same style as in Figure \ref{fig:CCMassSFR}.
Magenta lines indicate the best-fitting red sequence (solid) and its `blue edge' ($1\sigma$ lower limit) in \citet{Masters10}; background density maps show redshift-matched reference samples as described in Section \ref{sec:zMatchedResampling}.
Here we divide SLSN into hydrogen-deficient (SLSN-I) and hydrogen-rich (SLSN-II) subtypes.
This direct comparison of host luminosity and color across CC SN subtypes, and between CC SN hosts and the corresponding unweighted, redshift-matched samples, could be biased. We control for possible selection effects using luminosity-weighted reference samples in Figure \ref{fig:CCcolourMagOffset}.
CC SN hosts overlap with the blue cloud, but red sequence galaxies also host a fraction of CC SNe.
See Section \ref{sec:CCMassSFR} for discussion. 
}
\end{figure*}

Figure \ref{fig:CCcolourMag} shows the rest-frame absolute magnitudes and colours of CC SN hosts, grouped by subtypes. The greater sample size allows us to refine SLSN into hydrogen-deficient (SLSN-I) and hydrogen-rich (SLSN-II) subtypes.
We show the best-fitting red sequence of quiescent galaxies and its `blue edge' (i.e., $1\sigma$ lower limit) in \citet{Masters10} for reference. We choose the red sequence here rather than the blue cloud of star-forming galaxies because, in the mass-SFR plane, the SFMS is a more clearly defined structure than quiescent galaxies; however, in the galaxy colour-magnitude diagram (CMD), the red sequence is a more clearly defined structure than the blue cloud.
The photometric sample of CC SN hosts shows similar signatures as in the spectroscopic sample (Figure \ref{fig:CCMassSFR}).
Except for a few subtypes, CC SN hosts are mostly on the luminous side of the blue cloud. Those luminous star-forming galaxies may have more massive star progenitors and are thus preferred hosts (Section \ref{sec:WeightedResampling}).
Fainter star-forming galaxies, even more abundant in number density, do not contribute much observed CC SNe.
Even CC SN hosts are generally blue-cloud galaxies, their colour-magnitude distributions extend into and overlap with the red sequence in several subtypes (SN Ib*, Ic*, etc.).
Besides those quiescent CC SN hosts as we noted earlier (Figure \ref{fig:CCMassSFR}), galaxies can also appear redder due to dust extinction, whereas the true specific SFR could be higher than their colour implies. Therefore, red sequence galaxies can also host a fraction of CC SNe.

\begin{figure*}
\centering
\includegraphics[width=0.49\linewidth]{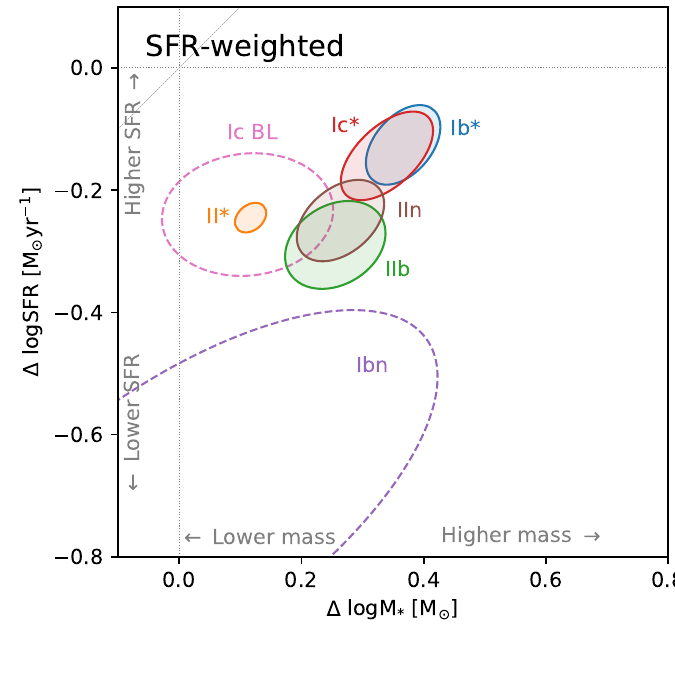}
\includegraphics[width=0.49\linewidth]{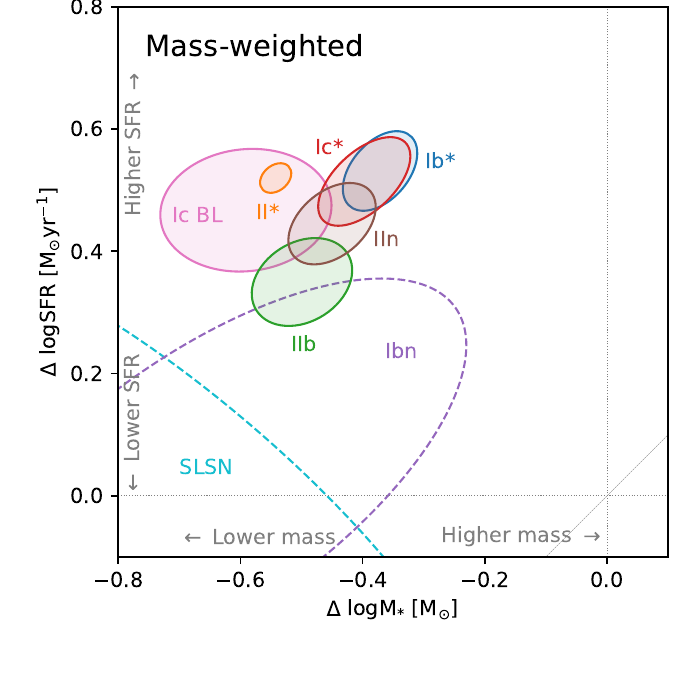} 
\caption{\label{fig:CCMassSFROffset}
Host stellar mass and SFR offsets of CC SN subtypes under the baseline of average massive stars (i.e., using SFR-weighted reference sample; left) and average stars (i.e., using mass-weighted reference sample; right).
The centroids of ellipses indicate the offsets of mean stellar mass and mean SFR of the observed hosts from the weighted reference samples, while the sizes of ellipses reflect the standard error of the mean and its covariance.
Solid ellipses represent subtypes with significant differences in mean host properties compared to the reference sample (above $2\sigma$ level under the Hotelling's $t^2$-test).
This implies that assuming SN rates are proportional to the weighting variable (mass or SFR) results in mean host mass and SFR that are statistically different from those observed.
Conversely, dashed ellipses represent subtypes with consistent mean host properties (i.e., insignificant differences) with the reference samples, i.e., assuming that SN rates are proportional to the weighting variable can produce the observed mean host properties.
For these subtypes, assuming that SN rates are proportional to the weighting variable yields mean host properties that are consistent with observations.
See Section \ref{sec:CCMassSFR} for discussion.
}
\end{figure*}

\begin{figure}
\vskip 5mm
\centering
\includegraphics[width=\linewidth]{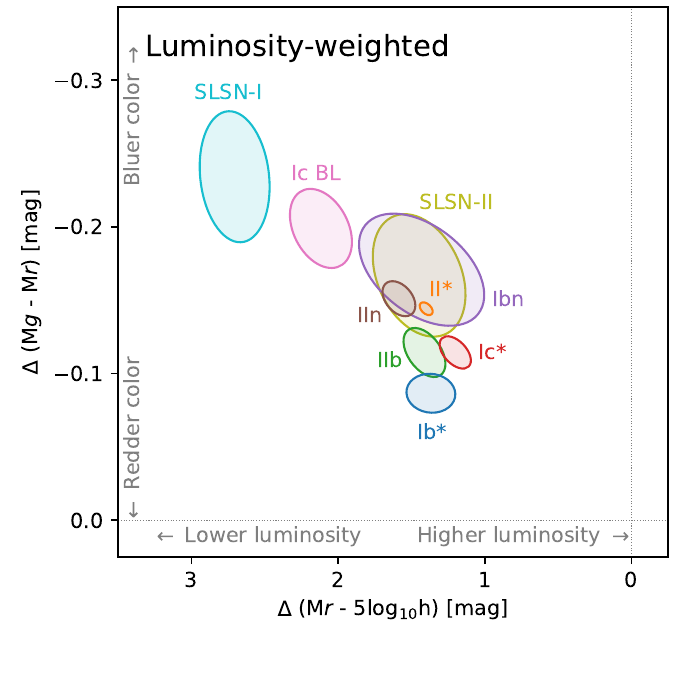}
\caption{\label{fig:CCcolourMagOffset}
Host rest-frame absolute magnitude and colour offsets of CC SN subtypes under the baseline of rest-frame $r$-band stellar light (i.e., using luminosity-weighted reference samples).
Ellipses follow the same style as in Figure \ref{fig:CCMassSFROffset}.
Here, solid ellipses imply that assuming SN rates are proportional to the r-band luminosity produces mean host colors and luminosities that are statistically different from those observed.
We revert both axes so that the right side indicates relatively higher luminosities, and the upper side indicates relatively bluer colours.
See Section \ref{sec:CCMassSFR} for discussion.
}
\end{figure}

We then focus on individual CC SN subtypes. 
As we discussed in Section 3.5, the direct comparison across CC SN subtypes in Figure 2 may suffer from selection biases. Therefore, we use the redshift-matched, SFR- or mass-weighted reference samples to further investigate the differences in host properties across these subtypes.
To aid the comparison of subtypes, besides the mass-SFR plane and CMD, we also summarize the offsets of mean host properties with respect to the mass-weighted and SFR-weighted reference samples in Tables \ref{tab:summarytablespec}, \ref{tab:summarytablephoto} and Figures \ref{fig:CCMassSFROffset}, \ref{fig:CCcolourMagOffset}.
Since hosts of CC SN subtypes substantially overlap with star-forming galaxies, the systematic differences in host properties across subtypes are subtle, especially in the smaller spectroscopic sample.
For example, the fraction of SFMS galaxies (i.e., those above the dashed line in Figure \ref{fig:CCMassSFR}) in the hosts of each subtype lies within the $95$ per cent confidence interval of each other.
Still, the comparison across subtypes reveals several interesting trends. 
The properties of SN Ib* and Ic* hosts are clearly distinct from SN II* hosts.
Compared to SN II* hosts, SN Ib* and Ic* hosts are more massive but consistent in sSFR under the baseline of average massive stars (Figure \ref{fig:CCMassSFROffset}, left); under the baseline of average stars (Figure \ref{fig:CCMassSFROffset}, right), SN Ib* hosts are also more quiescent (lower sSFR).
The photometric sample indicates redder colours but similar luminosities of SN Ib* and Ic* hosts than SN II* hosts under the baseline of average stellar light (Figure \ref{fig:CCcolourMagOffset}).
The fraction of red sequence galaxies is also higher in SN Ib* ($29.9_{-5.9}^{+6.4}$ per cent) and Ic* ($30.9_{-5.0}^{+5.4}$ per cent) hosts, compared to SN II* ($20.8_{-1.7}^{+1.8}$ per cent)\footnote{Unless noted otherwise, percentile fractions are not corrected for the redshift bias; error ranges indicate the $95$ per cent confidence intervals.}.
Neither spectroscopic nor photometric sample shows a significant difference between SN Ib* and Ic* hosts.

Certain SN Ib and Ic subtypes feature different mean host properties compared to their `normal' siblings.
Despite the limited sample size, SN Ibn tends to occur in bluer hosts compared to SN Ib*. For the spectroscopic sample, under the massive star baseline, the hosts of SN Ibn are different from SN Ib* in the mass-SFR plane, but the difference is not significant for individual variables.
Notably, the secondary peak in the clump of quenched galaxies (Figure \ref{fig:CCMassSFR}) is likely an artifact due to shot noise.
SN Ic-BL hosts feature similar properties as SN Ic* hosts under the baseline of average massive stars or average stars (Figure \ref{fig:CCMassSFROffset}); their host properties are also consistent with the `mean host properties' of average massive stars.
However, under the baseline of stellar light, their hosts show clearly bluer colours and lower luminosities than hosts of their sibling subtype SN Ic*, as well as hosts of SN Ib* and IIb (Figure \ref{fig:CCcolourMagOffset}).
Finally, neither spectroscopic nor photometric sample shows a significant difference between SN IIb and Ib* hosts.
SN II represents the most abundant SN subtype in a volume-limited sample (e.g., \citealt{Li11_frac, Shivvers17}).
Hosts of interacting SN IIn, the only spectroscopic subtype of SN II,\footnote{In this work, we do not further classify SN II into II-P and II-L. See Section \ref{sec:SubtypeNotes} for a discussion.} show no clear difference with their sibling SN II* in either spectroscopic or photometric samples.
Notably, the transitional subtype SN IIb, which we consider a subtype of SE SN, shows similar redder host colour than SN II* as its sibling SN Ib*.

The conclusions from comparisons of host mass and SFR between SN Ib*, Ic* and II*, SN Ib* and IIb, as well as SN II* and IIn, remain qualitatively the same under the baseline of average massive stars for SNe at $z<0.05$. However, when using only SNe detected by wide-field surveys under the same baseline, the previously higher host mass of SN Ib* and Ic* compared to SN II* is no longer statistically significant, potentially due to the reduced sample sizes.

SLSNe represent the extreme of luminous stellar transients. They are rare in the spectroscopic sample, while in the larger photometric sample, they show distinct host properties (Figure \ref{fig:CCcolourMag}).
SLSN-I hosts are fainter in luminosity than those of other CC SN subtypes (except for SN Ic-BL and IIn) under the baseline of stellar light; they are also bluer in colour than the hosts of SN Ib*, IIb, and Ic*.
Their hosts, on average, are likely the faintest and bluest among CC SN subtypes.
Conversely, SLSN-II hosts have similar mean luminosity and colour with other CC SN subtypes, except for SLSN-I whose hosts are of lower luminosity and bluer colour.
As a side note, despite the extreme host properties, the fraction of red sequence galaxies in SLSN hosts remains broadly consistent with other CC SN subtypes ($18.6_{-9.4}^{+13.5}$ per cent for SLSN-I, and $29.6_{-14.5}^{+18.6}$ per cent for SLSN-II).
Furthermore, the comparison of CC SN mean host properties under the baseline of average massive stars (Figure \ref{fig:CCMassSFROffset}, left), average stars (Figure \ref{fig:CCMassSFROffset}, right), and $r$-band stellar light (Figure \ref{fig:CCcolourMagOffset}) reveals common signatures.
Leaving out SN Ibn and SLSN whose sample sizes are limited, host galaxies of CC SN subtypes are almost always more massive and quiescent than those of average newly born stars (except for Ic-BL), or less massive and more star-forming than those of average stars. % <<< check?
The results indicate that the observed rates of most CC SN subtypes do not directly scale with the current SFR; either massive and high-SFR galaxies produce more \textit{observed} CC SN than low-mass and low-SFR galaxies at any redshift, or CC SN in low-mass and low-SFR galaxies are incomplete in our spectroscopic sample.
Hosts of CC SN subtypes are, on average, fainter in luminosity and bluer than the `host properties' of average stellar light, indicating that luminous and red galaxies are not the preferred environments for CC SN progenitors.

\subsection{Gas-phase metallicity of CC SN hosts} \label{sec:CCMassMetallicity}

\begin{figure*}
\includegraphics[width=\textwidth]{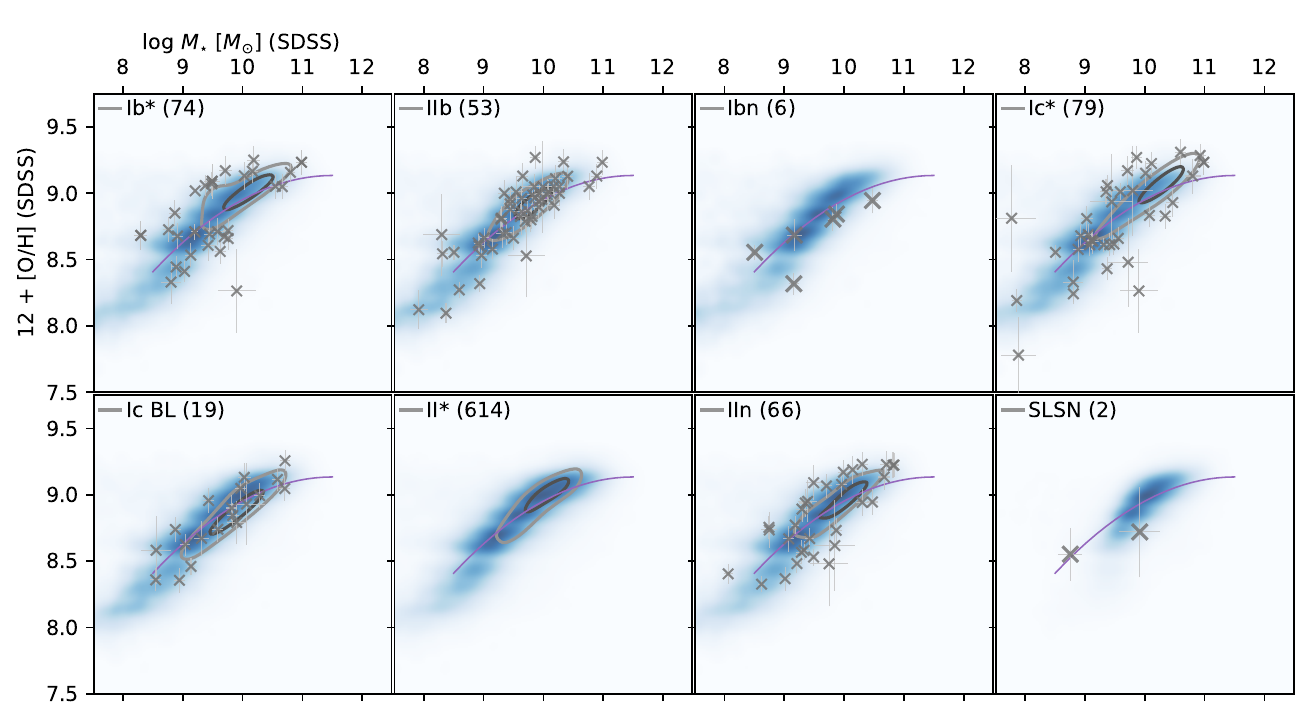}
\caption{\label{fig:CCMassMetallicity}
Host galaxy stellar mass and gas-phase metallicity of CC SN subtypes in the spectroscopic sample.
Contours and scatter points follow the same style as in Figure \ref{fig:CCMassSFR}; background density maps show redshift-matched reference samples (Section \ref{sec:zMatchedResampling}). Magenta lines indicate the median mass-metallicity relationship of \citet{Tremonti04}. See Section \ref{sec:CCMassMetallicity} for discussion.
This figure directly compares host mass and metallicity across SN subtypes and between SN hosts and unweighted redshift-matched reference samples, which could be biased. We control for possible redshift-driven selection effects in Figure \ref{fig:CCMassMetallicityOffset}.
}
\end{figure*}

\begin{figure*}
\centering
\vskip 0.5cm
\includegraphics[width=0.49\linewidth]{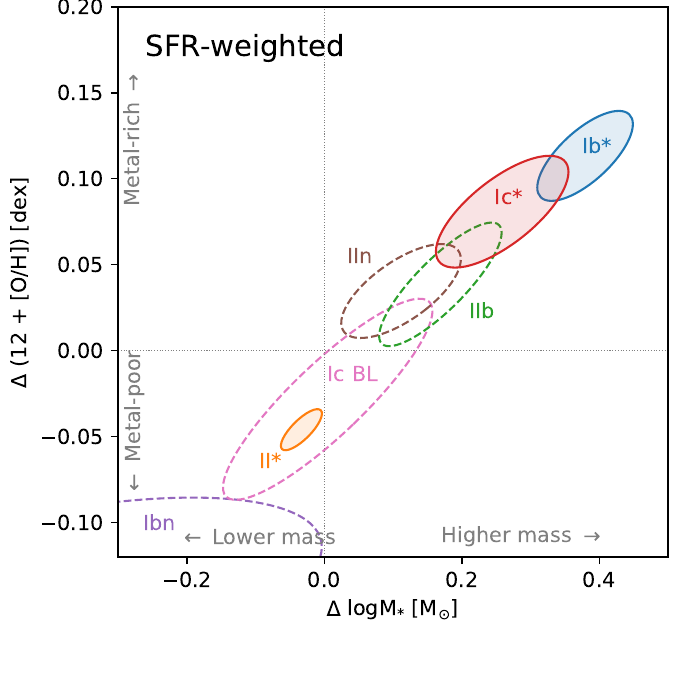}
\includegraphics[width=0.49\linewidth]{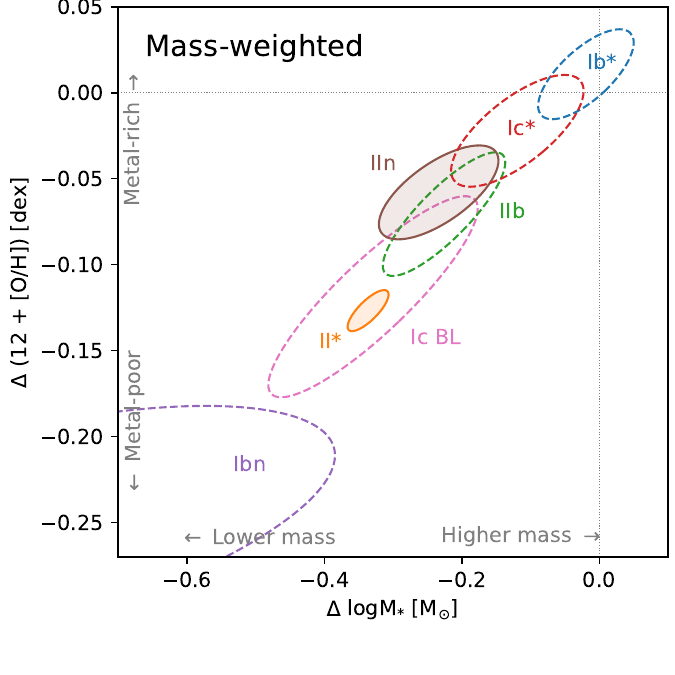}
\caption{\label{fig:CCMassMetallicityOffset}
Host stellar mass and gas-phase metallicity offsets of CC SN subtypes under the baseline of average massive stars (i.e., using SFR-weighted reference samples; left) and average stars (i.e., using mass-weighted reference samples; right).
Ellipses follow the same style as in Figure \ref{fig:CCMassSFROffset}.
Solid ellipses indicate that for the subset of SN with complete host mass and metallicity, assuming that the rates are proportional to the weighting variable results in mean host mass and metallicity that are statistically different from those observed; conversely, for dashed ellipses, the reference samples and observed hosts show consistent mean host mass and metallicity.
See Section \ref{sec:CCMassMetallicity} for discussion.
}
\end{figure*}

Metallicity is a critical factor that regulates the mass loss of massive stars (e.g., \citealt{Vink01, Heger03}), alters the initial mass function slope (e.g., \citealt{Kroupa01}), and affects the binary fraction in a stellar population (e.g., \citealt{ElBadry19, Moe19}).
Consequently, the outcome of stellar population evolution, and hence the relative frequencies of CC SN subtypes, depends on the metallicity of the host.
The spectroscopic sample has gas-phase metallicity (or simply `metallicity') for a subset of the hosts, represented as the relative abundance of oxygen ($12+\text{[O/H]}$), measured through the simultaneous fitting of multiple nebular emission lines \citep{Tremonti04}.
Assuming the measured metallicity trace the abundance of recently-born massive stars (i.e., homogeneous abundance within the galaxy), and assuming that the biases related to radial metallicity gradient (e.g., \citealt{Sanchez14, Ho15}), limited fibre aperture, and redshift evolution of chemical abundance are properly compensated with our resampling techniques (Section \ref{sec:SpecProperties}), we compare the host metallicity across CC SN subtypes. 
Any deviation in the measured metallicity could be indicative evidence for such dependence. 
Gas-phase metallicity follows a close relationship with stellar mass as a result of galaxy chemical abundance evolution (e.g., \citealt{Brooks07, Ma16}).
This `mass-metallicity relationship' (MMR; e.g., \citealt{Tremonti04, Kewley08, Mannucci10}) has been previously used to interpret the observed dependence of CC SN subtype on host stellar mass (e.g., \citealt{Graur17_rates, Schulze21}).
We also compare the metallicity of host galaxies across subtypes in the context of MMR. Figure \ref{fig:CCMassMetallicity} shows the mass-metallicity plane of CC SN hosts, grouped by subtypes, with the redshift-matched reference samples of each subtype and the best-fitting median MMR in \citet{Tremonti04} in the background for reference. 
Like star-forming galaxies in general, CC SN hosts also closely follow the MMR (Figure \ref{fig:CCMassMetallicity}). The sparse outliers either bear significant errors in their metallicity values or reside at the low-mass end where the MMR has a greater scatter. The consistency here indicates that their hosts are not special compared to typical star-forming galaxies in chemical abundance.
Notably, the host metallicity ranges are similar across CC SN subtypes, without sharp truncation at either lower or higher values side. 
The substantial overlap of host metallicity distributions indicates that metallicity is likely not the single dominating factor that determines the progenitor evolution for any specific subtype; there are either important, metallicity-insensitive factors or complex dependence of metallicity due to the interplay of several factors.
Nevertheless, the similarity of host metallicity ranges does not eliminate the possible dependence on metallicity given the vastly different stellar mass, close binary configuration, and chemical inhomogeneity within the same galaxy.
Despite the overall consistency with the MMR and similar ranges, there are tentative shifts in the mass-metallicity plane with respect to the median MMR line in \citet{Tremonti04}. For instance, the $1\sigma$ density contours of SN Ib*, Ic*, and II* are above the median MMR line, while SN Ic-BL and Ibn are possibly beneath the line.
The shifts are partly attributable to the SFR-dependence of the MMR (e.g., \citealt{Torrey18}), as relatively quiescent galaxies, such as SN Ib* and Ic* hosts (Section \ref{sec:CCMassSFR}), are usually `chemically evolved' than star-forming galaxies of similar mass and may thus locate above the median MMR line.
Also, the spectroscopic sample may have a different pattern of fibre aperture bias than the full \citet{Tremonti04} sample; we could have overestimated or underestimated the host metallicity with respect to the MMR in \citet{Tremonti04}, depending on the redshift range of the subtype.
Furthermore, the evolution of the MMR within the redshift range of our spectroscopic sample, albeit insignificant, could also contribute to the deviations across subtypes here.
To reveal the subtle differences in mean host metallicity as hinted in Figure \ref{fig:CCMassMetallicity}, we further control for the biasing factors related to redshift and any possible cosmic evolution using weighted reference samples (Figure \ref{fig:CCMassMetallicityOffset}).
Compared to the baseline of either average massive stars or average stars, 
the mean mass and metallicity of CC SN subtypes form a sequence aligned with their error ellipses (i.e., covariance for the standard error of the mean).
The sequence here indicates that CC SN subtypes have similar MMR slopes for their hosts, and the differences in host metallicity are qualitatively consistent with, but not necessarily due to, their differences in host stellar mass.

We then compare the offsets across subtypes.
The hosts of SN Ib* and Ic* are, on average, more metal-rich than those of SN II* under either baseline; they are also more metal-rich than those of average massive stars (but not average stars).
Conversely, the hosts of SN II* are relatively metal-deficient than those of average massive stars and average stars.
The higher metallicity of SN Ib* and Ic* hosts than SN II* hosts is broadly consistent with earlier results (e.g., \citealt{Anderson10, Modjaz11, Kelly12, Sanders12, Galbany16}).
We further show that the mean host metallicity of SN Ib* and Ic* is even higher than those of average massive stars; their hosts represent the metal-rich side of the sequence outlined in Figure \ref{fig:CCMassMetallicityOffset}. The relatively metal-deficient hosts of SN II* represent the other side of the sequence.
We note that under either baseline, SN Ib* and Ic* hosts show no difference in their mean host metallicity. Therefore, the degree of envelope stripping beyond the outermost hydrogen layer is insensitive to the progenitor metallicity.
The conclusion regarding the host metallicity of SN Ib*, Ic*, and II* holds for SNe at $z<0.05$;
for SNe detected by wide-field surveys, the difference between SN Ib* and II* hosts is the only one that remains statistically significant, although the signs of the offsets remain the same compared to our original conclusions.

The interpretation of the higher metallicity in SN Ib* and Ic* hosts than SN II* hosts can be intricate.
A possible scenario is that at least a fraction of the observed SNe Ib* and Ic* are substantially `delayed' compared to SNe II*, allowing their hosts to become more quiescent and chemically enriched than galaxies of similar stellar mass.
Population synthesis models predict delayed CC SN explosions via close binary mergers and engulfment (e.g., \citealt{Zapartas17}).
However, the predicted delay time is relatively short (up to $200$ Myr for the binary channel) compared to the timescale of galaxy chemical evolution (at the order of Gyr; e.g., \citealt{Torrey18}), and neither does this explain the higher host stellar mass of SN Ib* and Ic* than SN II*.
Nevertheless, the detection of SE SNe in quiescent galaxies (e.g., \citealt{Hosseinzadeh19}) implies that such delayed explosions remain a possible scenario.

A more likely scenario is that chemically enriched stellar populations yield intrinsically higher fractions of SE SNe, either via enhanced stellar winds, which lead to Wolf-Rayet stars and hydrogen-deficient SN spectra, or through envelope stripping during close binary interactions.
Higher metallicity allows single stars in a wider range of ZAMS mass to explode as SE SNe (e.g., \citealt{Heger03}), but whether stellar winds alone can effectively boost the fractions of SN Ib* and Ic* remains uncertain.
On the other hand, Galactic surveys of solar-type stars favour a decreased close binary fraction as the stellar metallicity (in [Fe/H]) increases \citep{ElBadry19, Moe19}.
Assuming that the deficiency of metal-rich close binaries also extrapolates to newly formed massive stars, then envelope stripping via close binary interaction is likely not a favourable reason for the preference of SN Ib* and Ic* for metal-rich hosts.
Recent studies further revealed a decreased close binary fraction for $\alpha$-enhanced (higher [$\alpha$/Fe]) stars upon the dependence on [Fe/H] \citep{Mazzola20}, suggesting a complex relationship between close binary fraction and chemical abundance patterns; the relative contribution of the close binary channel may also depend on the formation history of the host \citep{Thomas05, deLaRosa11}.
The progenitors of SN II* are red supergiants (e.g., \citealt{Smartt09IIP}).
Their endpoint of evolution is likely less metallicity-sensitive than single stars of higher ZAMS mass; they are also more abundant in number than those more massive stars.
Therefore, we expect SN II* hosts to have similar metallicity as those of average massive stars.
The negative metallicity offset under the baseline of massive stars (Figure \ref{fig:CCMassMetallicityOffset}, left) indicates that either there are selection effects that are not taken into account, or the assumption here is oversimplified.

We further discuss the sibling subtypes of SN Ib*, Ic*, and II*.
SN IIb hosts are more metal-rich than SN II* hosts under the massive star baseline; other than that, SN IIb hosts have consistent mean mass and metallicity with hosts of other subtypes under either baseline.
The conclusion holds for SNe at $z<0.05$, but is no longer significant for SNe detected by wide-field surveys.
The location of SN IIb in Figure \ref{fig:CCMassMetallicityOffset} also implies its nature as an intermediate subtype of SN II* and Ib*, despite the broad consistency of host properties between SN Ib* and IIb. 
Hosts of SN Ibn are metal-deficient compared to those of their sibling SN Ib*, as well as SN Ic*, under either baseline.  However, the difference with respect to SN Ic* hosts is only significant under the univariate $t$-test. 
The results suggest that SN Ibn hosts represent the metal-deficient end of the sequence (Figure \ref{fig:CCMassMetallicityOffset}), while larger samples are required to consolidate the results.
The metallicity difference between SN Ibn hosts and SN Ib*, Ic* hosts remains significant for SNe at $z<0.05$, but it is not significant for those detected by wide-field surveys due to the reduced sample sizes.
SN Ic-BL hosts are more metal-deficient than SN Ib* hosts under the baseline of massive stars; other than that, they have consistent mean metallicity than other CC SN subtypes. %
The difference is not statistically significant for the smaller samples of SNe at lower redshifts or detected by wide-field surveys.
Previous works reported a lower mean metallicity of SN Ic-BL hosts than SN Ic* hosts, but our results are only marginally consistent with the conclusion (metallicity difference of $-0.108\pm0.067$ and $-0.094\pm0.067$, under average massive star and average star baselines).
SN IIn hosts have a higher mean metallicity than the hosts of SN II* under either baseline, indicating that their progenitors, supposed to be luminous blue variables (LBVs; e.g., \citealt{GalYam07_05gl, Smith11_10jl}), are possibly more common in metal-rich environments. 
Yet, the conclusion is below statistical significance for SNe detected by wide-field surveys.
Under the baseline of average massive stars, SN IIn hosts are metal-deficient compared to SN Ib* hosts. % <<<
SN IIb, Ic-BL, and Ibn have consistent host mass and metallicity as those of average massive stars and average stars, but SN IIn is only consistent with average massive stars in these host properties.

Finally, it is noteworthy that we use the gas-phase oxygen abundance to trace the progenitor abundance, while stellar evolution models are usually based on iron abundance due to its closer relevance to opacity and stellar wind. The variation of [$\alpha$/Fe] could also affect the difference in gas-phase metallicity across subtypes.

\subsection{Stellar mass and SFR of SN Ia hosts} \label{sec:IaMassSFR}

\begin{figure*}
\includegraphics[width=\textwidth]{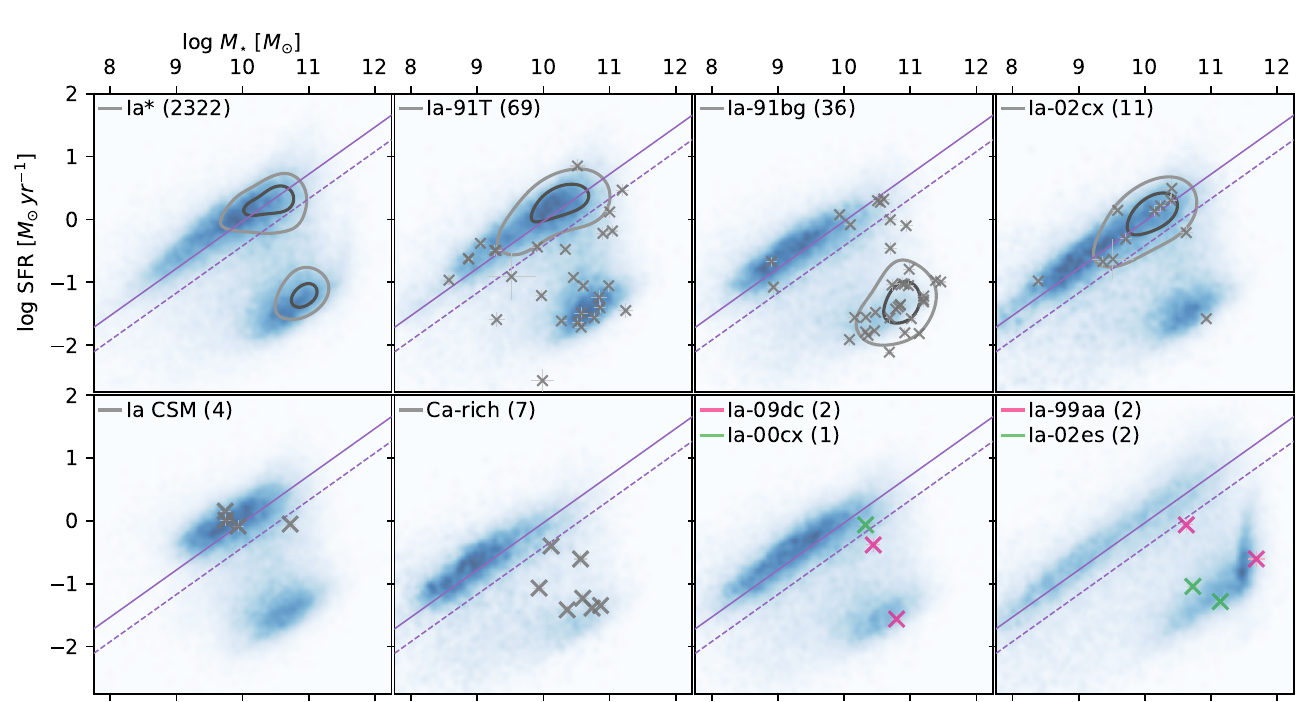}
\caption{
\label{fig:IaMassSFR}
Host stellar mass and SFR of SN Ia subtypes in the spectroscopic sample.
Contours and scatter points follow the same style as in Figure \ref{fig:CCMassSFR}.
Magenta lines represent the SFMS (solid) and its $1\sigma$ lower limit (dashed) in \citet{Chang15}; background density maps show the redshift-matched reference samples.
Multiple subtypes, if grouped in one panel, are indicated using different colours. Their reference samples are based on a general volumetric galaxy sample.
SN Ia occurs in both star-forming and quiescent galaxies; subtypes of SN Ia show greater heterogeneity in their host properties.
Here we directly compares the host stellar mass and SFR across SN Ia subtypes and between SN Ia subtypes and the unweighted, redshift-matched reference samples, which could be biased. We control for possible redshift-driven selection effects in Figure \ref{fig:IaMassSFROffset}.
See Section \ref{sec:IaMassSFR} for discussion.
}
\end{figure*}

\begin{figure*}
\includegraphics[width=\textwidth]{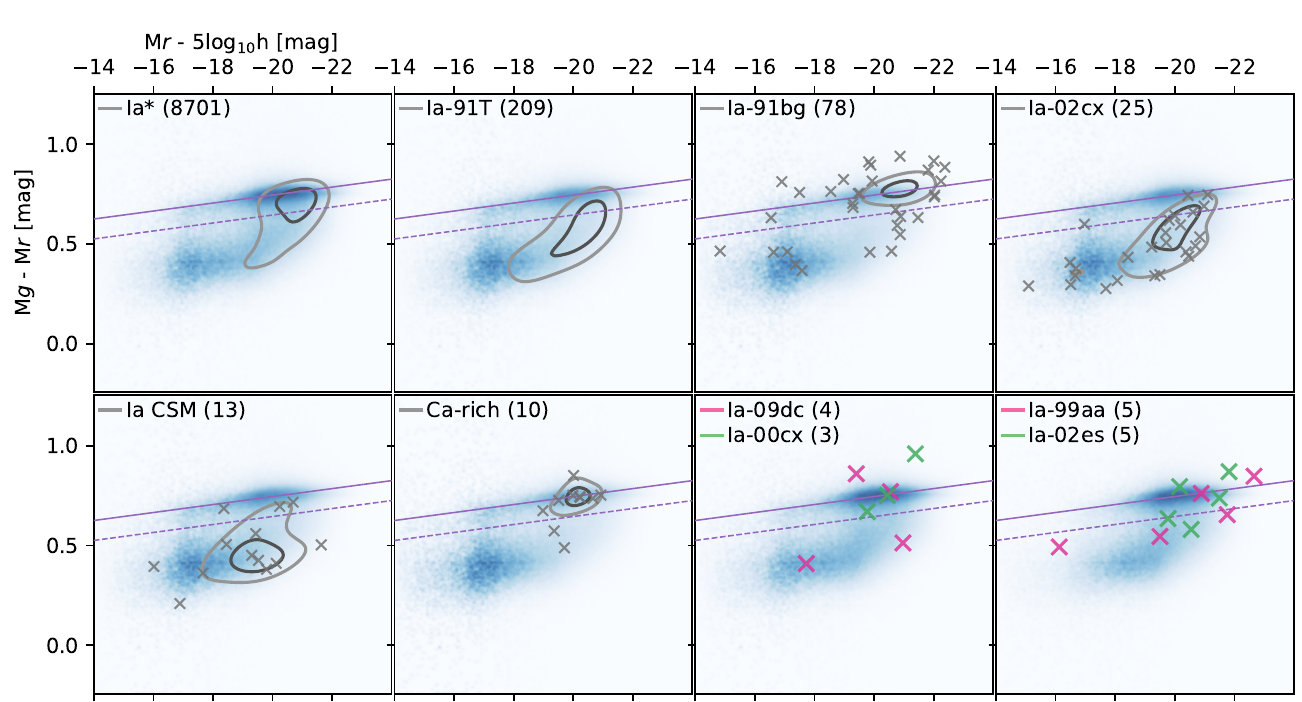}
\caption{\label{fig:IacolourMag}
Host galaxy rest-frame $r$-band absolute magnitude and $g$-$r$ colour of SN Ia subtypes in the photometric sample.
Contours and scatter points follow the same style as in Figure \ref{fig:CCMassSFR}.
Magenta lines indicate the best-fitting red sequence (solid) and its `blue edge' ($1\sigma$ lower limit) in \citet{Masters10}, while background density maps show redshift-matched reference samples as described in Section \ref{sec:zMatchedResampling}.
Normal SN Ia occurs in both blue cloud and red sequence galaxies, while SN Ia subtypes show significant heterogeneity in their host properties.
We compare the host luminosity and color of SN Ia subtypes, and between SN Ia hosts and the unweighted, redshift-matched samples here. We further control for possible redshift-driven selection effects in the comparison shown in Figure \ref{fig:IacolourMagOffset}.
See Section \ref{sec:IaMassSFR} for discussion.
}
\end{figure*}

\begin{figure*}
\centering
\includegraphics[width=0.495\linewidth]{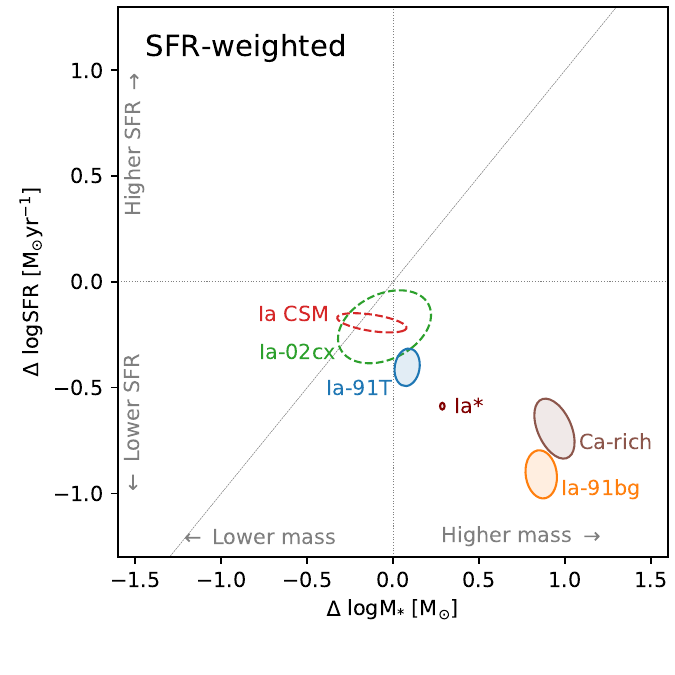}
\includegraphics[width=0.495\linewidth]{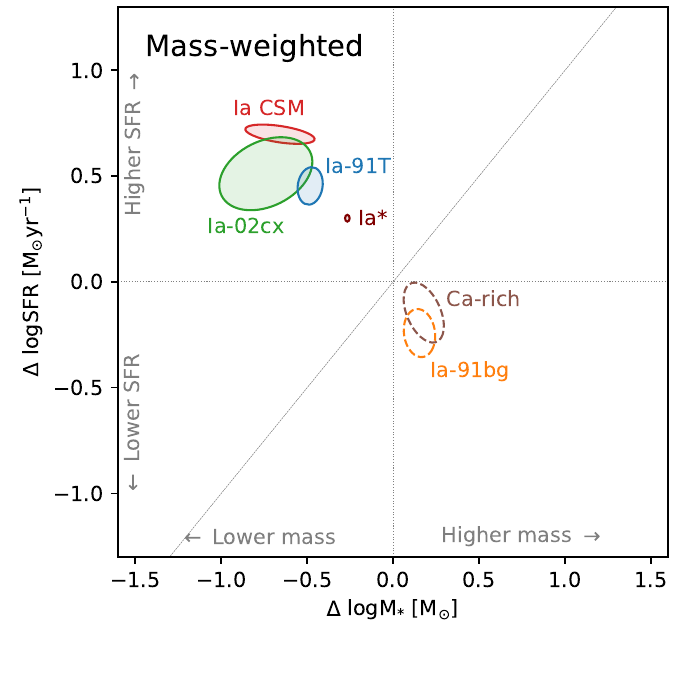} 
\caption{ \label{fig:IaMassSFROffset}
Host stellar mass and SFR offsets of SN Ia subtypes under the baseline of average massive stars (i.e., using SFR-weighted reference sample; left) and average stars (i.e., using mass-weighted reference sample; right).
The ellipses follow the same style as in Figure \ref{fig:CCMassSFROffset}.
Similarly, for subtypes in solid ellipses, assuming SN rates are proportional to the weighting variable (SFR or mass) leads to mean host mass and SFR that are significantly different from the observed population; conversely, for subtypes in dashed circles, this assumption results in mean host mass and SFR consistent with the observed population.
SN Ia-91T, Ia-02cx, and Ia-CSM hosts have consistent or similar properties with the `hosts' of average massive stars;
while SN Ia-91bg and Ca-rich transient hosts are consistent in properties with the `hosts' of average stars.
See Section \ref{sec:IaMassSFR} for discussion.
}
\end{figure*}

We show the stellar mass and SFR of SN Ia hosts in Figure \ref{fig:IaMassSFR}, grouped by subtypes. Similar as in Figure \ref{fig:CCMassSFR}, we show the best-fitting SFMS and its $1\sigma$ lower limit in \citet{Chang15} and the redshift-matched reference samples for each subtype.
Four subtypes with small sample sizes (SN Ia-09dc, Ia-00cx, Ia-99aa, and Ia-02es) are grouped into two panels. We use the general volumetric sample in Appendix \ref{appendix:Mockcatalogue} as the reference sample in these two panels, but in our later analysis, we use their corresponding distribution-matched reference samples.
Most SN Ia in our spectroscopic and photometric samples are unspecified SN Ia*, and the majority should be `normal' SN Ia rather than subtypes, as discussed earlier in Section \ref{sec:Subtypes}.
Those SN Ia* occurs in both star-forming and quiescent galaxies.
About $47.8\pm2.0$ per cent SN Ia* hosts are SFMS galaxies (i.e., above the $1\sigma$ lower limit in Figure \ref{fig:IaMassSFR}) in the spectroscopic sample; while about $40.7\pm1.0$ per cent SN Ia* hosts are red sequence galaxies (i.e., above the `blue edge' line in Figure \ref{fig:IacolourMag}).
Also, the mean host properties of SN Ia* progenitors lie in-between the mean `host properties' of average massive stars and average stars (Figure \ref{fig:IaMassSFROffset}).
Such heterogeneity of host properties implies a wide range of progenitor ages, as already pointed out in previous works (see the review of \citealt{Maoz12Review, Maoz14ARAA}).

\begin{figure}
\centering
\includegraphics[width=\linewidth]{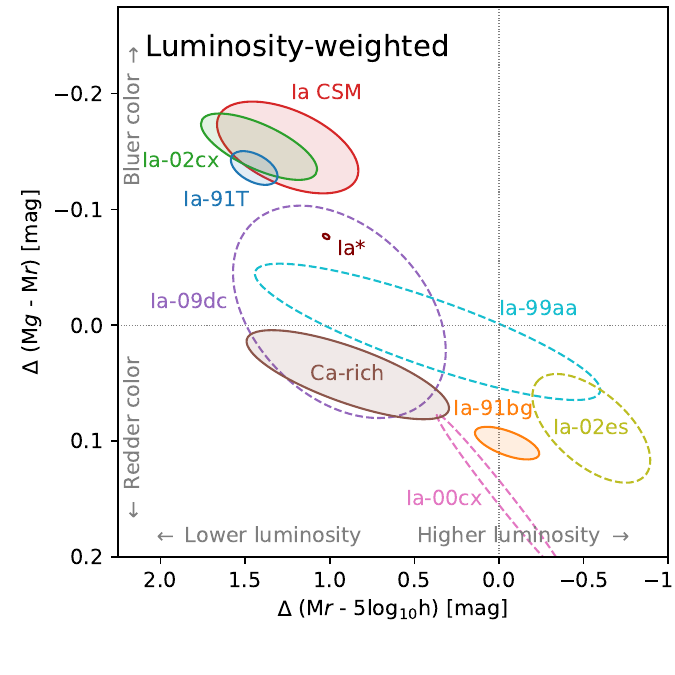}
\caption{\label{fig:IacolourMagOffset}
Host rest-frame absolute magnitudes and colours offsets of SN Ia subtypes under the baseline of rest-frame $r$-band stellar light (i.e., using luminosity-weighted reference samples).
Ellipses follow the same style as in Figure \ref{fig:CCMassSFROffset}.
Similar as in Figure \ref{fig:CCcolourMagOffset}, solid ellipses imply that assuming SN rates are proportional to the r-band luminosity produces mean host colors and luminosities that are statistically different from those observed; meanwhile, dashed circles indicate that the assumption yields consistent mean host luminosity and color with those observed.
We revert both axes so that the right side indicates relatively higher luminosity, and the upper edge indicates relatively bluer colour.
See Section \ref{sec:IaMassSFR} for discussion.
}
\end{figure}

We then compare host properties across SN Ia subtypes. 
In contrast to the similarity of host properties across CC SN subtypes, the host properties across SN Ia subtypes differ dramatically (Figures \ref{fig:IaMassSFR}, \ref{fig:IacolourMag}).
We begin with the host properties of SN Ia-91T and Ia-91bg, the two most abundant SN Ia subtypes in our sample with distinct host properties; then, we compare the host properties of other subtypes to normal SN Ia*, as well as these two subtypes.
Most SN Ia-91T hosts are on the SFMS, showing similar distribution as CC SN hosts. The fraction of SFMS galaxies in SN Ia-91T hosts ($66.7_{-11.6}^{+10.3}$ per cent) is higher than SN Ia*, while the fraction of red sequence hosts ($24.6_{-5.4}^{+6.1}$ per cent) is also lower than SN Ia*.
In contrast, SN Ia-91bg hosts reside in the clump of massive quiescent galaxies and at the luminous side of the red sequence; only $19.4_{-10.3}^{+15.0}$ per cent SN Ia-91bg hosts are in the SFMS, lower than SN Ia*; meanwhile, $84.8_{-9.1}^{+6.6}$ per cent SN Ia-91bg hosts are in the red sequence, higher than SN Ia*.
Notably, SN Ia-91T hosts are, on average, comparable in mass while more quiescent than the `hosts' of average massive stars (Figure \ref{fig:IaMassSFROffset}, left), indicating that the rates are not simply proportional to the current SFR; while the hosts of SN Ia-91bg are consistent in mass and SFR with the `hosts' of average stars (Figure \ref{fig:IaMassSFROffset}, right).
Under the baseline of average stellar light (Figure \ref{fig:IacolourMagOffset}), the hosts of SN Ia-91T are clearly bluer in colour and fainter in luminosity than the `host' of average stellar light, while the hosts of SN Ia-91bg are redder.
The differences in host mass and sSFR between SN Ia* and Ia-91T, as well as between SN Ia* and Ia-91bg, are significant. However, in the case of Ia-91T versus Ia*, the mass difference becomes statistically insignificant for low-redshift SNe, and the sSFR difference falls below significance for SNe detected by wide-field surveys.
The dramatic contrast of host properties between SN Ia-91T and Ia-91bg can be attributed to their different progenitor delay times which has been discussed in previous works (e.g., \citealt{Neill09, Taubenberger17, Barkhudaryan19, Hakobyan20}). % <<< check references!
SN Ia-91T has relatively short-lived progenitors contributed by recent and ongoing star formation; therefore, SN Ia-91T hosts have similar properties as CC SN hosts.
The sparsity of SN Ia-91T in quiescent galaxies indicates that their progenitors reach the condition of explosion before the host become `quenched' (i.e., the delay time is shorter than the timescale of galaxy evolution).
Conversely, SN Ia-91bg progenitors are relatively long-lived.
The preference of SN Ia-91bg for massive quiescent hosts implies that their explosions are substantially delayed -- long enough for their hosts to become `fully quenched' (i.e., the delay time is close to, or longer than, the timescale of galaxy evolution).
The down-sizing formation of galaxies \citep{Neistein06} predicts that such massive quiescent galaxies form early in the history of the universe.
SN Ia-91bg may, therefore, arise from the oldest stellar populations.
Besides the longer delay times, other mechanisms might also enhance the rates of SN Ia-91bg in massive quiescent galaxies.
For example, dynamical encounters with stars and interactions with supermassive black holes can harden binaries in certain configurations \citep{Heggie75}, increase the chance of close binary interactions, and boost the frequencies for certain stellar transients.
Hosts of SN Ia-02cx (or `Iax') prefer star-forming galaxies (Figures \ref{fig:IaMassSFR}, \ref{fig:IacolourMag}).
About $83.3_{-27.0}^{+13.0}$ per cent SN Ia-02cx hosts in the spectroscopic sample are SFMS galaxies, consistent with SN Ia-91T and higher than SN Ia*; only $20.0_{-11.9}^{+18.4}$ per cent SN Ia-02cx hosts in the photometric sample are red sequence galaxies, also consistent with SN Ia-91T while lower than SN Ia*.
Their host mass and SFR, on average, are broadly consistent with those of average massive stars (Figure \ref{fig:IaMassSFROffset}, left).
This conclusion remains unchanged for both the low-redshift and wide-field subsamples.
The preference of SN Ia-02cx for star-forming galaxies is pointed out in earlier works (see the review of \citealt{Jha17Iax}).
Their progenitors could also have short delay times like SN Ia-91T progenitors.
The interacting subtype SN Ia-CSM also clearly prefers star-forming and blue galaxies (Figures \ref{fig:IaMassSFR}, \ref{fig:IacolourMag}).
About $75.0_{-46.6}^{+22.2}$ per cent SN Ia-CSM hosts in the spectroscopic sample are SFMS galaxies; only $23.1_{-16.1}^{+26.7}$ per cent SN Ia-CSM hosts in the photometric sample are red sequence galaxies. Both fractions are consistent with SN Ia-91T and Ia* hosts.
Notably, their hosts are consistent in stellar mass and SFR with the `hosts' of average massive stars (Figure \ref{fig:IaMassSFROffset}, left). The conclusion remains the same for the wide-field sample, which still has a large enough sample for the test.
Given the limited sample size, they are only different from SN Ia-91bg and Ca-rich transients in their host mass and SFR under either baseline.
Under the stellar light baseline (Figure \ref{fig:IacolourMagOffset}), SN Ia-CSM hosts are also bluer than the hosts of SN Ia-91bg and Ca-rich transients, while the difference in luminosity is only significant compared to SN Ia-91bg.
The preference of SN Ia-CSM for blue and star-forming galaxies indicates that they originate from younger stellar populations with short progenitor delay times.

Ca-rich transients show strong Ca\RNum{2} lines during the nebular phase and are faint in general. We group Ca-rich transients under SN Ia because evidence suggests their thermonuclear origin (e.g., \citealt{Waldman11, De20}).
Hosts of Ca-rich transients are quiescent in general (Figure \ref{fig:IaMassSFR}, \ref{fig:IacolourMag}). None of their hosts in the spectroscopic sample are SFMS galaxies (with a $95$ per cent CI upper limit of $29.2$ per cent), consistent with SN Ia-91bg but lower than SN Ia*; while $72.7_{-29.2}^{+18.9}$ per cent of their hosts in the photometric sample are red sequence galaxies, consistent with SN Ia-91bg and higher than SN Ia*.
The host mass and sSFR difference between SN Ia* and Ca-rich transients remain significant for the low-redshift sample, which still has a large enough sample size for the test.
Similar to SN Ia-91bg, their hosts are consistent in stellar mass and SFR with average stars (Figure \ref{fig:IaMassSFROffset}, right).
Their hosts are more massive and quiescent on average than SN Ia-91T, Ia-02cx, and Ia-CSM hosts under the baseline of average stars; while under the baseline of average massive stars, they are also different from SN Ia* hosts with a lower stellar mass.
Like SN Ia-91bg, Ca-rich transients may also have long-lived progenitors.

Finally, we discuss the four minor subtypes with limited sample sizes: SN Ia-09dc, Ia-00cx, Ia-99aa, and Ia-02es.
As we noted earlier, these subtypes require ancillary data and human expertise to identify (Section \ref{sec:Subtypes}); the discussion here is limited by their sample sizes.
None of these subtypes reach the required minimal sample size to estimate the offsets and their error ellipses of spectroscopic properties. For the photometric sample, all these subtypes are broadly consistent in host absolute magnitude and colour with the `hosts' of average $r$-band stellar light (Figure \ref{fig:IacolourMagOffset}).
SN Ia-09dc (or `Ia-06gz', `Ia-03fg') feature higher peak luminosity than average SN Ia and are commonly denoted as `super-Chandrasekhar' SN Ia based on the estimate of ejecta mass.
The hosts of SN Ia-09dc can be either star-forming or quiescent. None of their two hosts in the spectroscopic sample are SFMS galaxies, putting a $95$ per cent CI upper limit of $66.7$ per cent; two of their four hosts in the photometric sample are red sequence galaxies ($50.0\pm37.7$ per cent).
Their host properties also show no significant difference with other Ia subtypes under either baseline.
The results here indicate that the hosts of SN Ia-09dc may not be homogeneous, and they could have diverse progenitor channels -- some could be the mergers of binary white dwarfs (e.g., \citealt{Taubenberger13}) with longer delay times, while some could be driven by interactions with circumstellar medium, with potentially similar preference as SN Ia-CSM.
SN Ia-00cx has similar and more extreme spectroscopic properties as SN Ia-91T (e.g., even stronger iron-peak lines and weaker intermediate element lines; \citealt{Li0100cx}).
The only SN Ia-00cx host in the spectroscopic sample lies within the SFMS ($95$ per cent CI lower limit of $14.7$ per cent); while the three SN Ia-00cx hosts in the photometric sample are all above the `blue edge' of the red sequence, indicating a $95$ per cent CI lower limit of $47.6$ per cent for the red sequence fraction, higher than SN Ia-91T and Ia-02cx.
Under the stellar light baseline, their hosts show redder colour than SN Ia-02cx and Ia-CSM, the two subtypes favoring star-forming galaxies.
We conclude that their hosts are more quiescent based on the fraction and the host colour offsets. 
Despite the similarity of spectroscopic properties, SN Ia-00cx could have a different origin than SN Ia-91T, as already hinted in \citet{Taubenberger17}.
SN Ia-99aa is another subtype with potential connections to SN Ia-91T; its intermediate spectroscopic properties bridge normal SN Ia and Ia-91T in the parameter space of SN Ia.
None of the two SN Ia-99aa hosts in the spectroscopic lies inside the SFMS, indicating a $95$ per cent CI upper limit of $66.7$ per cent for the SFMS fraction.
Meanwhile, two of the five SN Ia-99aa hosts are inside the red sequence ($40.0_{-30.6}^{+39.1}$ per cent).
Their hosts also show a broad range of luminosities.
Under the stellar light baseline, a $t$-test reveals redder colour than SN Ia-02cx hosts but not hosts of other subtypes.
Given the limited sample size, we cannot identify any possible systematic difference of SN Ia-99aa hosts with SN Ia* or Ia-91T hosts.
Finally, we focus on the subluminous and slowly-fading SN Ia-02es, a subtype in the underpopulated region in the SN Ia light curve parameter space.
None of the two SN Ia-02es hosts in our spectroscopic sample are in the SFMS region, indicating a $95$ per cent CI upper limit of $66.7$ per cent for the SFMS fraction. Meanwhile, three of the five SN Ia-02es hosts in the photometric sample are inside the red sequence, indicating a red sequence fraction of $60.0_{-39.1}^{+30.6}$ per cent, and one of the remaining two is on the `blue edge' line.
The result here is broadly consistent with the earlier speculation that SN Ia-02es hosts are mostly quiescent \citep{Taubenberger17}. % <<< conclusions?

\subsection{Rate dependence relationship of CC SN} \label{sec:kModelCC}

\begin{figure*}
\centering
\includegraphics[width=0.48\linewidth]{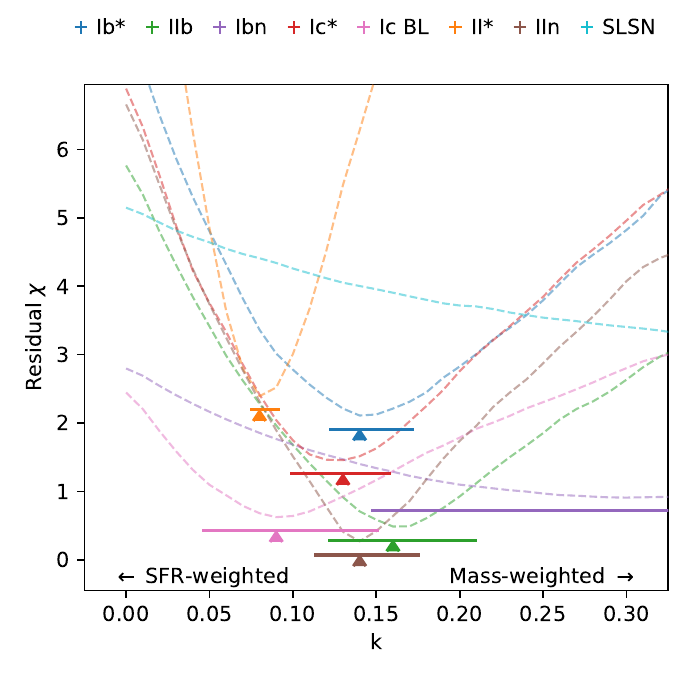}
\includegraphics[width=0.48\linewidth]{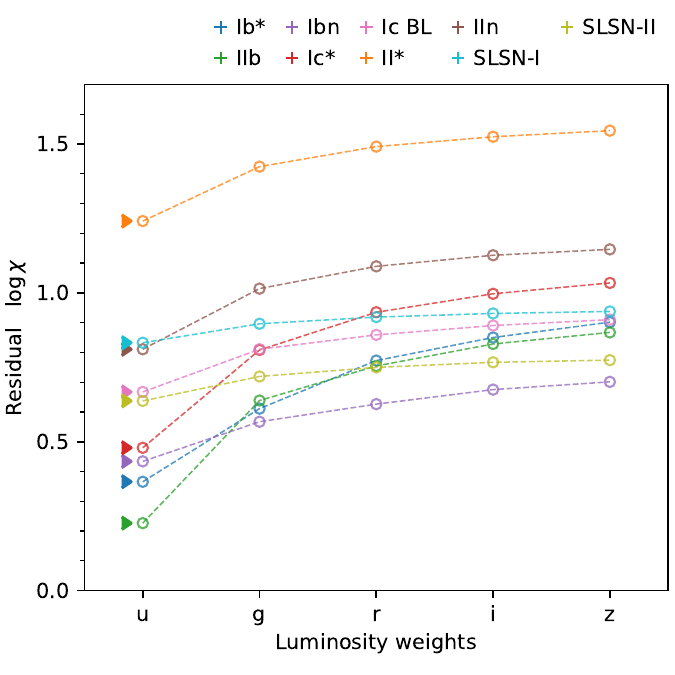}
\caption{\label{fig:kParamCC}
\textit{Left:} Minimizing the difference between the observed CC SN mean host spectroscopic properties ($\log M^*$, $\log\,\text{SFR}$) and predictions of a simple bilinear rate dependence model (`\textit{k}-model,' Equation \ref{eq:kmodel}; Section \ref{sec:WeightedResampling}). The best $k$-value for each subtype is indicated using a triangle symbol.
CC SN rates are closely dependent on, but not directly proportional to, their host SFR, given their small but non-zero $k$-values.
\textit{Right:} Minimizing the difference between the observed CC SN mean photometric properties (rest-frame $u$-$r$, $g$-$r$, $r$-$i$, $r$-$z$ colours) and the predicted host properties assuming that rate scales with the luminosity of a rest-frame band.
CC SN rates follow closer, or scale better with the luminosity in the rest-frame $u$-band than other bands.
In either panel, the differences in mean host properties are normalized by the error ellipse of the observed host galaxies.
}
\end{figure*}

How SN rates and hence the numbers of pre-explosion progenitors scale with galaxy properties gives valuable insights into the nature of their progenitors.
Absolute rate measurement (\citealt{Li11_rates, Graur17_rates}) requires time-controlled observation and well-understood selection effects, which is not possible with our dataset.
However, by assuming that the observed SN rates follow a bilinear function of host stellar mass and SFR like the empirical model in \citet{Sullivan06} or scales with the luminosity in a rest-frame band, we may test if the assumed rate dependence models could reproduce the observed mean properties of the host for a redshift-matched and rate-weighted sample. %

Figure \ref{fig:kParamCC} (left) shows the residual error ($\chi$) of $\log M^*$ and $\log\,\text{SFR}$ as a function of parameter $k$ in our simple bilinear rate dependence model (or `$k$-model;' Equation \ref{eq:kmodel}; Section \ref{sec:WeightedResampling}).
We calculate $\chi$ by normalizing the distance in the ($\log M^*$, $\log\,\text{SFR}$) plane with the error ellipses of the observed mean host properties, taking the covariance into consideration.
For clarity, we mark the $k$-values that minimize the residue error using triangle symbols; the associated errors for $k$, heuristically estimated from the minimum of residual error ($\chi_{\text{min}}$) by numerically solving $k'$ in $\chi^2 (k')=\chi^2_{\text{min}} + 1$, are indicated using solid horizontal bars.
As discussed in Section \ref{sec:WeightedResampling}, if SN rates are proportional to the current SFR, the $k$-values should be consistent with zero. Conversely, if SN rates are not exclusively contributed by recently-born stars, the rates may also depend on the stellar mass, and the $k$-values are expected to be non-zero.
CC SN subtypes have $k$-values close to but inconsistent with zero. Their observed rates closely depend on the current SFR of the hosts,
however, assuming that CC SN rates scale exactly with SFR (i.e., $k=0$) cannot explain the observed mean properties of CC SN hosts.
For subtypes including SN IIb, IIn, Ic-BL, the best-fitting $k$-values well describe the observed rates; their minima of $\chi(k)$ is under $1$ (i.e., reduced $\chi_r^2$ under $1$ for two observations, $\Delta\log M^*$ and $\Delta\log\,\text{SFR}$, and one free parameter, $k$), which implies an acceptable fit to the observed mean host properties.
The non-zero $k$-values indicate that there could be
(1) a population of delayed progenitors (i.e., with delay time longer than H$\alpha$ SFR indicator can trace), which show up as the dependency on host stellar mass;
(2) secondary dependent factors, such as metallicity, that affect the SN rate-to-SFR ratio;
(3) other selection effects (e.g., host dust extinction) that are not properly controlled for.
These scenarios are not exclusive and could work in conjunction with one another.
The best-fitting $k$-values also differ across certain CC SN subtypes.
SN II* has a lower $k$-value than SN Ib* and Ic*; the observed rates of SN II* show stronger dependence on host SFR than SN Ib* and Ic*.
Note that H$\alpha$ luminosity traces star formation over the past $10$ Myr (e.g., \citealt{FloresVelazquez21}), close to the longest delay time of SN Ib and Ic progenitors under the massive single-star scenario (based on the ZAMS mass range in \citealt{Heger03}), but shorter than the delay time of SN II if the progenitors are red supergiants.
Assuming that the difference in $k$-values can be attributable to a delay time effect and assuming massive single-star progenitors, the rates of SN Ib* and Ic* should follow closer (or scale better) with host SFR than SN II* does, in contrast to the results above.
A possible scenario is that a substantial fraction of SNe Ib* and Ic*, contributed by close binary progenitors rather than massive single stars, are delayed beyond the timescale of the H$\alpha$ SFR indicator, and even the average delay time of SN II*.
We find no significant difference in $k$-values between SN Ib* and Ic*.
SN IIn has a higher $k$-value than the sibling subtype SN II*, while the transitional SN IIb has a $k$-value consistent with SN Ib* but higher than SN II*.
This implies similar nature of SN Ib* and IIb progenitors but possibly different progenitors for SN IIn and II*.
SN Ic-BL may have a relatively lower $k$-value, but the difference is not statistically significant.
Due to the limited and likely biased samples, we do not discuss the results on SLSN and Ibn.

\begin{figure}
\centering
\includegraphics[width=\linewidth]{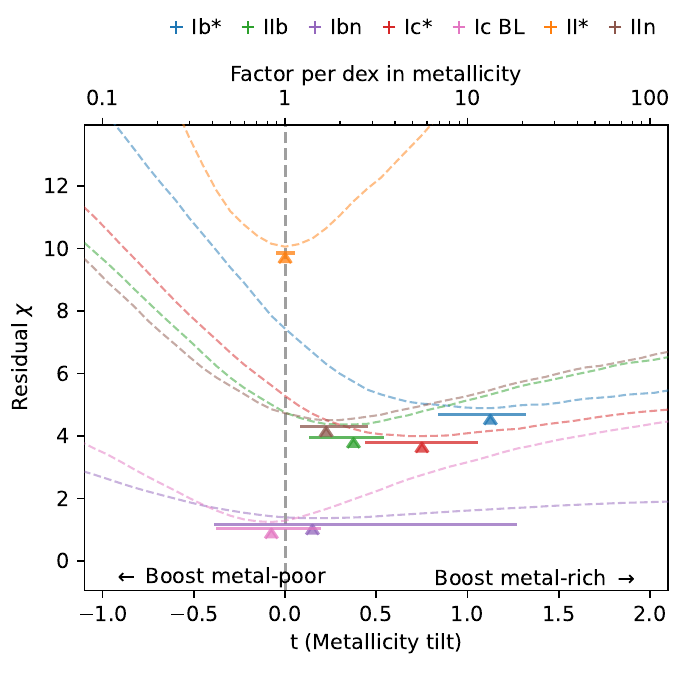}
\caption{\label{fig:zTiltCC}
Minimizing the difference between the observed CC SN mean host spectroscopic properties ($\log M^*$, $\log\,\text{SFR}$) and the predictions of the `metallicity tilt' rate dependence model (`\textit{t}-model,' Equation \ref{eq:tmodel}; Section \ref{sec:RateScaling}).
The differences in mean host properties are normalized using the error ellipse of the observed host galaxies.
The best-fitting $t$ parameter for each subtype and the corresponding factor per dex increase in metallicity is indicated using a triangular symbol.
Error ranges in $t$ are indicated as horizontal bars.
Rates of SN II* are insensitive to host metallicity ($t$-value close to zero), but SN Ib* and Ic* rates are enhanced in metal-rich galaxies ($t$-values are positive).
}
\end{figure}

Besides the possible difference in delay times,
the non-zero $k$-values could be a result of other secondary dependent factors such as host metallicity.
We further test the scenario in which the observed SN rate is proportional to the host SFR at a constant metallicity, while the SN production efficiency depends on the metallicity, represented as a `metallicity tilt' parameter $t$ (Equation \ref{eq:tmodel}; Section \ref{sec:RateScaling}).
This $t$ parameter represents the increase of SN production efficiency, in logarithmic scale, per dex increase in host metallicity. Figure \ref{fig:zTiltCC} shows the results.
Similar as in Figure \ref{fig:kParamCC}, we attempt to minimize the residual error $\chi(t)$ in the ($\log M^*$, $\log\text{SFR}$) plane.
We notice that SN Ib* and Ic* have positive $t$-values, indicating that their SN production efficiencies are substantially higher in metal-rich hosts, with a factor of about $10$ increase per dex increase in metallicity.
SN II*, however, has a $t$-value consistent with zero; the SN production efficiency of SN II* is independent of, or insensitive to, the host metallicity; the rates should also scale with host SFR.
Notably, the best-fitting $k$-value of SN II (Figure \ref{fig:kParamCC}) rules out the simple proportionality of SN II rates to SFR.
The discrepancy of the $k$-model and $t$-model could be attributed to the different subsamples analyzed; we only use hosts with metallicity measurements, which are mostly star-forming galaxies, when finding the $t$ parameter.
For other subtypes, the best-fitting $t$-values are broadly consistent with zero.
With the larger photometric sample, we further test the proportionality of SN rates to the rest-frame luminosity by minimizing the residual error of mean host colours in the four-dimensional space ($u$-$r$, $g$-$r$, $r$-$i$, $r$-$z$).
Instead of fitting a parameter like $k$, we choose the luminosity of each rest-frame band as weights and identify the rest-frame band that best minimizes the residual error $\chi$ (Figure \ref{fig:kParamCC}, right panel).
For all CC SN subtypes, rest-frame $u$-band luminosity best reduces the residual error in colour space, indicating that the rates trace the young star contents of the hosts.
CC SN subtypes including SN Ib*, IIb, Ibn, and Ic* have minimal residual error around or under $3$, indicating that the observed rates scale well with the $u$-band luminosity of the host. %
Except for SN Ibn, other subtypes show significantly better scaling\footnote{We consider $p<0.05$ as significant under a $\chi^2$ likelihood ratio test with four observations and one free parameter.} in the rest-frame $u$-band than the second best band, and all subtypes show significantly better scaling in the $u$-band compared to the worst band.

\subsection{Rate dependence relationship of SN Ia}

\begin{figure*}
\centering
\includegraphics[width=0.48\linewidth]{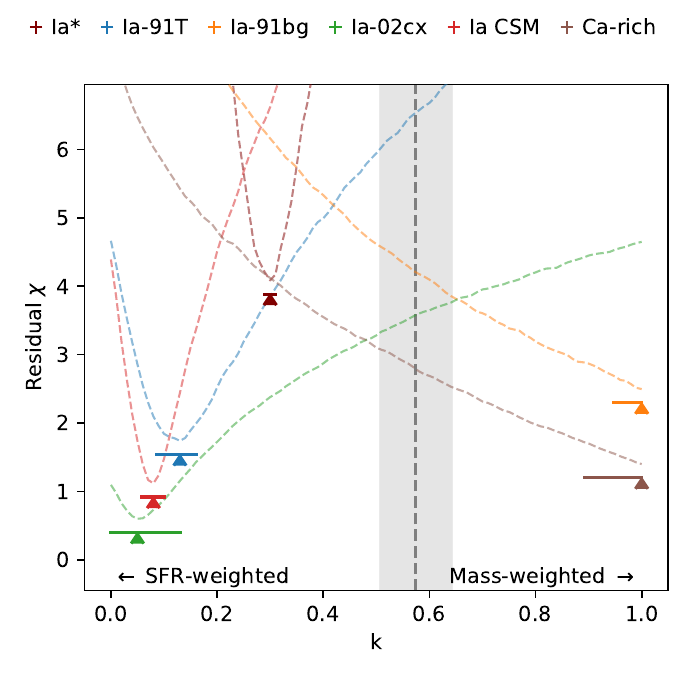}
\includegraphics[width=0.48\linewidth]{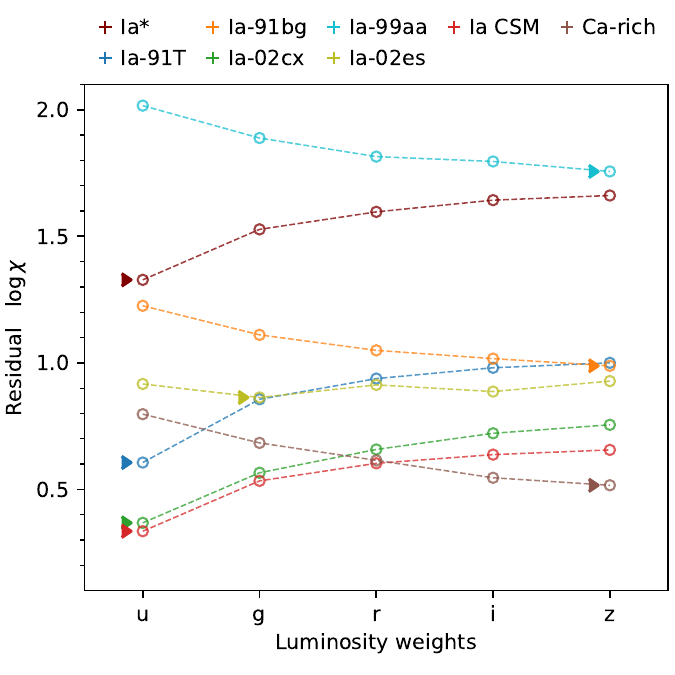}
\caption{\label{fig:kParamIa}
Similar as Figure \ref{fig:kParamCC}. \textit{Left:} Minimizing the differences between the observed SN Ia mean host spectroscopic properties ($\log M^*$, $\log\,\text{SFR}$) and the prediction of a simple bilinear rate dependence model (`\textit{k}-model;' Equation \ref{eq:kmodel}; Section \ref{sec:WeightedResampling}). The best $k$ parameter for each subtype is indicated using a triangle symbol.
The vertical dashed line and shade indicate the value and error of $k$ parameter in \citet{Sullivan06}.
Rates of SN Ia-91T, Ia-02cx, and Ia-CSM are closely dependent on, or even directly proportional to, their host SFR ($k$-value close to, or consistent with, zero); while the rates of SN Ia-91bg and Ca-rich are directly proportional to their host stellar mass ($k$ consistent with unity).
\textit{Right:} Minimizing the differences between the observed SN Ia mean host photometric properties (rest-frame $u$-$r$, $g$-$r$, $r$-$i$, $r$-$z$ colours) and the predicted host photometric properties assuming that rates scale with the luminosity in a rest-frame band.
Rates of SN Ia-91T, Ia-02cx, and Ia-CSM scale better with $u$-band luminosity, while rates of SN Ia-91bg and Ca-rich transients scale better with $z$-band luminosity.
In either panel, the differences in mean host properties are normalized using the error ellipse of the observed host galaxies.
}
\end{figure*}

We also examine if a simple combination of host stellar mass and SFR can describe the observed mean host properties ($\log M^*$, $\log\,\text{SFR}$) of SN Ia subtypes, using the simple bilinear rate dependence model ($k$-model; Equation \ref{eq:kmodel}; Section \ref{sec:WeightedResampling}) as we applied to CC SN subtypes (Section \ref{sec:kModelCC}).
Figure \ref{fig:kParamIa} (left) summarizes the best-fitting $k$-values for SN Ia subtypes. 

SN Ia subtypes have vastly diverse host properties, indicating a greater contrast in progenitor delay times than CC SN subtypes (Section \ref{sec:IaMassSFR}). The diversity of host properties and contrast in progenitor delay times are also evident in the best-fitting $k$-values for SN Ia subtypes.
Notably, SN Ia-CSM and Ia-91T have $k$-values very close to (but inconsistent with) zero, and SN Ia-02cx has a $k$-value consistent with zero; their observed rates and hence the numbers of pre-explosion progenitors are closely dependent on, or are even directly proportional to, the host SFR.
\footnote{For Ia-CSM, the conclusion based on $k$-value is different from our conclusion in Figure 10 due to the different modeling techniques and interpretation of statistical significance.}
The close dependence or direct proportionality of rates with host SFR is consistent with the fact that their hosts are mostly star-forming galaxies (Section \ref{sec:IaMassSFR}).
In contrast, SN Ia-91bg and Ca-rich transients have $k$-values consistent with unity; their observed rates and the numbers of pre-explosion progenitors are proportional to their host stellar mass and are insensitive to (or independent of) their host SFR.
The proportionality of rates with host stellar mass is also consistent with their preference for quiescent galaxies.
For SN Ia-02cx, Ia-91T, Ia-CSM, and Ca-rich transients, the residual errors ($\chi$) are under $2$, indicating that the simple $k$-model reasonably fits the observed mean host properties.
Finally, `normal' SN Ia* has an intermediary $k$-value of $0.30\pm0.01$;
the observed rate can be described by a simple combination of host SFR and stellar mass.
However, our estimate is inconsistent with the implied $k$-value based on the $A$, $B$ coefficients in \citet{Sullivan06} (vertical line and gray band). The $k$-value in \citet{Sullivan06} shows stronger dependence on stellar mass and weaker dependence on SFR, compared to our results.
Also, the residual error $\chi$ is above $4$ for SN Ia*, indicating a less optimal fit to the observed host properties than other SN Ia subtypes, likely due to the greater sample size and simplified assumption.

Using the photometric sample, we also test if the rates of SN Ia subtypes scale with the host luminosity in a specific rest-frame band.
We identify the band that minimizes the residual error in the rest-frame colour space ($u$-$r$, $g$-$r$, $r$-$i$, $r$-$z$), as we do earlier for CC SN subtypes (Section \ref{sec:kModelCC}).
The larger sample size allows us to analyze two more subtypes, SN Ia-02es and Ia-99aa.
As shown in Figure \ref{fig:kParamCC} (right), the observed rates of SN Ia-91T, Ia-02cx, and Ia-CSM follow closer (or scale better) with the luminosity in the $u$-band than in other bands; while the rates of SN Ia-91bg, Ia-99aa, and Ca-rich transients scale better with the $z$-band luminosity. Aside from that, SN Ia-02es show no specific preference for any band.
Except for SN Ia-CSM, Ca-rich, and Ia-02es, the `best-fitting' band of other subtypes show significantly better rate scaling ($p<0.05$ under a likelihood ratio test) than the second best band, and all subtypes show significantly better scaling in the best-fitting band compared to the worst band, even including SN Ia-02es.

\subsection{Summary of Results}

We further summarize the host galaxy properties and rate dependency relationships of CC SN and SN Ia subtypes in Tables \ref{tab:summarytablecc} and \ref{tab:summarytableia}.

\begin{landscape}
\begin{table}
\caption{Statistics of host physical properties by subtype.}
\begin{threeparttable}
\label{tab:summarytablespec}
\begin{tabular}{@{\extracolsep{4pt}}lrrrrrrrrrrrrrrrrr@{}}
\hline
Name & \multicolumn{2}{c}{Number} & \multicolumn{5}{c}{$\log$M}                                         & \multicolumn{5}{c}{$\log$SFR}                                                             & \multicolumn{5}{c}{12+[O/H]} \\
\cline{2-3} \cline{4-8} \cline{9-13} \cline{14-18}
     & $N_1$ & $N_2$              & $\mu$ & $\sigma$ & SEM & $\Delta_{\text{S}}$ & $\Delta_{\text{M}}$  & $\mu$ & $\sigma$ & SEM & $\Delta_{\text{S}}$ & $\Delta_{\text{M}}$    & $\mu$ & $\sigma$ & SEM & $\Delta_{\text{S}}$ & $\Delta_{\text{M}}$ \\
\multicolumn{18}{c}{} \\
\multicolumn{18}{c}{Core-collapse supernovae} \\
\hline
CC & 1472 & 907 & $9.982$ & $0.762$ & $0.020$ & $0.165$ & $-0.519$ & $-0.341$ & $0.708$ & $0.018$ & $-0.248$ & $0.492$ & $8.845$ & $0.263$ & $0.009$ & $-0.009$ & $-0.099$ \\
$\triangleright\;$SE & 382 & 237 & $9.989$ & $0.780$ & $0.040$ & $0.285$ & $-0.447$ & $-0.404$ & $0.721$ & $0.037$ & $-0.218$ & $0.466$ & $8.886$ & $0.253$ & $0.016$ & $0.060$ & $-0.038$ \\
$\quad$$\triangleright\;$Ib & 212 & 133 & $9.953$ & $0.761$ & $0.052$ & $0.320$ & $-0.430$ & $-0.458$ & $0.729$ & $0.050$ & $-0.217$ & $0.441$ & $8.879$ & $0.247$ & $0.021$ & $0.072$ & $-0.033$ \\
$\quad$$\quad$$\triangleright\;$Ib* & 117 & 74 & $9.992$ & $0.659$ & $0.061$ & $0.367$ & $-0.372$ & $-0.378$ & $0.704$ & $0.065$ & $-0.129$ & $0.536$ & $8.924$ & $0.225$ & $0.026$ & $0.118$ & $0.012$ \\
$\quad$$\quad$$\triangleright\;$IIb & 84 & 53 & $9.891$ & $0.753$ & $0.082$ & $0.251$ & $-0.500$ & $-0.533$ & $0.659$ & $0.072$ & $-0.292$ & $0.349$ & $8.839$ & $0.262$ & $0.036$ & $0.035$ & $-0.072$ \\
$\quad$$\quad$$\triangleright\;$Ibn & 12 & 7 & $9.873$ & $1.468$ & $0.424$ & $-0.005$ & $-0.658$ & $-0.778$ & $1.162$ & $0.335$ & $-0.735$ & $0.024$ & $8.662$ & $0.205$ & $0.078$ & $-0.165$ & $-0.260$ \\
$\quad$$\triangleright\;$Ic & 154 & 100 & $9.986$ & $0.829$ & $0.067$ & $0.294$ & $-0.441$ & $-0.364$ & $0.770$ & $0.062$ & $-0.159$ & $0.503$ & $8.883$ & $0.288$ & $0.029$ & $0.059$ & $-0.040$ \\
$\quad$$\quad$$\triangleright\;$Ic* & 126 & 81 & $10.004$ & $0.846$ & $0.075$ & $0.338$ & $-0.404$ & $-0.365$ & $0.814$ & $0.073$ & $-0.143$ & $0.514$ & $8.899$ & $0.293$ & $0.033$ & $0.082$ & $-0.023$ \\
$\quad$$\quad$$\triangleright\;$Ic BL & 28 & 19 & $9.907$ & $0.741$ & $0.140$ & $0.112$ & $-0.591$ & $-0.360$ & $0.531$ & $0.100$ & $-0.236$ & $0.460$ & $8.818$ & $0.255$ & $0.059$ & $-0.025$ & $-0.118$ \\
$\triangleright\;$II & 1120 & 687 & $9.985$ & $0.755$ & $0.023$ & $0.152$ & $-0.530$ & $-0.325$ & $0.705$ & $0.021$ & $-0.245$ & $0.499$ & $8.833$ & $0.263$ & $0.010$ & $-0.031$ & $-0.115$ \\
$\quad$$\triangleright\;$II* & 785 & 494 & $10.010$ & $0.716$ & $0.026$ & $0.116$ & $-0.548$ & $-0.278$ & $0.679$ & $0.024$ & $-0.247$ & $0.517$ & $8.829$ & $0.266$ & $0.012$ & $-0.044$ & $-0.127$ \\
$\quad$$\triangleright\;$IIn & 127 & 67 & $10.006$ & $0.805$ & $0.071$ & $0.262$ & $-0.447$ & $-0.403$ & $0.750$ & $0.067$ & $-0.253$ & $0.450$ & $8.878$ & $0.224$ & $0.027$ & $0.036$ & $-0.057$ \\
$\triangleright\;$SLSN & 4 & 2 & $10.156$ & $0.938$ & $0.469$ & $-0.259$ & $-0.691$ & $-0.630$ & $0.818$ & $0.409$ & $-1.045$ & $-0.009$ & $8.636$ & $0.084$ & $0.059$ & $-0.334$ & $-0.361$ \\
$\quad$$\triangleright\;$SLSN-I & 2 & 1 & $9.733$ & $0.979$ & $0.693$ & $-0.501$ & $-0.973$ & $-0.635$ & $1.092$ & $0.772$ & $-0.878$ & $0.028$ & $8.552$ & $0.000$ & $0.000$ & $-0.403$ & $-0.437$ \\
$\quad$$\triangleright\;$SLSN-II & 2 & 1 & $10.579$ & $0.664$ & $0.469$ & $0.010$ & $-0.386$ & $-0.626$ & $0.382$ & $0.270$ & $-1.177$ & $-0.053$ & $8.720$ & $0.000$ & $0.000$ & $-0.264$ & $-0.284$ \\
\multicolumn{18}{c}{} \\
\multicolumn{18}{c}{Thermonuclear supernovae} \\
\hline
Ia & 2469 & 888 & $10.417$ & $0.650$ & $0.013$ & $0.293$ & $-0.269$ & $-0.414$ & $0.774$ & $0.016$ & $-0.586$ & $0.309$ & $8.904$ & $0.217$ & $0.007$ & $-0.012$ & $-0.071$ \\
$\triangleright\;$Ia* & 2329 & 831 & $10.424$ & $0.647$ & $0.013$ & $0.291$ & $-0.269$ & $-0.408$ & $0.773$ & $0.016$ & $-0.585$ & $0.311$ & $8.906$ & $0.216$ & $0.007$ & $-0.010$ & $-0.070$ \\
$\triangleright\;$Ia-91T & 69 & 36 & $10.182$ & $0.609$ & $0.073$ & $0.081$ & $-0.479$ & $-0.253$ & $0.738$ & $0.089$ & $-0.400$ & $0.461$ & $8.873$ & $0.223$ & $0.037$ & $-0.050$ & $-0.106$ \\
$\triangleright\;$Ia-91bg & 36 & 6 & $10.630$ & $0.549$ & $0.092$ & $0.864$ & $0.146$ & $-1.069$ & $0.683$ & $0.114$ & $-0.910$ & $-0.239$ & $8.906$ & $0.262$ & $0.107$ & $0.054$ & $-0.039$ \\
$\triangleright\;$Ia-09dc & 2 & 0 & $10.616$ & $0.176$ & $0.124$ & $0.806$ & $0.115$ & $-0.972$ & $0.590$ & $0.417$ & $-0.867$ & $-0.156$ & $-$ & $-$ & $-$ & $-$ & $-$ \\
$\triangleright\;$Ia-02cx & 12 & 8 & $9.734$ & $0.937$ & $0.271$ & $-0.037$ & $-0.745$ & $-0.327$ & $0.598$ & $0.173$ & $-0.200$ & $0.509$ & $8.875$ & $0.287$ & $0.101$ & $0.013$ & $-0.070$ \\
$\triangleright\;$Ia-00cx & 1 & 0 & $10.325$ & $0.000$ & $0.000$ & $0.009$ & $-0.430$ & $-0.062$ & $0.000$ & $0.000$ & $-0.374$ & $0.593$ & $-$ & $-$ & $-$ & $-$ & $-$ \\
$\triangleright\;$Ia-99aa & 2 & 0 & $11.155$ & $0.529$ & $0.374$ & $1.102$ & $0.453$ & $-0.335$ & $0.271$ & $0.192$ & $-0.457$ & $0.417$ & $-$ & $-$ & $-$ & $-$ & $-$ \\
$\triangleright\;$Ia-02es & 2 & 0 & $10.933$ & $0.208$ & $0.147$ & $1.299$ & $0.534$ & $-1.163$ & $0.121$ & $0.085$ & $-0.879$ & $-0.288$ & $-$ & $-$ & $-$ & $-$ & $-$ \\
$\triangleright\;$Ia CSM & 4 & 3 & $10.034$ & $0.403$ & $0.202$ & $-0.121$ & $-0.644$ & $0.009$ & $0.091$ & $0.046$ & $-0.185$ & $0.699$ & $8.957$ & $0.030$ & $0.017$ & $0.015$ & $-0.028$ \\
$\triangleright\;$Ca-rich & 7 & 2 & $10.453$ & $0.310$ & $0.117$ & $0.954$ & $0.160$ & $-1.065$ & $0.375$ & $0.142$ & $-0.680$ & $-0.159$ & $8.969$ & $0.030$ & $0.022$ & $0.269$ & $0.125$ \\
\hline
\end{tabular}
\begin{tablenotes}
\item Columns: $N_1$: numbers of hosts with mass and SFR; $N_2$: numbers of hosts with mass, SFR, and metallicity; $\mu$, $\sigma$, SEM: mean, standard deviation, and standard error of the mean; $\Delta_{\text{S}}$, $\Delta_{\text{M}}$: mean offsets with respect to average massive star and average star baselines.
\end{tablenotes}
\end{threeparttable}
\end{table}
\end{landscape}
% \begin{landscape}
\begin{table*}
\caption{Statistics of host photometric properties by subtype.}
\begin{threeparttable}
\label{tab:summarytablephoto}
\begin{tabular}{@{\extracolsep{4pt}}lrrrrrrrrr@{}}
\hline
Name & Number & \multicolumn{4}{c}{$M_r$} &\multicolumn{4}{c}{$M_g-M_r$} \\
\cline{3-6} \cline{7-10}
     &        & \multicolumn{1}{c}{$\mu$} & \multicolumn{1}{c}{$\sigma$} & \multicolumn{1}{c}{SEM} & \multicolumn{1}{c}{$\Delta_{\text{L}}$} & \multicolumn{1}{c}{$\mu$} & \multicolumn{1}{c}{$\sigma$} & \multicolumn{1}{c}{SEM} & \multicolumn{1}{c}{$\Delta_{\text{L}}$} \\
\multicolumn{10}{c}{} \\
\multicolumn{10}{c}{Core-collapse supernovae} \\
\hline
CC & 3888 & $-19.239$ & $2.073$ & $0.033$ & $0.246$ & $0.504$ & $0.214$ & $0.003$ & $-0.103$ \\
$\triangleright\;$SE & 887 & $-19.085$ & $2.010$ & $0.068$ & $-0.320$ & $0.530$ & $0.211$ & $0.007$ & $-0.083$ \\
$\quad$$\triangleright\;$Ib & 415 & $-18.938$ & $2.188$ & $0.107$ & $-0.954$ & $0.533$ & $0.207$ & $0.010$ & $-0.078$ \\
$\quad$$\quad$$\triangleright\;$Ib* & 216 & $-18.999$ & $2.398$ & $0.163$ & $-1.256$ & $0.551$ & $0.195$ & $0.013$ & $-0.062$ \\
$\quad$$\quad$$\triangleright\;$IIb & 172 & $-18.861$ & $1.871$ & $0.143$ & $-0.895$ & $0.519$ & $0.218$ & $0.017$ & $-0.084$ \\
$\quad$$\quad$$\triangleright\;$Ibn & 28 & $-18.880$ & $2.259$ & $0.427$ & $0.371$ & $0.465$ & $0.202$ & $0.038$ & $-0.156$ \\
$\quad$$\triangleright\;$Ic & 378 & $-19.068$ & $1.881$ & $0.097$ & $-0.374$ & $0.510$ & $0.204$ & $0.010$ & $-0.102$ \\
$\quad$$\quad$$\triangleright\;$Ic* & 300 & $-19.264$ & $1.834$ & $0.106$ & $-0.751$ & $0.528$ & $0.190$ & $0.011$ & $-0.084$ \\
$\quad$$\quad$$\triangleright\;$Ic BL & 78 & $-18.314$ & $1.866$ & $0.211$ & $1.114$ & $0.442$ & $0.238$ & $0.027$ & $-0.174$ \\
$\triangleright\;$II & 3016 & $-19.297$ & $2.090$ & $0.038$ & $0.291$ & $0.498$ & $0.212$ & $0.004$ & $-0.110$ \\
$\quad$$\triangleright\;$II* & 2190 & $-19.477$ & $2.075$ & $0.044$ & $0.456$ & $0.495$ & $0.213$ & $0.005$ & $-0.109$ \\
$\quad$$\triangleright\;$IIn & 333 & $-18.849$ & $2.022$ & $0.111$ & $0.345$ & $0.490$ & $0.216$ & $0.012$ & $-0.130$ \\
$\triangleright\;$SLSN & 73 & $-18.395$ & $1.727$ & $0.202$ & $1.951$ & $0.436$ & $0.265$ & $0.031$ & $-0.202$ \\
$\quad$$\triangleright\;$SLSN-I & 43 & $-17.894$ & $1.562$ & $0.238$ & $2.464$ & $0.415$ & $0.293$ & $0.045$ & $-0.223$ \\
$\quad$$\triangleright\;$SLSN-II & 27 & $-19.200$ & $1.642$ & $0.316$ & $1.197$ & $0.484$ & $0.217$ & $0.042$ & $-0.156$ \\
\multicolumn{10}{c}{} \\
\multicolumn{10}{c}{Thermonuclear supernovae} \\
\hline
Ia & 9159 & $-19.710$ & $1.839$ & $0.019$ & $0.523$ & $0.583$ & $0.224$ & $0.002$ & $-0.050$ \\
$\triangleright\;$Ia* & 8784 & $-19.723$ & $1.832$ & $0.020$ & $0.552$ & $0.583$ & $0.224$ & $0.002$ & $-0.049$ \\
$\triangleright\;$Ia-91T & 211 & $-18.959$ & $2.005$ & $0.138$ & $0.876$ & $0.506$ & $0.212$ & $0.015$ & $-0.116$ \\
$\triangleright\;$Ia-91bg & 79 & $-20.291$ & $1.690$ & $0.190$ & $-1.872$ & $0.735$ & $0.125$ & $0.014$ & $0.131$ \\
$\triangleright\;$Ia-09dc & 4 & $-19.662$ & $1.257$ & $0.628$ & $0.097$ & $0.638$ & $0.183$ & $0.092$ & $0.018$ \\
$\triangleright\;$Ia-02cx & 25 & $-19.022$ & $1.714$ & $0.343$ & $-0.742$ & $0.488$ & $0.143$ & $0.029$ & $-0.128$ \\
$\triangleright\;$Ia-00cx & 3 & $-20.535$ & $0.662$ & $0.382$ & $-1.948$ & $0.794$ & $0.122$ & $0.070$ & $0.158$ \\
$\triangleright\;$Ia-99aa & 5 & $-20.186$ & $2.282$ & $1.021$ & $-2.241$ & $0.660$ & $0.132$ & $0.059$ & $0.040$ \\
$\triangleright\;$Ia-02es & 5 & $-20.758$ & $0.779$ & $0.348$ & $-3.887$ & $0.724$ & $0.105$ & $0.047$ & $0.149$ \\
$\triangleright\;$Ia CSM & 13 & $-19.087$ & $1.506$ & $0.418$ & $0.854$ & $0.485$ & $0.143$ & $0.040$ & $-0.133$ \\
$\triangleright\;$Ca-rich & 11 & $-19.353$ & $1.984$ & $0.598$ & $-1.654$ & $0.676$ & $0.127$ & $0.038$ & $0.085$ \\
\hline
\end{tabular}
\begin{tablenotes}
\item Columns: $\mu$, $\sigma$, SEM: mean, standard deviation, and standard error of the mean; $\Delta_{\text{L}}$: mean offsets with respect to the average stellar light baseline.
\end{tablenotes}
\end{threeparttable}
\end{table*}
% \end{landscape}

% \begin{landscape}
\begin{table*}
\caption{Subtype-wise difference of host spectroscopic properties.}
\begin{threeparttable}
\label{tab:typewisetablespec}
\begin{tabular}{@{\extracolsep{4pt}}rlccrrrr@{}}
\hline
\multicolumn{2}{c}{Subtypes} & $t^2_{\text{MS}}$ & $t^2_{\text{MZ}}$ & \multicolumn{1}{c}{$\Delta\log M$} & \multicolumn{1}{c}{$\Delta\log \text{SFR}$} & \multicolumn{1}{c}{$\Delta\log \text{sSFR}$} & \multicolumn{1}{c}{$\Delta$ (12+[O/H])} \\
\hline
\multicolumn{8}{c}{} \\
\multicolumn{8}{c}{SFR-weighted reference sample (average massive star baseline)} \\
\hline
Ib* & Ic* & $\times$ & $\times$ & ${0.030\pm0.097}\,\times$ & ${0.014\pm0.097}\,\times$ & ${-0.017\pm0.098}\,\times$ & ${0.035\pm0.042}\,\times$ \\
Ib* & II* & $\checkmark$ & $\checkmark$ & ${0.252\pm0.066}\,\checkmark$ & ${0.118\pm0.069}\,\times$ & ${-0.143\pm0.074}\,\times$ & ${0.162\pm0.029}\,\checkmark$ \\
Ic* & II* & $\checkmark$ & $\checkmark$ & ${0.222\pm0.080}\,\checkmark$ & ${0.104\pm0.076}\,\times$ & ${-0.125\pm0.078}\,\times$ & ${0.126\pm0.035}\,\checkmark$ \\
Ib* & IIb & $\times$ & $\times$ & ${0.116\pm0.102}\,\times$ & ${0.163\pm0.097}\,\times$ & ${0.014\pm0.117}\,\times$ & ${0.082\pm0.044}\,\times$ \\
Ib* & Ibn & $\checkmark$ & $\checkmark$ & ${0.372\pm0.428}\,\times$ & ${0.606\pm0.342}\,\times$ & ${0.238\pm0.323}\,\times$ & ${0.283\pm0.082}\,\checkmark$ \\
Ic* & Ic BL & $\times$ & $\times$ & ${0.225\pm0.159}\,\times$ & ${0.093\pm0.124}\,\times$ & ${-0.127\pm0.186}\,\times$ & ${0.108\pm0.067}\,\times$ \\
II* & IIn & $\times$ & $\times$ & ${-0.146\pm0.076}\,\times$ & ${0.006\pm0.071}\,\times$ & ${0.127\pm0.081}\,\times$ & ${-0.080\pm0.030}\,\checkmark$ \\
Ia-91T & Ia-91bg & $\checkmark$ & $\times$ & ${-0.784\pm0.117}\,\checkmark$ & ${0.510\pm0.144}\,\checkmark$ & ${1.283\pm0.189}\,\checkmark$ & ${-0.104\pm0.113}\,\times$ \\
Ia-91T & Ia-02cx & $\times$ & $\times$ & ${0.118\pm0.280}\,\times$ & ${-0.200\pm0.194}\,\times$ & ${-0.330\pm0.314}\,\times$ & ${-0.063\pm0.108}\,\times$ \\
Ia-91T & Ia CSM & $\times$ & $\times$ & ${0.202\pm0.214}\,\times$ & ${-0.214\pm0.100}\,\times$ & ${-0.422\pm0.254}\,\times$ & ${-0.064\pm0.041}\,\times$ \\
Ia-91T & Ca-rich & $\checkmark$ & $\times$ & ${-0.874\pm0.138}\,\checkmark$ & ${0.280\pm0.167}\,\times$ & ${1.145\pm0.245}\,\checkmark$ &  \\
Ia-91bg & Ia-02cx & $\checkmark$ & $\times$ & ${0.901\pm0.286}\,\checkmark$ & ${-0.710\pm0.207}\,\checkmark$ & ${-1.612\pm0.332}\,\checkmark$ & ${0.041\pm0.147}\,\times$ \\
Ia-91bg & Ia CSM & $\checkmark$ & $\times$ & ${0.986\pm0.221}\,\checkmark$ & ${-0.725\pm0.123}\,\checkmark$ & ${-1.705\pm0.276}\,\checkmark$ & ${0.039\pm0.108}\,\times$ \\
Ia-91bg & Ca-rich & $\times$ & $\times$ & ${-0.090\pm0.149}\,\times$ & ${-0.230\pm0.182}\,\times$ & ${-0.138\pm0.268}\,\times$ &  \\
Ia-02cx & Ia CSM & $\times$ & $\times$ & ${0.084\pm0.337}\,\times$ & ${-0.015\pm0.178}\,\times$ & ${-0.092\pm0.373}\,\times$ & ${-0.002\pm0.103}\,\times$ \\
Ia-02cx & Ca-rich & $\checkmark$ & $\times$ & ${-0.991\pm0.295}\,\checkmark$ & ${0.480\pm0.223}\,\times$ & ${1.474\pm0.367}\,\checkmark$ &  \\
Ia CSM & Ca-rich & $\checkmark$ & $\times$ & ${-1.076\pm0.233}\,\checkmark$ & ${0.495\pm0.149}\,\checkmark$ & ${1.566\pm0.317}\,\checkmark$ &  \\
\multicolumn{8}{c}{} \\
\multicolumn{8}{c}{Mass-weighted reference sample (average star baseline)} \\
\hline
Ib* & Ic* & $\times$ & $\times$ & ${0.032\pm0.097}\,\times$ & ${0.022\pm0.097}\,\times$ & ${-0.028\pm0.098}\,\times$ & ${0.035\pm0.042}\,\times$ \\
Ib* & II* & $\checkmark$ & $\checkmark$ & ${0.176\pm0.066}\,\checkmark$ & ${0.019\pm0.069}\,\times$ & ${-0.185\pm0.074}\,\checkmark$ & ${0.138\pm0.029}\,\checkmark$ \\
Ic* & II* & $\times$ & $\checkmark$ & ${0.144\pm0.080}\,\times$ & ${-0.003\pm0.076}\,\times$ & ${-0.157\pm0.078}\,\times$ & ${0.103\pm0.035}\,\checkmark$ \\
Ib* & IIb & $\times$ & $\times$ & ${0.128\pm0.102}\,\times$ & ${0.187\pm0.097}\,\times$ & ${0.007\pm0.117}\,\times$ & ${0.084\pm0.044}\,\times$ \\
Ib* & Ibn & $\times$ & $\checkmark$ & ${0.286\pm0.428}\,\times$ & ${0.513\pm0.342}\,\times$ & ${0.220\pm0.323}\,\times$ & ${0.272\pm0.082}\,\checkmark$ \\
Ic* & Ic BL & $\times$ & $\times$ & ${0.188\pm0.159}\,\times$ & ${0.054\pm0.124}\,\times$ & ${-0.121\pm0.186}\,\times$ & ${0.094\pm0.067}\,\times$ \\
II* & IIn & $\times$ & $\times$ & ${-0.100\pm0.076}\,\times$ & ${0.067\pm0.071}\,\times$ & ${0.150\pm0.081}\,\times$ & ${-0.070\pm0.030}\,\checkmark$ \\
Ia-91T & Ia-91bg & $\checkmark$ & $\times$ & ${-0.625\pm0.117}\,\checkmark$ & ${0.700\pm0.144}\,\checkmark$ & ${1.323\pm0.189}\,\checkmark$ & ${-0.067\pm0.113}\,\times$ \\
Ia-91T & Ia-02cx & $\times$ & $\times$ & ${0.266\pm0.280}\,\times$ & ${-0.049\pm0.194}\,\times$ & ${-0.299\pm0.314}\,\times$ & ${-0.036\pm0.108}\,\times$ \\
Ia-91T & Ia CSM & $\times$ & $\times$ & ${0.165\pm0.214}\,\times$ & ${-0.238\pm0.100}\,\checkmark$ & ${-0.398\pm0.254}\,\times$ & ${-0.078\pm0.041}\,\times$ \\
Ia-91T & Ca-rich & $\checkmark$ & $\times$ & ${-0.639\pm0.138}\,\checkmark$ & ${0.620\pm0.167}\,\checkmark$ & ${1.272\pm0.245}\,\checkmark$ &  \\
Ia-91bg & Ia-02cx & $\checkmark$ & $\times$ & ${0.891\pm0.286}\,\checkmark$ & ${-0.749\pm0.207}\,\checkmark$ & ${-1.623\pm0.332}\,\checkmark$ & ${0.031\pm0.147}\,\times$ \\
Ia-91bg & Ia CSM & $\checkmark$ & $\times$ & ${0.790\pm0.221}\,\checkmark$ & ${-0.938\pm0.123}\,\checkmark$ & ${-1.721\pm0.276}\,\checkmark$ & ${-0.011\pm0.108}\,\times$ \\
Ia-91bg & Ca-rich & $\times$ & $\times$ & ${-0.014\pm0.149}\,\times$ & ${-0.080\pm0.182}\,\times$ & ${-0.051\pm0.268}\,\times$ &  \\
Ia-02cx & Ia CSM & $\times$ & $\times$ & ${-0.101\pm0.337}\,\times$ & ${-0.190\pm0.178}\,\times$ & ${-0.099\pm0.373}\,\times$ & ${-0.042\pm0.103}\,\times$ \\
Ia-02cx & Ca-rich & $\checkmark$ & $\times$ & ${-0.905\pm0.295}\,\checkmark$ & ${0.669\pm0.223}\,\checkmark$ & ${1.571\pm0.367}\,\checkmark$ &  \\
Ia CSM & Ca-rich & $\checkmark$ & $\times$ & ${-0.804\pm0.233}\,\checkmark$ & ${0.858\pm0.149}\,\checkmark$ & ${1.670\pm0.317}\,\checkmark$ &  \\
\hline
\end{tabular}
\begin{tablenotes}
\item Columns: \textit{subtypes}: the two subtypes compared for each column; 
$t^2_{\text{MS}}$ and $t^2_{\text{MZ}}$: the statistical significance for the difference in the mass--SFR plane and mass--metallicity plane under the Hotelling's $t^2$-test;
$\Delta \log$M, $\Delta \log$SFR, $\Delta \log$sSFR, and $\Delta$(12+[O/H]): the difference (the first subtype minus the second subtype) in the corresponding host properties, with the statistical significance under the student's $t$-test indicated;
Tick symbols represent statistically significant ($2\sigma$ level) differences, while cross symbols represent insignificant differences.\\
This table is available in its entirety in machine-readable form.
\end{tablenotes}
\end{threeparttable}
\end{table*}
% \end{landscape}
% \begin{landscape}
\begin{table}
\caption{Subtype-wise difference of host photometric properties.}
\begin{threeparttable}
\label{tab:typewisetablephot}
\begin{tabular}{rlcrr}
\hline
\multicolumn{2}{c}{Subtypes} & $t^2$ & \multicolumn{1}{c}{$\Delta M_r$} & \multicolumn{1}{c}{$\Delta (M_g - M_r)$} \\
\hline
\multicolumn{5}{c}{} \\
\multicolumn{5}{c}{Luminosity-weighted reference sample (stellar light baseline)} \\
\hline
Ib* & Ic* & $\times$ & ${0.149\pm0.198}\,\times$ & ${0.031\pm0.017}\,\times$ \\
Ib* & II* & $\checkmark$ & ${0.004\pm0.172}\,\times$ & ${0.065\pm0.014}\,\checkmark$ \\
Ic* & II* & $\checkmark$ & ${-0.145\pm0.115}\,\times$ & ${0.034\pm0.012}\,\checkmark$ \\
Ib* & IIb & $\times$ & ${-0.022\pm0.220}\,\times$ & ${0.029\pm0.021}\,\times$ \\
Ib* & Ibn & $\times$ & ${-0.049\pm0.458}\,\times$ & ${0.089\pm0.040}\,\checkmark$ \\
Ic* & Ic BL & $\checkmark$ & ${-0.929\pm0.237}\,\checkmark$ & ${0.088\pm0.029}\,\checkmark$ \\
II* & IIn & $\times$ & ${-0.200\pm0.121}\,\times$ & ${0.001\pm0.013}\,\times$ \\
SLSN-I & SLSN-II & $\checkmark$ & ${1.306\pm0.396}\,\checkmark$ & ${-0.069\pm0.061}\,\times$ \\
Ia-91T & Ia-91bg & $\checkmark$ & ${1.502\pm0.235}\,\checkmark$ & ${-0.236\pm0.020}\,\checkmark$ \\
Ia-91T & Ia-02cx & $\times$ & ${0.062\pm0.370}\,\times$ & ${0.019\pm0.032}\,\times$ \\
Ia-91T & Ia CSM & $\times$ & ${0.198\pm0.440}\,\times$ & ${0.021\pm0.042}\,\times$ \\
Ia-91T & Ca-rich & $\checkmark$ & ${0.577\pm0.614}\,\times$ & ${-0.176\pm0.041}\,\checkmark$ \\
Ia-91bg & Ia-02cx & $\checkmark$ & ${-1.440\pm0.392}\,\checkmark$ & ${0.255\pm0.032}\,\checkmark$ \\
Ia-91bg & Ia CSM & $\checkmark$ & ${-1.304\pm0.459}\,\checkmark$ & ${0.257\pm0.042}\,\checkmark$ \\
Ia-91bg & Ca-rich & $\times$ & ${-0.925\pm0.628}\,\times$ & ${0.060\pm0.041}\,\times$ \\
Ia-02cx & Ia CSM & $\times$ & ${0.137\pm0.540}\,\times$ & ${0.002\pm0.049}\,\times$ \\
Ia-02cx & Ca-rich & $\checkmark$ & ${0.515\pm0.689}\,\times$ & ${-0.195\pm0.048}\,\checkmark$ \\
Ia CSM & Ca-rich & $\checkmark$ & ${0.379\pm0.730}\,\times$ & ${-0.197\pm0.055}\,\checkmark$ \\
\hline
\end{tabular}
\begin{tablenotes}
\item Columns are similar as in Table \ref{tab:typewisetablespec}, but comparing host photometric properties instead;
$t^2$ indicates the statistical significance of the difference in the ($M_r$, $M_g-M_r$) plane.\\
This table is available in its entirety in machine-readable form.
\end{tablenotes}
\end{threeparttable}
\end{table}
% \end{landscape}
\begin{landscape}
\begin{table}
\caption{Summary of Host Properties for Core-collapse Supernovae}
\begin{threeparttable}
\label{tab:summarytablecc}
\begin{tabular}{@{\extracolsep{4pt}}p{0.05\linewidth}p{0.21\linewidth}p{0.27\linewidth}p{0.27\linewidth}@{}}
\hline
Type    & SN signature           & Host properties & Rate Dependencies \\
\hline

Ib* &
H-deficient CC SN with He signatures. &
More massive and metal-rich than SN II* hosts; also redder and more luminous than SN II* hosts. &
Rate not proportional to host SFR; higher SN production efficiency in metal-rich hosts. \\

IIb &
Similar to SN Ib*, but with early-time H signatures. &
Broadly consistent with Ib hosts in spectroscopic and photometric properties; redder color than SN II* hosts. &
Similar to Ib*, except for weak or no metallicity dependence in SN production efficiency. \\

Ibn &
Similar to SN Ib*, with narrow He emission lines due to circumstellar interaction (CSI). &
Bluer than SN Ib* hosts; metal-poor compared to SN Ib* and Ic* hosts. &
(Limited sample size.) \\

Ic* &
H-deficient CC SN without He signatures. &
Indistinguishable from SN Ib* hosts in spectroscopic and photometric properties; similar to SN Ib* when compared to SN II* hosts. &
Similar to Ib*. \\

Ic-BL &
Similar to SN Ic*, but with high-velocity spectral signatures, sometimes associated with LGRBs. &
Bluer color and lower luminosity compared to SN Ib*, IIb, and Ic* hosts; metal-poor compared to hosts of SN Ib*, and possibly, SN Ic*. &
(Limited sample size.) \\

II* &
H-rich CC SN. &
Mostly star-forming galaxies. &
Rates not proportional to SFR; weak or no metallicity dependence in SN production efficiency. \\

IIn &
SN II with strong and narrow H emission lines due to CSI. &
Broadly consistent with SN II* hosts in mass, SFR, color, and luminosity; more metal-rich compared to SN II* hosts. &
Similar to SN II*. \\

SLSN-I &
H-deficient CC SN with high peak luminosity. &
Lower luminosity than most CC SN subtypes, bluer than SN Ib*, IIb, and Ic* hosts. &
(Limited sample size.) \\

SLSN-II &
H-rich CC SN with high peak luminosity. &
Broadly consistent in luminosity and color with most CC SN subtypes, except for SLSN-I. &
(Limited sample size.) \\

\hline
\end{tabular}
\begin{tablenotes}
\item Comparison is primarily made with SN II*. We focus on spectroscopic properties such as host mass, SFR, and metallicity when available.
\end{tablenotes}
\end{threeparttable}
\end{table}
\end{landscape}

\begin{landscape}
\begin{table}
\caption{Summary of Host Properties for Thermonuclear Supernovae.}
\begin{threeparttable}
\label{tab:summarytableia}
\begin{tabular}{@{\extracolsep{4pt}}p{0.05\linewidth}p{0.25\linewidth}p{0.25\linewidth}p{0.25\linewidth}@{}}
\hline
Type    & SN signature           & Host properties & Rate Dependencies \\
\hline

Ia* &
Thermonuclear explosions of WDs. &
Both star-forming and quiescent galaxies. &
Rates proportional to neither SFR nor mass. \\

Ia-91T &
High luminosity, weak intermediate-mass element (IME) lines but strong Fe III lines around the peak due to high ionizing temperatures. &
Mostly star-forming galaxies, similar to CC SN hosts; higher (s)SFR, lower mass and luminosity, and bluer color compared to SN Ia hosts. &
Rates closely follow, but are not proportional to, SFR. \\

Ia-91bg &
Low luminosity, strong IME signatures and Ti II lines due to low ionizing temperatures. &
Mostly massive and quiescent galaxies; higher mass and luminosity, lower (s)SFR, and redder color than SN Ia hosts. &
Rates proportional to host mass. \\

Ia-09dc &
High luminosity, slow light curve evolution and strong C signatures, potentially super-Chandrasekhar SN Ia. &
Broadly consistent with SN Ia hosts in luminosity and color. (Limited sample size.) &
(Limited sample size.) \\

Ia-02cx &
Low luminosity, low ejecta velocity, and high ionizing temperatures. &
Mainly star-forming hosts; higher sSFR and bluer color compared to SN Ia hosts, consistent spectroscopic and photometric properties with Ia-91T hosts; lower mass, luminosity, higher (s)SFR, bluer color than Ia-91bg hosts. &
Rates proportional to SFR. \\

Ia-00cx &
Similar to Ia-91T, but with more persistent Fe III lines and higher-velocity Ca II signatures. &
Broadly consistent with SN Ia hosts in luminosity and color. (Limited sample size.) &
(Limited sample size.)  \\

Ia-99aa &
Intermediate of Ia-91T and normal Ia. &
Broadly consistent with most SN Ia subtypes in luminosity and color. (Limited sample size.) &
Rates better scale with the $z$-band luminosity. (Limited sample size.) \\

Ia-02es &
Similar to Ia-91bg in spectral signatures and luminosity, but with slower light curve evolution. &
Mostly massive and quiescent hosts; more luminous and redder hosts compared to SN Ia, Ia-91T, Ia-02cx, and Ia-CSM, but consistent in luminosity and color with Ia-91bg hosts. &
Rates better scale with the $z$-band luminosity. \\

Ia-CSM &
Strong, narrow H lines due to CSM interaction, high luminosity, and slow light curve evolution. &
Mainly star-forming hosts; higher SFR than SN Ia hosts, consistent mass and SFR with Ia-91T and Ia-02cx hosts, lower (s)SFR and mass compared to Ia-91bg and Ca-rich hosts. &
Rates closely follow, but are not proportional to, SFR. \\

Ca-rich &
Strong Ca lines at late time, low luminosity. &
Lower mass, lower sSFR, and redder color compared to SN Ia, Ia-91T, and Ia-02cx hosts; consistent with Ia-91bg hosts in mass, (s)SFR, luminosity, and color. &
Rates proportional to host mass. \\

\hline
\end{tabular}
\begin{tablenotes}
\item Comparison is primarily made with SN Ia*. We focus on spectroscopic properties such as host mass, SFR, and metallicity when available.
\end{tablenotes}
\end{threeparttable}
\end{table}
\end{landscape}

\section{Conclusions} \label{sec:Conclusions}

We investigate the host galaxy spectroscopic and photometric properties of major SN subtypes. We focus on two questions: How do host galaxy properties differ across SN subtypes? How do the observed SN rates depend on host properties?
We select SN host galaxies from the transient host database of \citet{Qin22} and assemble two samples: 
a spectroscopic sample with host stellar mass, star formation rate (SFR), and gas-phase metallicity of $4128$ SNe (Section \ref{sec:SpecProperties}), and a photometric sample with $k$-corrected absolute magnitudes and rest-frame colours in the \textit{ugriz} bands derived from SDSS photometry for $12943$ SNe (Section \ref{sec:PhotoProperties}).

To control for potential redshift-driven biases and selection effects, we resample the parent galaxy catalogues to match the redshift distribution of the observed hosts of each subtype, calculate the offsets of mean host properties with respect to the redshift-matched reference samples, and compare the offsets across SN subtypes (Section \ref{sec:zMatchedResampling}).
SNe prefer massive, high-SFR, or luminous hosts due to the greater number of progenitor stars, but dwarf galaxies are far more abundant in a cosmic volume; to eliminate this intrinsic bias and set robust baselines of comparison, the redshift-matched reference samples are weighted by the stellar mass, SFR, and luminosity of the host (Section \ref{sec:WeightedResampling}).
Therefore, the offsets of host properties are measured with respect to a hypothetical SN subtype whose observed rates are proportional to the host stellar mass, SFR, or luminosity. 
From another perspective, we compare the observed host properties to the `host galaxy properties' of an average star (mass-weighted reference samples), an average massive star (SFR-weighted reference samples), and average stellar light (luminosity-weighted reference samples) in the parent galaxy catalogue.
We tested the redshift-sampling techniques by selecting a subsample of low-redshift SNe ($z<0.05$), where the completeness of both SNe and their host galaxies is higher, and redshift-driven bias is less of a concern compared to the full sample.
We then compared some of the key results with those of the full sample. The conclusions are qualitatively the same, although some conclusions become statistically insignificant due to the reduced sample sizes.

We summarize the host galaxy spectroscopic and photometric properties of SN subtypes in Tables \ref{tab:summarytablespec}, \ref{tab:summarytablephoto}; the differences of mean host properties across subtypes under these baselines and the corresponding statistical significance are summarized in Tables \ref{tab:typewisetablespec}, \ref{tab:typewisetablephot}.

Despite the general preference for star-forming galaxies, the mean host properties of CC SN subtypes differ subtly across subtypes (i.e., confidence level above 2$\sigma$ in a $t$-test or $t^2$-test; Sections \ref{sec:CCMassSFR}, \ref{sec:CCMassMetallicity}).
Hosts of SN Ib and Ic are more massive, metal-rich, and redder compared to the hosts of SN II.
Certain SN subtypes could also have different host properties than their normal siblings.
For example, hosts of SN Ibn are bluer in colour compared to normal SN Ib. Hosts of SN Ic-BL are not just bluer, but also fainter in luminosity than hosts of normal SN Ic.
SN IIb hosts have consistent photometric and spectroscopic properties with SN Ib hosts, but they are redder than SN II hosts like SN Ib hosts, indicating a similar nature of SN Ib and IIb progenitors and a disparity of SN IIb and II progenitors.
Similarly, SN IIn hosts have consistent photometric and spectroscopic properties with SN II hosts, except that SN IIn hosts are more metal-rich than SN II hosts.
Hydrogen-deficient SLSNe reside in bluer and lower luminosity hosts than most CC SN subtypes, marking them a group of transients with extreme host properties.
In contrast, the photometric host properties of hydrogen-rich SLSNe are broadly consistent with other CC SN subtypes, except for the bluer colour than SN Ib hosts.

SNe Ia occur in both star-forming and quiescent galaxies (Section \ref{sec:IaMassSFR}). The diversity of SN Ia host properties indicates a wide dynamic range of progenitor delay times, namely the time duration from star formation to SN explosion.
SN Ia subtypes have dramatically different host properties, loosely grouped into two categories.
SN Ia subtypes including Ia-91T, Ia-02cx, and Ia-CSM, prefer star-forming hosts; other subtypes, including Ia-91bg, Ca-rich, Ia-02es, and Ia-00cx, favor quiescent hosts.
The distinct host properties across SN Ia subtypes indicate a great contrast of progenitor delay times.
SN Ia subtypes with shorter delay times tend to explode in star-forming hosts like CC SN subtypes; they disfavor quiescent galaxies due to the absence of star formation and hence progenitors therein.
SN Ia subtypes with long delay times (e.g., comparable to the Hubble timescale) are preferably seen in galaxies with old stellar ages, i.e., those most massive and quiescent ones, as predicted by the downsizing formation of galaxies.

The literature-compiled dataset includes SNe discovered by both galaxy-targeted and wide-field (or untargeted) surveys.
We find that limiting the analysis to a few wide-field surveys that are relatively unbiased with respect to host properties does not qualitatively alter the conclusions.
Any remaining biases, including those related to host galaxy extinction and the timescales of transients, might be mitigated by more uniform and complete samples from time-domain surveys with consistent target selection strategies,
such as the ZTF Census of the Local Universe Survey \citep{De20} and the Bright Transient Survey \citep{Fremling20, Perley20}, as well as the forthcoming larger Rubin/LSST transient sample, potentially including photometric classification.

To further characterize how the observed SN rates depend on host properties, we compare the mean stellar mass and SFR of observed hosts with rate-weighted, redshift-matched reference samples and test if the assumed rate dependence model reproduces the observed properties (Section \ref{sec:RateScaling}).
We consider two simple rate dependence models, one assumes that the observed rate is a bilinear function of host stellar mass and SFR ($k$-model); the other assumes that at a constant metallicity, the rate is proportional to the host SFR; meanwhile, the SN production efficiency (i.e., the number of SN explosions per unit stellar mass formed) has a power-law dependency on metallicity ($t$-model).
In complementarity to the $k$-model above, we also test if the observed rate scales with the luminosity in a specific rest-frame band by comparing the predicted mean colours with the observed.

The observed rates of CC SN subtypes are closely dependent on the host galaxy SFR, as the best-fitting coefficients of the stellar mass term ($k$-values) are close to zero (Section \ref{sec:kModelCC}); their rates also follow closer (or scale better) with the rest-frame $u$-band luminosity, compared to other bands.
However, the observed CC SN rates are not directly proportional to SFR, since the $k$-values are inconsistent with zero for all subtypes.
SN Ib and Ic also have higher $k$-values than SN II, indicating a stronger dependence of observed rates on host stellar mass.
Besides possible secondary selection effects (e.g., host dust extinction), the interpretation for the higher $k$-values of SN Ib and Ic could be multifold.
One scenario is that a fraction of SNe Ib and Ic progenitors are long-lived (e.g., longer than than the H$\alpha$ SFR indicator can trace). Such a substantially delayed population may lead to a non-negligible dependence on the host stellar mass, representing those long-lived progenitors.
Another scenario is that the observed rates of certain CC SN subtypes depend on other host properties, such as metallicity.
For the latter scenario, using a simple metallicity-dependent rate dependence model, we show that the SN production efficiency of SN II is insensitive to, or independent of, host metallicity (the power law index, or $t$-value, is consistent with zero); however, the SN production efficiencies of SN Ib and Ic are higher for metal-rich hosts, with a factor of about $10\times$ increase per dex increase in metallicity.

Similar to the dramatic contrast in their host properties, the best-fitting rate dependence model of SN Ia subtypes also shows two extremes.
On one side, the rates of SN Ia-02cx are proportional to the host SFR ($k$-value is consistent with zero); SN Ia-91T and Ia-CSM rates are closely dependent on, but not exactly proportional to, their host SFR, with $k$-values close to but inconsistent with zero.
Like CC SN subtypes, their rates follow closer, or scale better, with the rest-frame $u$-band luminosity than other bands.
On the other side, the rates of SN Ia-91bg and Ca-rich transients are proportional to the host stellar mass, showing no dependence on SFR ($k$-values consistent with one). Their rates also scale better with the rest-frame $z$-band luminosity.
Similar as the observed host properties, the best-fitting rate dependence model of SN Ia subtypes also reflects the dramatic difference in their progenitor delay times.

Finally, the observed rates of normal SN Ia do not closely (or solely) depend on either host SFR or host stellar mass ($k$-value is inconsistent with zero or one); instead, a linear combination of them can fit the rates ($k=0.30\pm0.01$).
Compared to the empirical `A+B' model of \citet{Sullivan06}, our best-fitting $k$-value of SN Ia indicates stronger dependence of rates on host SFR than stellar mass.

We discuss the different host properties across SN subtypes and the implications on their rate dependence relationships.
However, the discussion in this work is based on the central assumption that selection effects are mostly redshift-driven and can be controlled by matching the redshift distributions. The impact of other possible secondary dependent factors, such as host galaxy dust extinction, SN-host angular separation, and host axis ratio or inclination, should be assessed in future studies.
We use rest-frame colours and absolute magnitudes to proxy host galaxy physical properties. The larger photometric sample ensures robust statistics for several minor subtypes, but interpreting the photometric properties could be a non-trivial task. Performing spectral energy distribution (SED) modeling to obtain detailed physical properties could be beneficial to consolidate our conclusions.
Furthermore, we use the gleaned and standardized SN classification in the database, but the criteria and technique for classification could be inconsistent across those original reference sources. Evaluating the consistency of classification would be necessary.

The redshift-matching technique we developed could be used for similar datasets where the actual selection functions are hard to characterize but presumably driven by one dominating factor. Also, for transient surveys with known selection functions, matching the observed and resampled host properties could be a test of their rate dependence models.

\section*{Acknowledgements}

We thank the anonymous referee for suggestions that improved this paper. Nathan Smith, Peter Behroozi, and David Sand also provided helpful comments during this work. 
Funding for the Sloan Digital Sky Survey IV has been provided by the Alfred P. Sloan Foundation, the U.S. Department of Energy Office of Science, and the Participating Institutions. SDSS-IV acknowledges support and resources from the Center for High-Performance Computing at the University of Utah. The SDSS web site is www.sdss.org.

SDSS-IV is managed by the Astrophysical Research Consortium for the Participating Institutions of the SDSS Collaboration including the 
Brazilian Participation Group, the Carnegie Institution for Science, 
Carnegie Mellon University, the Chilean Participation Group, 
the French Participation Group, Harvard-Smithsonian Center for Astrophysics, 
Instituto de Astrof\'isica de Canarias, The Johns Hopkins University, 
Kavli Institute for the Physics and Mathematics of the Universe (IPMU) / 
University of Tokyo, the Korean Participation Group, Lawrence Berkeley National Laboratory, 
Leibniz Institut f\"ur Astrophysik Potsdam (AIP),  
Max-Planck-Institut f\"ur Astronomie (MPIA Heidelberg), 
Max-Planck-Institut f\"ur Astrophysik (MPA Garching), 
Max-Planck-Institut f\"ur Extraterrestrische Physik (MPE), 
National Astronomical Observatories of China, New Mexico State University, 
New York University, University of Notre Dame, 
Observat\'ario Nacional / MCTI, The Ohio State University, 
Pennsylvania State University, Shanghai Astronomical Observatory, 
United Kingdom Participation Group,
Universidad Nacional Aut\'onoma de M\'exico, University of Arizona, 
University of colourado Boulder, University of Oxford, University of Portsmouth, 
University of Utah, University of Virginia, University of Washington, University of Wisconsin, 
Vanderbilt University, and Yale University.

%%%%%%%%%%%%%%%%%%%%%%%%%%%%%%%%%%%%%%%%%%%%%%%%%%
\section*{Data Availability}

The data underlying this article are available in the article and in its online supplementary material.

%%%%%%%%%%%%%%%%%%%% REFERENCES %%%%%%%%%%%%%%%%%%

% The best way to enter references is to use BibTeX:

\bibliographystyle{mnras}
\bibliography{main} % if your bibtex file is called example.bib

% Alternatively you could enter them by hand, like this:
% This method is tedious and prone to error if you have lots of references
%\begin{thebibliography}{99}
%\bibitem[\protect\citeauthoryear{Author}{2012}]{Author2012}
%Author A.~N., 2013, Journal of Improbable Astronomy, 1, 1
%\bibitem[\protect\citeauthoryear{Others}{2013}]{Others2013}
%Others S., 2012, Journal of Interesting Stuff, 17, 198
%\end{thebibliography}

%%%%%%%%%%%%%%%%%%%%%%%%%%%%%%%%%%%%%%%%%%%%%%%%%%

%%%%%%%%%%%%%%%%% APPENDICES %%%%%%%%%%%%%%%%%%%%%

\appendix

\section{Confusion Matrix of Classification} \label{appendix:ConfusionMatrix}

\begin{figure*}
\centering
\includegraphics[width=0.625\textwidth]{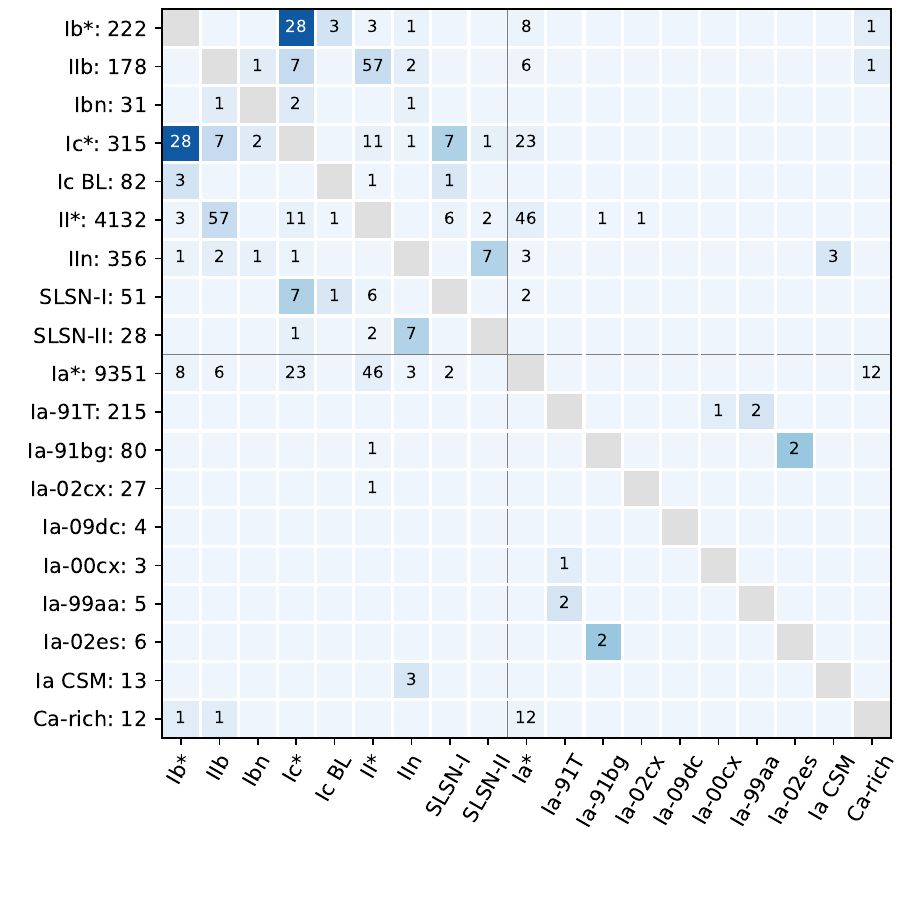}
\caption{\label{fig:TypeConfusionMatrix}
Confusion matrix of CC SN and SN Ia subtypes.
Numbers indicate the SNe under each subtype (row) that are also classified into another subtype (column) due to the inconsistency across our multiple reference sources.
Darker colours indicate a higher fraction of common records between the row and column subtypes.
Most confusion occurs between subtypes with similar spectroscopic or photometric properties.
See Appendix \ref{appendix:ConfusionMatrix} for detailed discussion.
}
\end{figure*}

SN classification in this work is compiled across various reference sources. There are sometimes multiple subtypes assigned by different reference sources to the same SN.
Rather than selecting a unique subtype for each SN, we list such multiply-classified SNe under every subtype they are classified into.
We summarize the overlap of SN subtypes, or the `confusion matrix' of SN subtypes analyzed in this work, in Figure \ref{fig:TypeConfusionMatrix}.

Most confusion of SN classification happens within CC SN subtypes. Specifically, there is a $\sim10$ per cent overlap of SN Ib* and Ic* given the challenges in distinguishing these two types in certain circumstances; similarly, about $7$ per cent SN Ic* are also classified as SN Ia*.
Nearly $30$ per cent of SN IIb are classified as SN II* simultaneously. Besides the subtle differences of these two subtypes, the overlap of SN IIb and II is also due to our choice to consider SN IIb as a subtype of SN Ib, rather than SN II.
There is also overlap of SLSN-I and SN Ic, as well as SLSN-II and SN IIn, because we consider SLSN as a parallel subtype of SE SN (including SN Ic) and SN II (including SN IIn), rather than grouping them under SE SN and SN II.
There are also overlap in SN Ia-99aa and Ia-91T, as well as SN Ia-02es and Ia-91bg, due to their similar observational signatures.

\section{Mock Parent catalogue for the Photometric Sample} \label{appendix:Mockcatalogue}

Generating reference samples requires complete redshift data in the parent catalogue. 
Most galaxies in the SDSS photometric catalogue have no spectroscopic redshift.
Photometric redshifts and the associated absolute magnitudes, although provided by multiple reference sources (e.g., \citealt{Brescia14, Beck16}), are neither accurate nor precise for the purpose here.
Given the lack of redshift data, the SDSS photometric catalogue cannot serve as the parent catalogue. 
Therefore, we choose an alternative approach to create the reference samples for our photometric sample.
We assume that the population evolution of galaxies within the redshift interval of our SN hosts is negligible ($90$ per cent hosts of the photometric sample lie under $z\sim0.35$).
We construct a \textit{volumetric} sample of galaxies, namely a representative sample of galaxies with proper weights to reflect their number density as a function of absolute magnitudes and rest-frame colours in a comoving cosmic volume.
In other words, we create a distribution function of galaxy parameterized with absolute magnitudes and colours.
We then resample this volumetric sample or distribution function of galaxies with a redshift-dependent cutoff on absolute magnitudes imposed to imitate the flux-limited selection function of the photometric sample. %

Here, we choose the SDSS MGS \citep{Strauss02}, a nearly flux-limited spectroscopic galaxy survey, to construct the volumetric sample.
First, we use the spectroscopic redshifts of galaxies, in combination with their extinction-corrected circular Petrosian magnitudes from the official flux-based matches (`\texttt{photoMatchPlate}' catalogue in DR16), to estimate their absolute magnitudes and rest-frame colours under the same k-correction procedure as described in Section \ref{sec:PhotoProperties}.
Next, we calculate the maximal visible redshift $z_{\text{max}}$ of each galaxy in this sample and the corresponding enclosed comoving volume $V_{\text{max}}$.
We estimate $z_{\text{max}}$ by finding the root of
$$M[0.67 \mu m \times (1+z)] + \mu(z) + 2.5\log(1+z) = 17.8$$
where $0.67\,\mu m$ is the central wavelength of SDSS r-band filter, $17.8$ is the extinction-corrected $r$-band limiting magnitude of SDSS spectroscopic targets, and finally, $M(\lambda_{\text{rest}})$ is the absolute magnitude as a function of rest-frame wavelength $\lambda_{\text{rest}}$ (i.e., the rest-frame spectral energy distribution of the galaxy), interpolated from the k-corrected absolute magnitudes in \textit{ugriz} bands.
\footnote{Generally, the function $M(\lambda_{\text{rest}})$ is evaluated with an on-the-fly k-correction, rather than interpolation \citep{Blanton03lumfunc}.}
Using the root of the equation, we calculate the enclosed comoving volume $V_{\text{max}}$ for the maximal visible redshift of each galaxy.
The proper weights that convert the galaxy sample into a volumetric one can be then derived using the classical $1/V_{\text{max}}$ method \citep{Schmidt68}.
Galaxies in this sample, when weighted by their $1/V_{\text{max}}$, reflect the distribution function of galaxy properties in a low-redshift comoving cosmic volume.

To resample this volumetric sample, we further estimate the \textit{photometric} maximal visible redshift of each galaxy under the limiting magnitudes of the SDSS photometric catalogue ($r\sim21.9$, based on the $50$ per cent completeness magnitude limit in \citealt{Rykoff15}), using the same root-finding procedure above.
We ignore Galactic dust extinction here because the impact is minor. Based on the dust map and extinction coefficients of \citet{Schlegel98} and \citet{Schlafly11}, the median reddening at the position of our selected host galaxies is $E(B-V)\sim0.027$, which translates to a median $r$-band Galactic extinction of only $A_r=0.06$ mag.
To resample this volumetric sample at a given redshift, we first determine which galaxies are visible at this redshift and then use the weights of these visible galaxies to draw galaxies.
We also optionally weight these remaining visible galaxies by their luminosity in a given rest-frame band so that the probability to be drawn at a fixed redshift also scales with their luminosity.
The volumetric sample technique provides a mock parent galaxy catalogue for the photometric sample, while the assumptions we made also imply some limitations.
First, we consider the SDSS MGS as an ideal flux-limited sample and use the classical $1/V_{\text{max}}$ method to construct the volumetric sample. Yet, there are other selection effects not taken into account, such as fibre collision and surface brightness cut.
The mock parent catalogue could underestimate the true number density of galaxies that prefer high-density environments (e.g., massive elliptical galaxies).
Second, we consider Galactic foreground extinction a negligible factor when deriving the photometric maximal visible redshifts; we also match redshift distribution only, rather than a joint redshift-extinction distribution.
The redshift-matched reference samples, as a result, may extend to lower luminosity limits than they should be if the observed hosts have higher Galactic extinction.
Finally, we ignore the population evolution of galaxies across the redshift range of our SNe, including their number density, luminosity function, and SED. A careful evaluation of the impact is necessary when applying this technique to a transient sample extending to higher redshifts or with better-controlled selection effects. 

%%%%%%%%%%%%%%%%%%%%%%%%%%%%%%%%%%%%%%%%%%%%%%%%%%

% Don't change these lines
\bsp	% typesetting comment
\label{lastpage}
\end{document}